\documentclass[useAMS,usenatbib]{mn2e}
\usepackage{latexsym}
\usepackage{amsmath}
\usepackage{amssymb}
\usepackage{float}
\usepackage{subcaption}
\usepackage{xcolor}
\usepackage{caption}
\usepackage{graphicx}
\usepackage{tikz,xcolor,hyperref}

\definecolor{lime}{HTML}{A6CE39}
\DeclareRobustCommand{\orcidicon}{
	\begin{tikzpicture}
	\draw[lime, fill=lime] (0,0) 
	circle [radius=0.16] 
	node[white] {{\fontfamily{qag}\selectfont \tiny ID}};
	\draw[white, fill=white] (-0.0625,0.095) 
	circle [radius=0.007];
	\end{tikzpicture}
	\hspace{-2mm}
}
\foreach \x in {A, ..., Z}{\expandafter\xdef\csname orcid\x\endcsname{\noexpand\href{https://orcid.org/\csname orcidauthor\x\endcsname}
			{\noexpand\orcidicon}}
}

\title[Core formation via filament fragmentation and the impact of ambient pressure on it]{Core formation via filament fragmentation and the impact of ambient pressure on it}
\author[S. V. Anathpindika and Di Francesco, James]{S. V. Anathpindika\orcidA{}$^{1,}$ $^{2}$\thanks{E-mail: sumed$\_$k@yahoo.co.in (SVA)} and Di Francesco, J\orcidB{}$^{3}$\\
  $^{1}$Indian Institute of Technology, Kharagpur, West Bengal, India\\
  $^{2}$University Observatory M$\ddot{\mathrm{u}}$nchen, Schneirstrasse 1, 81679, M$\ddot{\mathrm{u}}$nchen, Germany \\
  $^{3}$National Research Council of Canada, Herzberg, Astronomy \& Astrophysics Research Centre, \\
  5071 West Saanich Road, Victoria (BC), Canada V9E 2E7 \\}
\begin{document}

\date{Accepted 0000 December 00. Received 0000 December 00; in original form 0000 October 00}

\pagerange{\pageref{firstpage}--\pageref{lastpage}} \pubyear{2002}

\maketitle

\label{firstpage}

\begin{abstract}
Prestellar cores are generally spheroidal, some of which appear oblate while others appear prolate. Very few of them appear circular in projection. Little, however, is understood about the processes or the physical conditions under which prolate/oblate cores form. %Starting with the end result of realisations developed in the prequel to this work, here we study further the impact of ambient pressure on the formation of cores. 
We find that an initially sub-critical filament experiencing relatively low pressure ($\lesssim 10^{4}$ K cm$^{-3}$) forms prolate cores (i.e., those with axial ratios in excess of unity) via gradual accumulation of gas in density crests. Meanwhile, a filament that is initially transcritical and experiences pressure similar to that in the Solar neighbourhood (between $\mathrm{few}\ \times 10^{4}$ K cm$^{-3}$ -  $\mathrm{few}\ \times 10^{5}$ K cm$^{-3}$) forms oblate cores (i.e., those with axial ratios less than unity) via \emph{Jeans like} fragmentation. At higher pressure, however, fragments within the filament do not tend to survive as they rebound soon after formation. We also argue that quasi-oscillatory features of velocity gradient observed along the filament axis, and in the direction orthogonal to the axis, are integral to the filament evolution process and arise due to the growth of corrugations on its surface. The axial component of the velocity gradient, in particular, traces the gas-flow along the filament length. We therefore posit that it could be used to constrain the filament-formation mechanism. The magnitude of the respective components of velocity gradients increases with increasing external pressure.
\end{abstract}


\section{Introduction}
Prestellar cores - or hereafter, simply cores - are rightly known as the cradles of star formation as individual cores spawn stars. Morphology, i.e., the 3-D structure of cores is naturally a subject of interest as it could hold crucial clues about the dynamic processes that would have formed them. Furthermore, ground-based observations over the last four decades (e.g., Schneider \& Elmegreen 1979), and more detailed \emph{Herschel} observations in recent years  (e.g., Andr{\' e} \emph{et al.} 2010, Molinari \emph{et al.} 2010, Jackson \emph{et al.} 2010, Menschikov \emph{et al.} 2010) show that cores are often located within dense filamentary structures in Molecular Clouds ({\small MC}s), a result that further constrains models that attempt to reconcile core formation. 
\subsection{Core morphology}
While the spheroidal nature of cores is not in doubt, despite several studies over the last two decades, it has not been possible to establish conclusively the morphologies of cores, i.e., if cores are preferentially broader than their natal filaments, or if the converse is true, i.e., if cores are largely "pinched". Although it is tempting to describe broadness of a core as it being prolate, and likewise an oblate core as being pinched (i.e., core size smaller than the width of its natal filament), such descriptions would be over simplifications as the intrinsic shape of a core is essentially determined by its axial ratio (e.g. Tassis 2007). \\\\
Previously, cores identified by Myers \emph{et al.} (1991), Ryden (1996), and Curry (2000) in their respective surveys were found to be preferentially prolate. On the contrary, cores identified by Jones \emph{et al.} (2001), Jones \& Basu (2002), Goodwin et al. (2002), Tassis (2007) and Tassis \emph{et al.} (2009) were largely oblate. More recent continuum maps of filaments in the Taurus region with \emph{Herschel} data show that cores are thinner than their natal filaments (e.g. Marsh \emph{et al.} 2014). Similar findings were also reported by Palmeirim \emph{et al.} (2013). By contrast, observations of {\small L1495/B213} filaments in the Taurus region of N$_{2}$H$^{+}$ and C${}^{18}\textrm{O}$ emission revealed cores that were generally broader than their natal filaments (e.g., Hacar \emph{et al.} 2013, Tafalla \& Hacar 2015). \\\\
Interestingly, however, there is definitely a dearth of truly spherical cores, i.e., cores with a roughly circular projection on the plane of the sky. For example, only 3 cores in a sample of 27 in Lupus I analysed using \emph{Herschel} data were found to have an aspect ratio similar to that of a spherical object (Poidevin \emph{et al.} 2014). Similarly, Tritsis \emph{et al.} (2015) found only 3 approximately spherical cores out of a sample of 27 drawn from a number of nearby clouds. Determining the core radius is a naturally complex business. For example, transitions of different molecular species have different critical densities and so, molecular emission represents essentially the dynamical state of gas at those densities. Thus, while N$_{2}$H$^{+}$ is an excellent tracer of dense gas, for it freezes out only at densities upward of $\sim 10^{7}$ cm$^{-3}$ (Tielens 2005), C${}^{18}\textrm{O}$ on the other hand, traces relatively less dense gas as it freezes out at much lower densities, $\gtrsim 10^{4}$ cm$^{-3}$. Consequently, observational evidence with regard to the morphology of cores is naturally biased.  \\\\
Numerical models of core-formation are equally inconclusive about core morphology. Early magnetohydrodynamic realisations by Gammie \emph{et al.} (2003) and Li \emph{et al.} (2004) showed the formation of spheroidal, but preferentially prolate cores. Other realisations by Basu \& Ciolek (2004) and Ciolek \& Basu (2006), however, showed preferentially oblate cores. Recent work, such as that by Chen \& Ostriker (2014), demonstrates approximately simultaneous formation of filaments and cores via convergence of magnetised gas-flows, but is also similarly inconclusive over their morphological tendencies. \\\\
More recent numerical work on the subject, however, has largely concentrated on the evolution of individual filaments. For example, Gritschneder \emph{et al.} (2017) observed that initial perturbations superposed on the surface of a filament amplified with time and eventually caused it to fragment. This observation was described as \emph{geometrical fragmentation} by these authors. Heigl \emph{et al.} (2018), in particular, showed that an initially sub-critical filament ($f_{cyl}$\footnote{$f_{cyl} = \Big(\frac{M_{l}}{M_{l_{crit}}}\Big)$; $M_{l_{crit}}$ = 16.4 M$_{\odot}$ pc$^{-1}$, the critical filament linemass at 10 K.} = 0.2) evolved to form a \emph{broad} core. An initially transcritical filament ($f_{cyl}$ = 0.8), by contrast, formed a \emph{pinched} core. In the prequel to this paper (Anathpindika \& Di Francesco 2021 - hereafter Paper I), we extended the ambit of this recent work to study the impact of external pressure on filament fragmentation. We concluded that while the magnitude of external pressure bears upon the evolution of a filament, the filament is in general susceptible to a \emph{sausage - type} instability irrespective of the magnitude of external pressure. Furthermore, we observed that a filament in ambience similar to that in the Solar neighbourhood tends to form pinched fragments. At higher pressure, typically upward of $\sim 10^{6}$ K cm$^{-3}$, however, the filament disrupted. \emph{Fragments reported in Paper I were relatively large clumps that will likely fragment further to form cores and so, the question about the possible  impact of ambient conditions on core morphology is still open.} 
\subsection{Filament velocity gradients}
Typical filaments also exhibit a velocity gradient along their length as well as in the radial direction, i.e., along the direction perpendicular to the filament axis. Velocity gradients along the filament axis, for example, have been observed in the Serpens South Cluster-forming region (Kirk \emph{et al.} 2013, Friesen \emph{et al.} 2013). Similar evidence has also been found in {\small DR21}, a massive elongated clump (Csengeri \emph{et al.} 2011). These observed velocity gradients have been variously interpreted as evidence for gas flows along the filament towards filament hubs that are sites of active star-formation (e.g.,  Kirk \emph{et al.} 2013, Peretto \emph{et al.} 2014). Alternatively, these motions could be due to the projection of turbulent gas flows or indeed, converging gas flows as in the case of {\small DR21} (Csengeri \emph{et al.} 2011), or even due to oscillations within over-densities or starless cores in filaments (e.g. Hacar \emph{et al.} 2013). \\\\
Interestingly, the magnitude of the radial component of the velocity gradient is usually an order of magnitude higher than its axial component (Fern{\' a}ndez - L{\' o}pez \emph{et al.} 2014). In the Serpens Main - South region the axial component of the velocity gradient has been interpreted as being due to the motion of the entire filament within the local gravitational potential well (e.g. Dhabal \emph{et al.} 2018). More recently, Chen \emph{et al.} (2020) reported such velocity gradients in a number of velocity coherent structures in the {\small NGC1333} region in Perseus molecular cloud. Relatively sophisticated numerical models such as those developed by Chen \& Ostriker(2014) support the hypothesis that the observed velocity gradients along the length of filaments and in the direction orthogonal to their axis must represent gas flows along the axis and in the direction of gas accreted by filaments, respectively. 
\subsection{Context of this paper}
Classical analytic models such as those by Nagasawa (1987) or Inutsuka \& Miyama (1992;1997) that studied the dynamical stability of an individual filament only investigated the gravitational stability of individual filaments and their propensity to fragment to form cores. In particular, Nagasawa (1987) showed that a filament could be affected either by a \emph{sausage-type} deformation instability or a \emph{Jeans-like} compressional instability. The former type occurs when the radius of a filament is smaller than its scale height while the latter is triggered when the filament radius is greater than its scale height. In a more recent contribution by Fischera \& Martin (2012) for example, the authors argued that while infinitely long cylinders with $f_{cyl} < 1$ are susceptible to gravitational instability, those having intermediate linemass, i.e., $f_{cyl}\gtrsim $0.5, are likely to fragment to form cores via compression, with core separation typically on the order of five times the filament width. Filaments with higher linemass, i.e., $f_{cyl}\sim$ 0.9, however, are likely to be rapidly destabilised due to growth of gravitational instability, leading to fragmentation and the fragments so formed being unlikely to be massive enough to collapse further to form stars. Likewise, Motiei \emph{et al.} (2021) showed that an intermediate magnetic field can stabilise thin filaments, of the kind ubiquitously found in nearby star forming {\small MCs}. \\\\
The observed variety in core morphology and the respective components of velocity gradient along and orthogonal to filament axes begs that further investigation be made into the dynamical evolution of filaments. While extending the work reported in paper I, we are now particularly interested in studying the possible impact of external pressure on the formation of cores, and on their morphology. In addition, we will also explore the origins of gradients in the axial and radial components of velocity. The plan of this paper is as follows - the numerical method and the set-up for the proposed realisations is described in \S 2. The results are presented in \S 3 while their broader implications on the picture of core formation and their morphology is discussed in \S 4. We summarise our conclusions in \S 5.  
%=======================
\section[]{Numerical method and initial conditions}
\subsection{Numerical method}
Realisations discussed in this paper were developed using the Smoothed Particle Hydrodynamics ({\small SPH}) code {\small SEREN}. Salient features of this code and the details of gas thermodynamics were discussed in Paper I so we avoid repeating them here, suffice to add that {\small SPH} particles denser than a certain threshold ($10^{6}$ cm$^{-3}$) were replaced with \emph{sink} particles. Since it is not our extant interest to collate statistics of putative star particles, evolution of \emph{sinks} was not followed in these realisations. Calculations were therefore terminated soon after they appeared in the filament. As in Paper I, we used the prescription of adaptive smoothing length for gas particles to ensure each particle has exactly 50 neighbours to minimise inaccuracies in the estimation of density (e.g. Attwood \emph{et al.} 2007).
%This is a well tested code (Hubber \emph{et al.} 2011), and includes contribution due to artificial thermal conductivity prescribed by Price (2008) to promote mixing between {\small SPH} particles in regions that exhibit a density gradient and thus avoids problems related to resolution of discontinuities and shocks. We have also included the carbon and hydrogen chemistry described by Bate \& Keto (2015). Under the scheme gravitational potential released during contraction contributes to heating besides contributions due to cosmic rays, photoelectric heating, and heating due to the formation of molecular hydrogen. These latter are characterised by the respective  heating functions $\Gamma_{cr}$, $\Gamma_{pe}$ and $\Gamma_{H_{2},g}$. We note that the respective heating functions, $\Gamma_{cr}$ and $\Gamma_{pe}$ are held constant while $\Gamma_{H_{2},g}$ is calculated by explicitly solving for the change in hydrogen fraction at each time step. \\\\
\begin{table*}
  \centering
 \begin{minipage}{160mm}
  \caption{Physical parameters for realisations developed in this work.}
  \begin{tabular}{@{}lp{20mm}p{15mm}p{20mm}p{20mm}p{30mm}ll@{}}
  \hline
  Sr No.     &   $M_{fil}$ & $f_{cyl}$ & $n_{fil}$\footnote{Average initial density of gas in the filamentary fragment} & $\mathcal{M}_{gas}$ & $T_{{\small ICM}}$\footnote{Initial temperature of {\small ICM} particles}  &$\Big(\frac{P_{int}}{k_{B}}\Big)$\footnote{Internal pressure supporting the filament initially} & $M_{res}$\\
  &            &          &          &  &    & & \\
  \hline
  & [M$_{\odot}$] &           & [cm$^{-3}$] &  & [K] & [K cm$^{-3}$] & [M$_{\odot}$]\\
  \hline
  & \multicolumn{5}{c}{Initial gas temperature, $T_{init}$ = 10 K; $\Big(\frac{P_{ICM}}{k_{B}}\Big)$\footnote{Constant external pressure acting on the filamentary fragment} = 6.0$\cdot10^{3}$ K cm$^{-3}$} & \\
 \hline
 1 & 23.8 & 0.2 & 69  & 3.5 & 173.5 & 1.12$\cdot$10$^{4}$ & 0.012\\
 2 & 59.6 & 0.5 & 173 & 4.2 & 69.4 & 3.84$\cdot$10$^{4}$ & 0.03\\
 3 & 107.3 & 0.9 & 311 & 5.4 & 38.6 & 1.11$\cdot$10$^{5}$ & 0.05 \\
 \hline
& \multicolumn{5}{c}{Initial gas temperature, $T_{init}$ = 12 K; $\Big(\frac{P_{ICM}}{k_{B}}\Big)$ = 7.3$\cdot10^{4}$ K cm$^{-3}$} & \\
 \hline
 4 & 23.8 & 0.2 & 69 & 9.7 & 2111 & 7.8$\cdot$10$^{4}$ & 0.012\\
 5 & 59.6 & 0.5 & 173 & 7.06 & 844.6 & 1.05$\cdot$10$^{5}$ & 0.03 \\
 6 & 107.3 & 0.9 & 311 & 6.8 & 469 & 1.8$\cdot$10$^{5}$ & 0.05\\
 \hline
& \multicolumn{5}{c}{Initial gas temperature, $T_{init}$ = 85 K; $\Big(\frac{P_{ICM}}{k_{B}}\Big)$ = 6.5$\cdot10^{5}$ K cm$^{-3}$} & \\
 \hline
 7 & 169 & 0.2 & 490 & 4.6 & 2654 & 9.1$\cdot$10$^{5}$ & 0.08\\
 8 & 422 & 0.5 & 1225 & 4.6 & 1061.7 & 2.3$\cdot$10$^{6}$ & 0.2 \\
 9 & 760 & 0.9 & 2204 & 5.5 & 590 & 5.91$\cdot$10$^{6}$ & 0.38\\
 \hline
& \multicolumn{5}{c}{Initial gas temperature, $T_{init}$ = 387 K; $\Big(\frac{P_{ICM}}{k_{B}}\Big)$ = 2.2$\cdot10^{6}$ K cm$^{-3}$} & \\
\hline
 10 & 777 & 0.9 & 2256 & 6.2 & 590 & 8.1$\cdot$10$^{6}$ & 0.39\\
\hline
  \end{tabular}
 \end{minipage}
\end{table*}
%Factors contributing to gas cooling on the other hand, are emissions due to electron recombination, singly ionised oxygen and the carbon fine-structure cooling characterised by the respective cooling functions $\Lambda_{rec}$, $\Lambda_{OI}$ and $\Lambda_{C^{+}}$. We also include cooling due to molecular line emission characterised by the cooling function, $\Lambda_{line}$. The thermal interaction between gas and dust is accounted for by including the rate for gas-dust interaction, $\Lambda_{gd}$, which may contribute to net heating or cooling depending on whether the dust is warmer or cooler than the gas, respectively. Equations of thermal balance for gas and dust are then iteratively solved to calculate the respective equilibrium temperature. 
\begin{figure}
\label{Fig. 1}
  \vspace*{1pt}
  \centering
   \includegraphics[angle=270,width=0.48\textwidth]{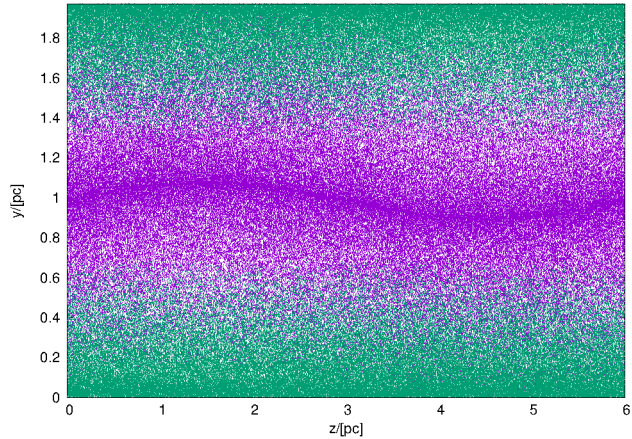}
 \caption{%\emph{Upper-panel :} Cartoon showing the schematic set-up of the fragment. Numerous elliptical features and the central circular region represent the density distribution within a typical fragment as was seen in paper I. \emph{Lower-panel :} 
Cross - section through the mid-plane of the assembled section of a filament and the box of {\small ICM} particles. Purple dots represent the gas particles assembled in the fragment while the green dots represent the {\small ICM} particles.}
\end{figure}
\subsection{Initial conditions}
In Paper I we witnessed the propensity of a filament to the \emph{sausage-type} instability irrespective of the magnitude of external pressure. In this work we therefore consider just a section of this filament. We note, however, that this sinusoidal section was not extracted from the fragmented filament in the earlier work, but was assembled from scratch by randomly positioning particles. Specifically, we wish to explore here whether the global contraction of a filament dominates the growth of axial perturbations. So a sinusoidal perturbation was initially superposed on the fragment. That is, with the filament axis aligned along the $z$ - direction, a sinusoidal perturbation was applied along the $y$ - direction. The perturbed $y$ - coordinate is given as -
\begin{equation}
  y'_{i} = y_{i} + \frac{r_{fil}A}{\pi}\sin\Big(\frac{2\pi z_{i}}{L_{fill}}\Big),
\end{equation}
where $A$ = 0.1 is the amplitude of density perturbation, $r_{fil}$ is the filament radius and, $L_{fil}$ is its length. The subscript $i$ refers to each gas particle. The $x$ and $z$ coordinates of the fragment are obtained in the usual fashion for cylindrical geometry. We note that particle positions are symmetric across the $y-z$ mid-plane. The assembled filamentary fragment is then placed in a periodic box of particles representing the intercloud medium ({\small ICM}). The number of {\small ICM} particles is calculated such that the particle number density is approximately similar across the gas - {\small ICM} interface. Fig. 1 shows a cross - section of the assembled filament along with the box of {\small ICM} particles jacketing it. Periodic boundary conditions were applied in all three directions. \\\\
With similar initial set-ups, authors in the past have experimented with both periodic boundary conditions (e.g., Gritschneder \emph{et al.} 2017) and semi-periodic conditions, i.e., open boundaries in directions orthogonal to the filament axis (e.g., Heigl \emph{et al.} 2016, Heigl \emph{et al.} 2018). Interestingly, the observed filament evolution in all these works was qualitatively similar. Furthermore, in Paper I we similarly adopted periodic boundary conditions and found that the observed filament evolution was also consistent with that reported in these previous works. While periodic boundaries are likely to prevent the edge effects expected in a cylindrical geometry, we need not be concerned about them here for we are primarily studying the fragmentation of the filament itself. In the interest of maintaining consistency with our choice of initial conditions, we retain periodic boundary conditions in this work.\\\\
We define the thickness of the envelope of {\small ICM} particles in terms of a parameter $\eta$, the integer multiple of the average initial smoothing length, $h_{avg}$; $\eta$ = 3 for all realisations. The average initial smoothing length, $h_{avg}$, is defined in the usual manner -
\begin{equation}
h_{avg} = \Big(\frac{3 r_{fil}^{2}L_{fil}}{32\pi N_{gas}}\Big)^{\frac{1}{3}},
\end{equation}  
where $N_{gas} = 2\cdot10^{5}$, i.e, the number of gas particles that represent the fragment. The initial radius and length of the fragment are respectively $r_{fil}$ = 0.9 pc and $L_{fil}$ = 6 pc. We used $\sim$350,000 particles to represent the {\small ICM} in the box enclosing the fragment. The minimum gas temperature achievable in these realisations is 7 K. This choice of $N_{gas}$ corresponds to an average smoothing length $\sim$0.009 pc, or equivalently a spatial resolution $\sim$0.018 pc (i.e. 2$h_{avg}$) which is good enough to resolve individual cores. The maximum resolvable density, $n_{max}$, is calculated using the Hubber criterion (Hubber \emph{et al.} 2006), which is the {\small SPH} equivalent of the Truelove criterion for optimal resolution in grid-based codes. With the current choice of $N_{gas}$, the maximum resolvable density, $n_{max}\sim 10^{6}$ cm$^{-3}$ at 7 K. \\\\
The initial density contrast between the average gas density and the {\small ICM} is assumed to be 10 in all realisations. Gas within the fragment is initially supported by a combination of thermal pressure and pressure due to turbulence. The initial gas turbulence within the fragment is characterised by the Mach number, $\mathcal{M}_{gas}$. We require approximate pressure balance at the gas-{\small ICM} interface, so if $P_{ICM}$ represents the pressure due to the {\small ICM} then,
\begin{eqnarray}
  P_{int} \equiv (\mu m_{H})n_{frag}a_{gas}^{2}(1 + \mathcal{M}_{gas}^{2}) = \\ \nonumber
     n_{{\small ICM}}(a_{{\small ICM}}^{2} + \sigma_{{\small ICM}}^{2}) + P_{grav},
\end{eqnarray}
where $\mu$ = 2.3 is the mean molecular mass of the gas in the fragment, and $m_{H}$ is the mass of a proton; $a_{gas}$ and $a_{{\small ICM}}$ are respectively the sound speeds for gas in the fragment and the {\small ICM}. Pressure due to self gravity of the fragment is denoted by $P_{grav}\equiv\frac{G M_{fil}^{2}}{\pi r_{fil}^{4}}$. The initial velocity dispersion for the {\small ICM} is obtained by inverting the above equation.\\\\
%Equation 3 reflects the fact that the filamentary fragment is set-up such that the internal pressure (combination of thermal pressure and pressure due to turbulence) within it balances the pressure due to the {\small ICM} and the pressure due to the self-gravity of the filamentary fragment. 
As in Paper I, the respective turbulent velocity fields in the filamentary fragment and the {\small ICM} were modelled with a relatively steep power spectrum ($\sim k^{-4}$). Listed in Table 1 are the various parameters used to develop the realisations discussed here. This choice of parameters spans a range of pressure between a few times $10^{3}$ K cm$^{-3}$ to a few times $\sim 10^{6}$ K cm$^{-3}$, as is typically reported in the outer regions of a disk galaxy and in the Solar neighbourhood, respectively. The disruption of a filament at pressures upward of $\sim 10^{6}$ K cm$^{-3}$ has already been demonstrated in Paper I. We have therefore avoided developing a realisation for such an extreme choice of external pressure. Observe, however, that our physical parameter space covers a fairly wide range of physical conditions as it spans three orders of magnitude in external pressure and roughly two orders of magnitude in the initial average gas density of the filamentary fragment. \\\\
The minimum resolvable mass, i.e., the smallest possible fragment mass in a realisation is given as, $M_{res} = M_{fil}\Big(\frac{2N_{neibs}}{N_{gas}}\Big)$ (Bate \& Burkert 1997), where $N_{neibs}$ = 50, is the number of neighbours in an {\small SPH} realization. As can be seen from Col. 8 in Table 1, $M_{res}$ in this work varies between 0.01 $M_{\odot}$ - 0.4 $M_{\odot}$ which is considerably smaller than the Jeans mass for our choice of minimum temperature and the maximum resolvable density in Cases 1 - 8, and comparable to it in Cases 9 - 10, so that we can be sure about the physical nature of observed filament evolution. Finally, to test numerical convergence, the set of realisations listed in Table 1 was repeated with another choice of a random number seed used to generate the initial turbulent velocity field (see \S 3.7 below). 
\section{Results}
\subsection{Evolution of the filamentary fragment}
%We briefly revisit the global evolution of the sinusoidal fragment as the evolution of a filament was discussed at length in paper I. 
Various panels of Fig. 2 show the rendered density images of our Case 1 and 3 models. Like the one discussed in Paper I, the filaments in these respective cases also initially contract in the radial direction and acquire a thin filamentary structure ($t=0.9$ Myrs). Then having acquired a centrally peaked density-profile these thin filaments rebound. The accompanying animation shows the entire sequence of evolution of the filament in Case 1, a few snapshots of which have been shown on panels in Fig. 2(a). The cycle of contraction and expansion of the filament repeats several times over the course of its evolution and acquires a centrally peaked density profile at least one more time (at $t=6.5\ \mathrm{Myr}$). \\\\
\begin{figure*}
\label{Fig. 2}
  \begin{subfigure}{160mm}
   \vspace*{1pt}   
   \includegraphics[angle=270,width=\textwidth]{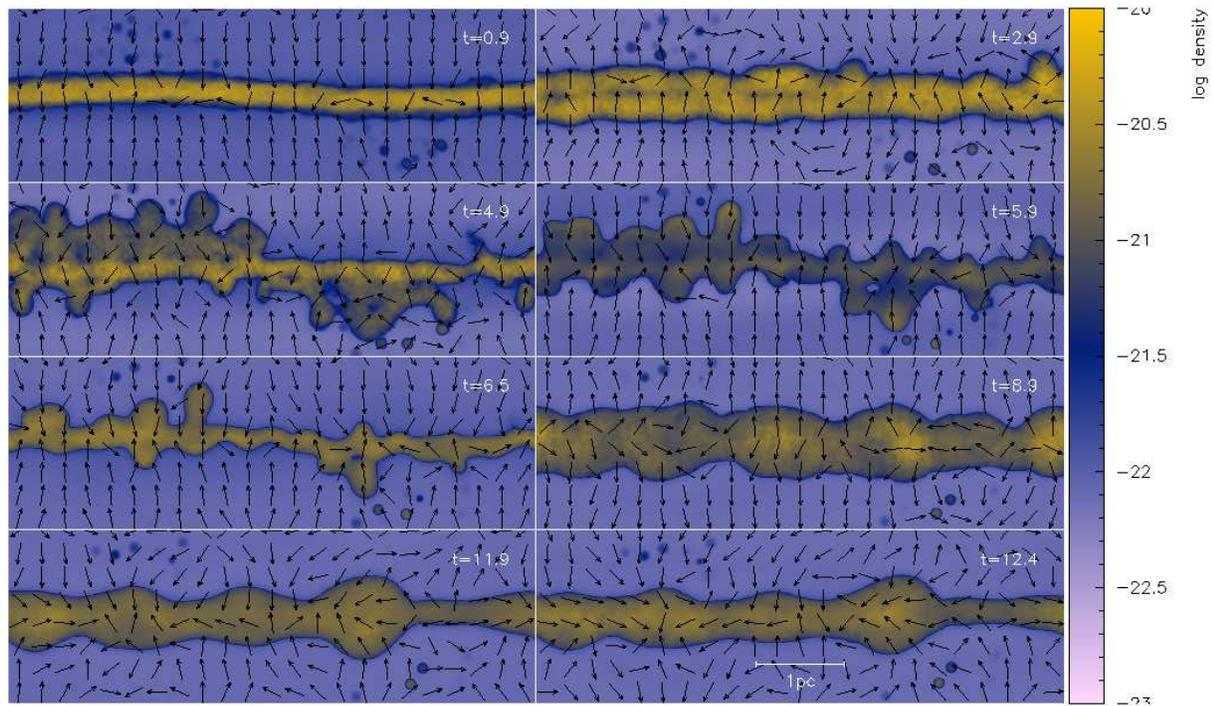}
   \caption{Case 1. }
  \end{subfigure}
  \begin{subfigure}{160mm}
    \vspace*{1pt}
    \includegraphics[angle=270,width=\textwidth]{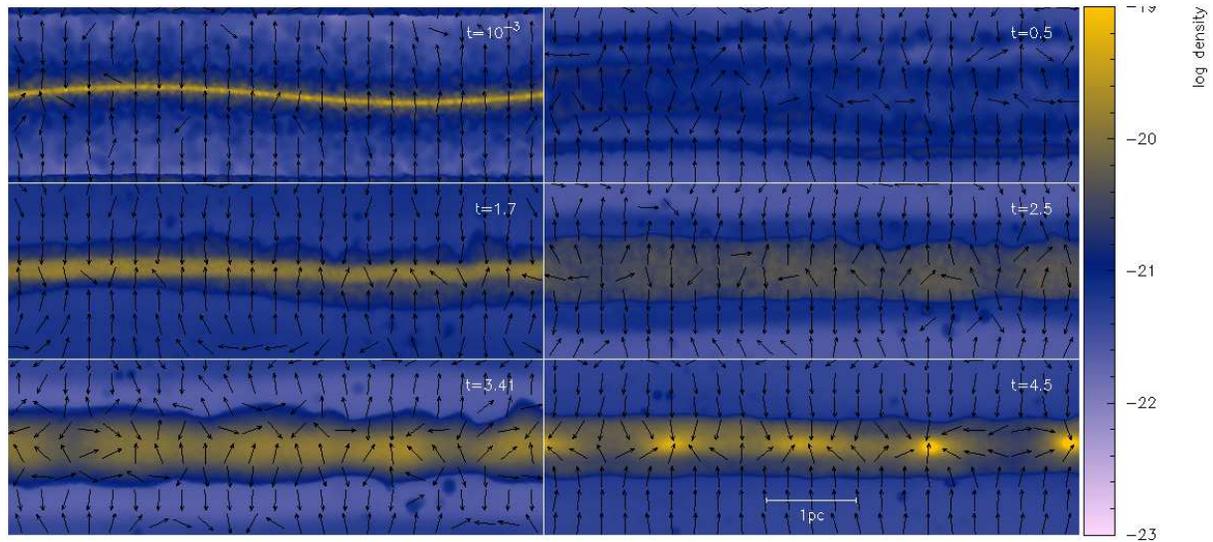}
    \caption{Case 3.}
  \end{subfigure}
  \caption{Temporal sequence of rendered density images showing the cross-section through the mid-plane of the filament. Density is marked in units of g cm$^{-3}$. Time in units of Myrs has been marked in the top right-hand corner of each panel. Note that since particles in the filament were originally arranged in concentric layers, particles close to the central axis naturally move towards the central axis on a timescale shorter than that for particles further away. Particles at the farthest extreme therefore lag behind and appear as tiny blobs in the vicinity of the filament in these rendered images. This is more readily visible in the associated animation for Case 1. Arrows overlaid on these rendered images represent the direction of local gas flow.} %\emph{Both these are low resolution rendered images, presently supplied in compliance with the submission requirements of this journal.}}
\end{figure*}
As is evident from these rendered images, fragmentation leading to the formation of individual cores is not as straightforward as suggested by analytic models (e.g. Inutsuka \& Miyama  1992). Instead, the oscillatory cycle of contraction and expansion is also associated with the growth of perturbations on the surface of the thin filament which further destabilise it. Growth of these perturbations is associated with transfer of momentum as gas flows into the crests and out of the troughs of these corrugations. This process is analogous to the classic features of the thin shell instability (e.g. Vishniac 1983; Anathpindika 2011a). Clearly, the filament does not simply break-up into fragments. Instead, spheroidal cores are assembled in regions where gas accumulates and in fact, the cores so assembled are typically broader than their natal filament. \\\\
Thus, in the present instance, i.e., in Case 1, we witness the formation of cores via what may be described as the \emph{Collect - and - Collapse} mode. Such behaviour is akin to that described by Wuensch \& Palou$\breve{\mathrm{s}}$ (2001) for fragments condensing out of shells of dense molecular gas driven by energetic supernova or ionising radiation from massive stars (e.g., Anathpindika 2011b). By borrowing this terminology here, we associate the temporal growth of local perturbations with a steady accumulation of mass in the density crests. Indeed, features of this mode of core-assembly are visible in the panels corresponding to $t$ = 8.9 Myr, 11.9 Myr and 12.4 Myr of Fig. 2(a). A core so assembled will collapse only after it has become sufficiently massive. In this work, however, we are not concerned with the statistical details of such collapsing objects. \\\\
\begin{figure*}
\label{Fig. 3}
  \vspace*{1pt}
  \includegraphics[angle=270,width=\textwidth]{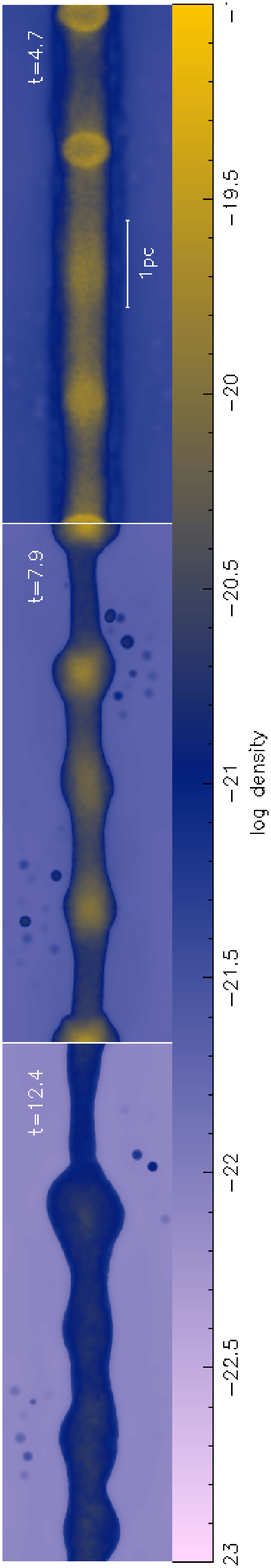}
  \caption{Similar to images shown on various panels of Figs. 2(a) and 2(b), but panels from left to right here show only the epoch when realisations were terminated in respectively Cases 1, 2 and 3. Notice the difference in core morphology and the difference in the timescale.}
\end{figure*}
Now the length of the fastest growing disturbance in a filament is
\begin{equation}
  \lambda_{frag} = \frac{22.1 a_{0}}{(4\pi G\rho_{c})^{1/2}},
\end{equation}
where $a_{0}$ and $\rho_{c}$ are respectively the average isothermal sound speed, and the average central density of the filament. $G$ is the gravitational constant. From this expression it follows that the fragmentation timescale is,
\begin{equation}
  t_{frag} = \frac{22.1}{(4\pi G\rho_{c})^{1/2}}.
\end{equation}
For Case 1, this timescale turns out to be $\sim$8 Myrs which is consistent with the epoch when condensations appear in the filament (panel corresponding to $t\sim$8.9 Myrs in Fig. 2(a)). It would also be instructive to compare the observed epoch of core-formation with the so-called \emph{e-folding} time, i.e., the timescale of growth of over densities, given as -
\begin{equation}
t_{e-fold} \sim 2.95(4\pi G\rho_{c})^{-1/2}
\end{equation}
(Nagasawa 1987), which in the present case turns out to be $\sim$1.05 Myrs. As can be seen, $t_{e-fold}$ is significantly shorter than $t_{frag}$. The significant difference between these respective timescales suggests, the \emph{collect-and-collapse} mode of core formation is fundamentally different from the mode in which cores condense via the rapid growth of over densities. We will further explore this point with other realisations developed with a higher external pressure and a higher initial average gas density.\\\\
\begin{figure*}
\label{Fig. 4}
  \begin{subfigure}{160mm}
   \vspace*{1pt}   
   \includegraphics[angle=270,width=\textwidth]{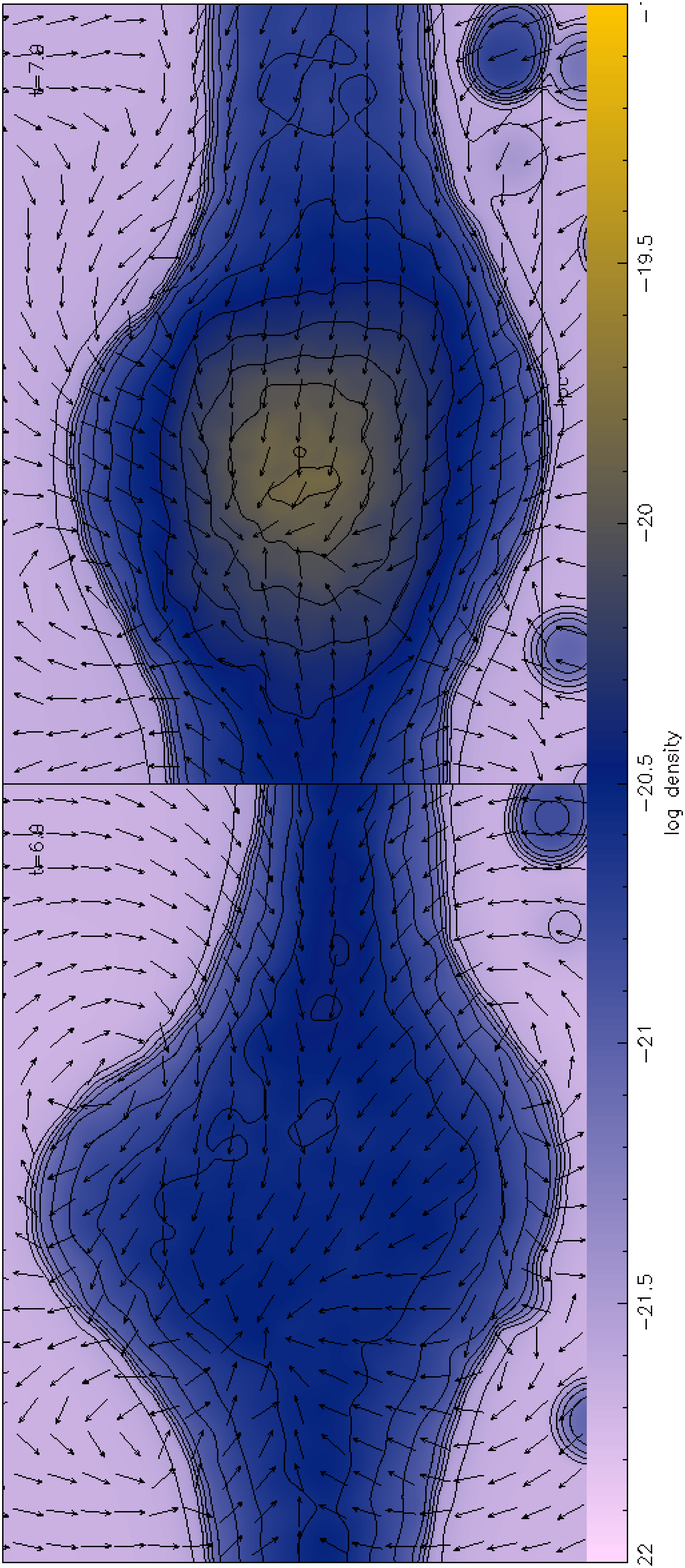}
   \caption{Case 2 ($t=$ 6.9 Myrs, $t$ = 7.9 Myrs for pictures on respectively the left and the right hand panel.). }
  \end{subfigure}
  \begin{subfigure}{160mm}
    \vspace*{1pt}
    \includegraphics[angle=270,width=\textwidth]{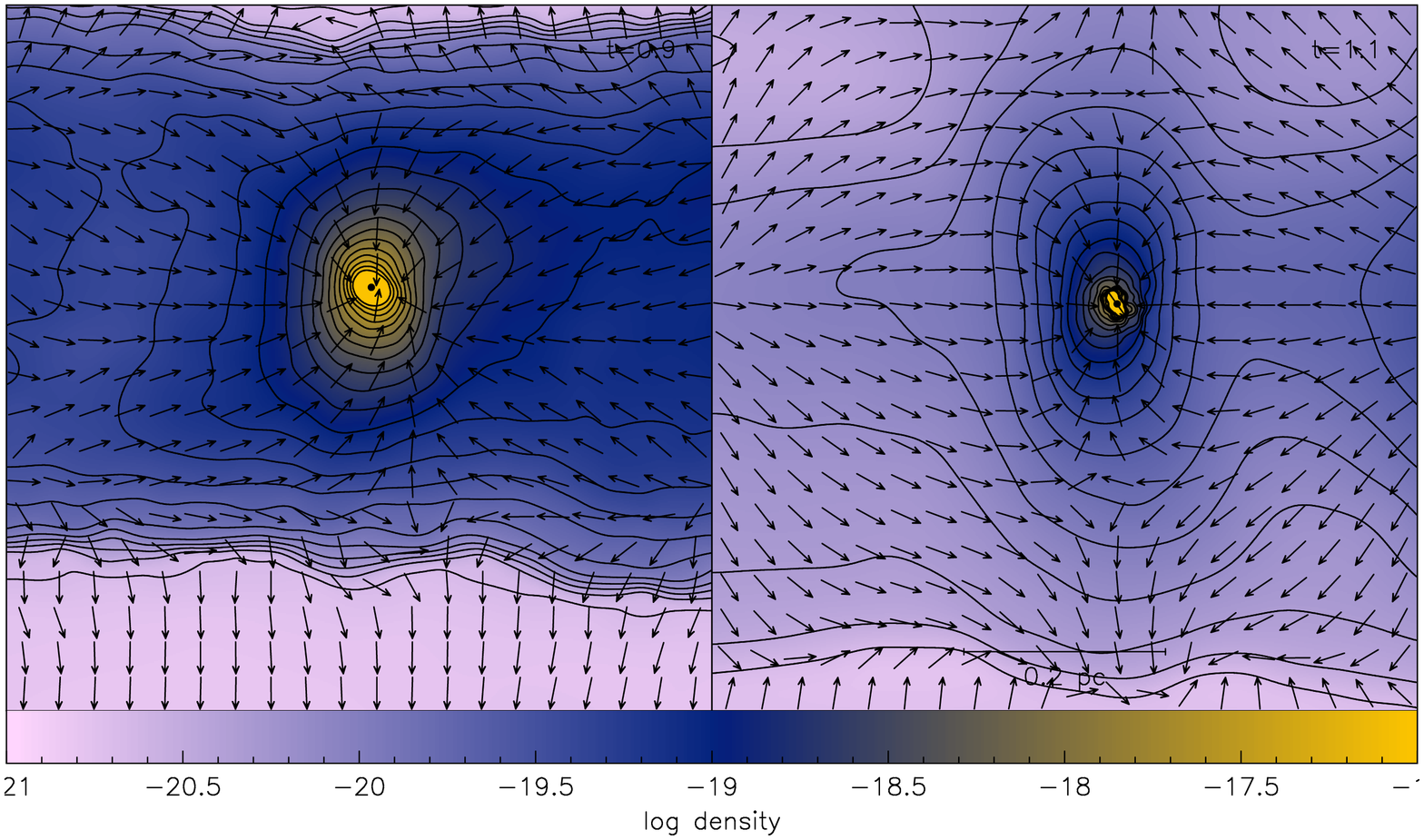}
    \caption{Case 9 ($t=$ 0.9 Myrs, $t$ = 1.1 Myrs for pictures on respectively the left and the right hand panel.). The thick black blob indicates the position of a sink particle.}
  \end{subfigure}
  \caption{Rendered density images similar to those plotted earlier, but now overlaid with density contours and velocity vectors showing the direction of gas-flow within the natal filament. Shown in these panels is a core soon after its formation and at a slightly later epoch. As usual time is marked in units of Myrs in the top right-hand corner of each panel.}
\end{figure*}
To avoid being repetitive, we do not explicitly discuss here the evolution of the fragment in Case 2 that was also initially sub-critical and therefore evolved in a manner similar to that in Case 1. Note, however, the rendered images shown on the left and the central panel of Fig. 3. It can be seen that not only is the core morphology similar in these two cases (i.e., Cases 1 and 2), but even the formation of cores is similarly delayed in Case 2. The rendered images on panels in Fig. 2(b) show the temporal evolution of the filament in Case 3 that was initially \emph{transcritical}. While the general evolutionary sequence of the filament is similar to that in Case 1, a comparison between Cases 1 and 3 highlights two main differences. First, the core-formation timescale in Case 3 is still relatively long, though more than a factor of 2 smaller than that in Case 1, but still within a factor of 2 of $t_{frag}$ given by Eqn. (5). Second, the cores in Case 3 are \emph{pinched}, i.e., they are smaller than their natal filament, and appear to have formed due to \emph{Jeans-like} fragmentation. The right-hand panel of Fig. 3 shows the terminal epoch of the fragmented filament in Case 3. A comparison of this latter image with the one on the lower right-hand panel of Fig. 2(b) shows that the filament expands a little in the radial direction even after fragmentation. More crucially, however, the fragments in Case 3 did not collapse either.
%It is thus clear that a sub-critical filament in a relatively low pressure environment is likely to evolve on a fairly long timescale and produce prolate (or even somewhat spherical) cores. A transcritical filament, by contrast, evolves much faster and is likely to produce oblate (i.e., \emph{pinched}) cores.\\\\
%==========================================
\subsection{Gas-flow within the natal filament : Implications for core - formation}
The dynamic nature of filament evolution and the direction of gas-flow towards cores within a filament have important implications for the orientation of outflows from a young stellar object ({\small YSO}) relative to its natal filament. Panels in Figs. 4(a) and 4(b) show a close-up of the core that formed respectively in Cases 2 and 9. We note that there is nothing special about the choice of these cases as they have been chosen merely for illustrative purposes. They represent realisations where the filament is initially subcritical and transcritical, respectively, and the fragments within are readily visible. It is interesting to note that in either case the gas delivered to the individual cores flows along the axial direction of the natal filament. In fact, the direction of the gas flow does not change even after the filament expands, most notably in Case 9, as illustrated by the picture on the right-hand panel of Fig. 4(b). \\\\
A core acquires angular momentum from the inflowing gas which is then ejected in the form of jets that are roughly orthogonal to the direction of the inflow (see seminal reviews by for e.g., Bodenheimer 1992; Boss 1993, Clarke 1992). In other words, a putative {\small YSO} in a scenario similar to the one shown in images in Figs. 4(a) and 4(b) is likely to launch outflows roughly orthogonal to the axis of its natal filament. This favoured orientation is consistent with the scenario of core-formation via fragmentation of a self-gravitating filament (e.g. Whitworth \emph{et al.} 1995). It is pertinent to note that the gas flow in the filament is largely axial despite the growth of oscillatory perturbations along its length. Nevertheless, radial oscillations of the natal filament could slightly alter the orientation of putative outflows relative to the natal filament.
%This is actually one for a fragment in case 1, but marked as case 2 due to their similarity.
\subsubsection{Density profile of prolate cores}
The radial density distribution of the core shown on the right-hand panel of Fig. 4(a) has been plotted in Fig. 5. Overlaid on this density distribution is the density profile of a thermally supported Bonnor - Ebert ({\small BE}) sphere having radius, $\xi_{B}$ = 6. Recall that the isothermal {\small BE} sphere with a radius less than the critical radius, $\xi_{crit}$ = 6.45 is stable. We note that the density distribution of this extracted core is qualitatively similar to that of well-known starless cores such as {\small B68} (e.g. Alves \emph{et al.} 2001), {\small L694-2} (e.g., Lee \& Myers 1999), and {\small L1517B} (e.g., Kirk \emph{et al.} 2005). Besides a density distribution, this extracted core also shows an inwardly directed velocity field as is often reported in typical starless cores, e.g., Maret \emph{et al.} (2007) towards the starless core {\small B68}, and Tafalla \emph{et al.} (1998) for the starless core {\small L1544}.
\begin{figure}
\label{Fig 5}
  \vspace*{1pt}
  \includegraphics[angle=270,width=0.48\textwidth]{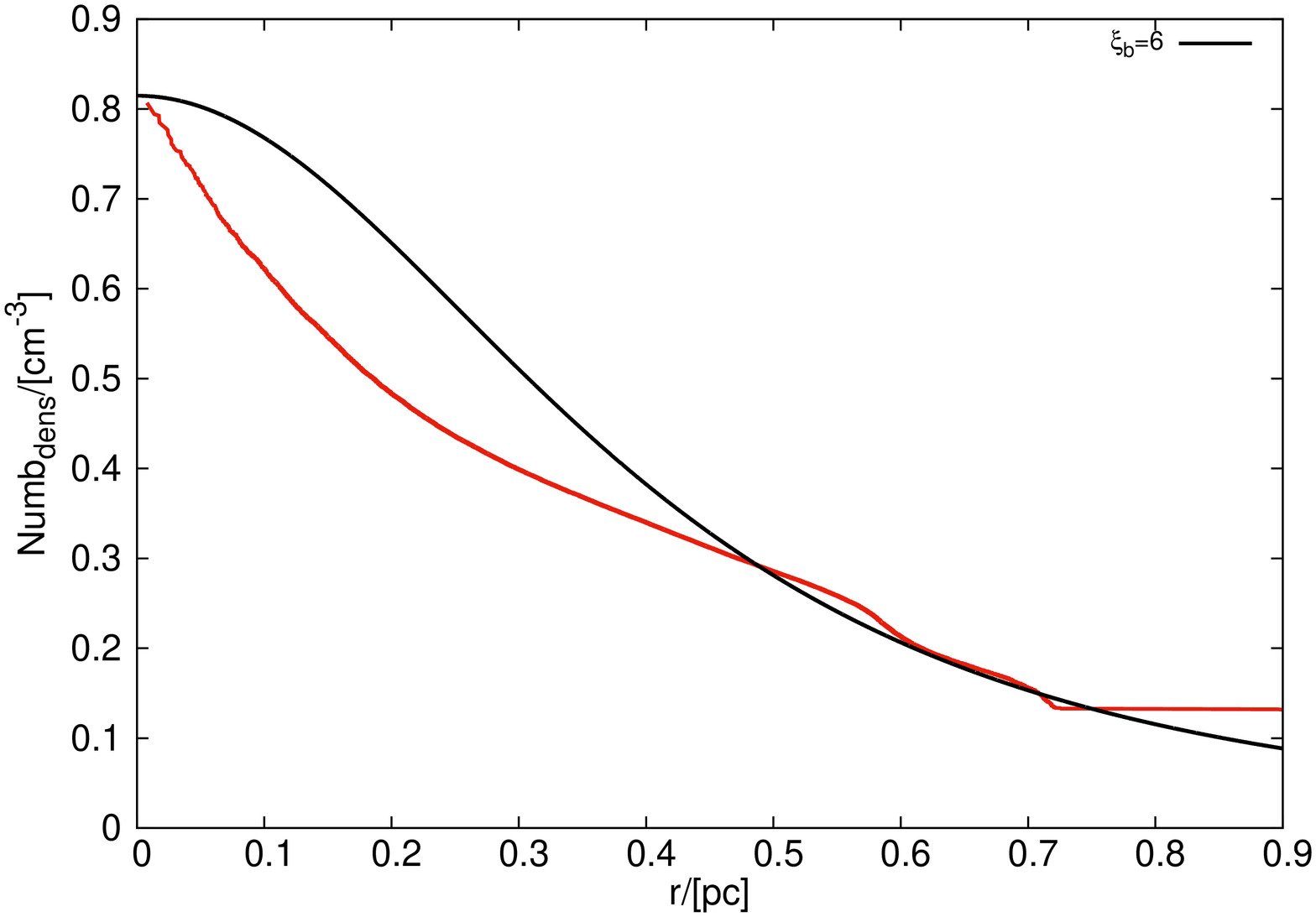}
  \caption{The radial distribution of gas density within the core shown on the right hand panel of Fig. 4(a); $r < r_{fil}$ is the radial coordinate within the core. Overlaid is the density profile of a {\small BE} sphere having radius $\xi_{b}$ = 6.}
\end{figure}
%==================================================
\begin{figure*}
\label{Fig 6}
  \begin{subfigure}{160mm}
   \vspace*{1pt}   
   \includegraphics[angle=270,width=\textwidth]{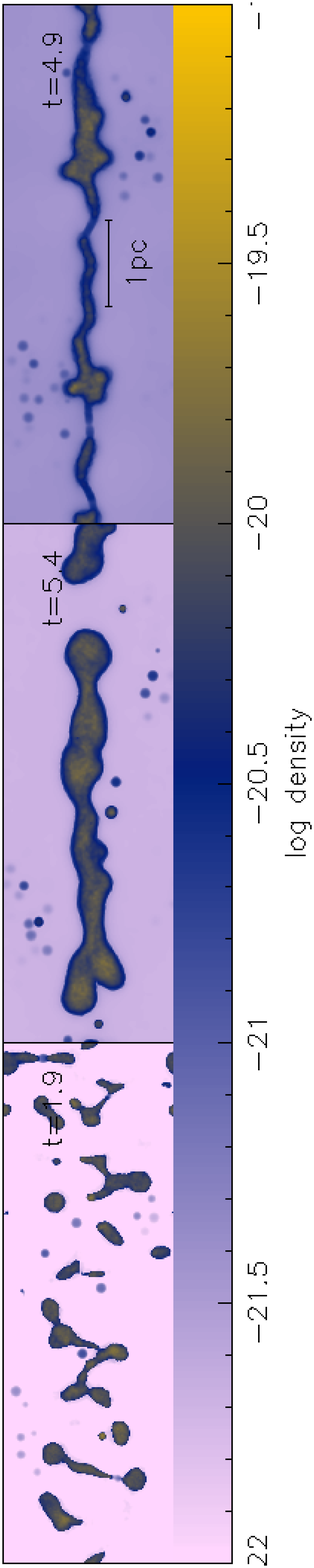}
   \caption{Respectively Cases 4, 5 and 6 from left to right.}
  \end{subfigure}
  \begin{subfigure}{160mm}
    \vspace*{1pt}
    \includegraphics[angle=270,width=\textwidth]{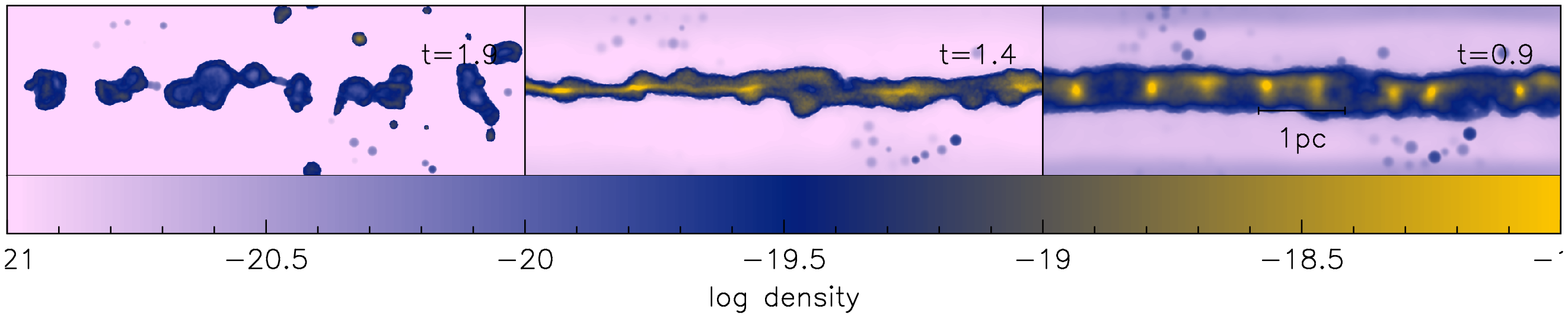}
    \caption{Respectively Cases 7, 8 and 9 from left to right.}
  \end{subfigure}
  \caption{Rendered density images similar to those plotted earlier, but now showing the terminal epoch of the fragment in respective cases. As usual time is marked in units of Myrs in the top right-hand corner of each panel.}
\end{figure*}
\subsection{Impact of higher pressure on filament fragmentation and subsequent formation of cores}
So far we have described results from the first three realisations that spanned the low to intermediate range of external pressure. Note that the picture of a fragment in Case 9 was shown in Fig. 4(b) merely to illustrate the direction of gas-flow towards a collapsing core in a filament. As in previously discussed cases, we saw in the remaining cases that the filament contracted radially also to form a thin dense filamentary structure which then evolved through cycles of contraction/expansion, followed by subsequent fragmentation. Shown on respective panels of Fig. 6(a) are rendered density images of the fragmented filament, i.e., the terminal epoch of the filament, in Cases 4, 5 and 6. Similarly, respective panels of Fig. 6(b) show the terminal epoch of the filament in a third set, i.e., in Cases 7, 8 and 9. Together, these respective sets of realisations span a range of pressure between a few $\times 10^{4}$ K cm$^{-3}$ and $\gtrsim 10^{5}$ K cm$^{-3}$.\\\\
The following features are visible from the set of images shown in Figs.  6(a) and 6(b) - \textbf{(i)} the subcritical filament ($f_{cyl}$=0.2) ends up as a wispy structure irrespective of the magnitude of external pressure as is evident from the respective images showing the final stage of the filament in Cases 4 and 7. In fact, the filament in the former Case acquires striated-like structure (i.e., the intermittent appearance of  elongated but condensed structure roughly perpendicular to the original filament axis), while in the latter, the end product appears like a fuzzy string of beads; \textbf{(ii)} the sub-critical filament ($f_{cyl}$ = 0.5) in Case 5, on the other hand, ends up as an elongated structure with broad cores, though none of them went on to collapse. On the contrary, the filament in Case 8  that was similarly sub-critical ($f_{cyl}$ = 0.5) ended up as a thin filamentary structure with pinched cores; \textbf{(iii)} and finally, in the initially transcritical filament ($f_{cyl}$ = 0.9) in Case 6 only two broad core-like structures are visible along a thin filament. By contrast, oblate cores (but still smaller than the natal filament, i.e., \emph{pinched}) can be readily seen in the transcritical filament ($f_{cyl}$ = 0.9) in Case 9, some of which did eventually collapse as is visible in the right-hand panel of Fig. 4(b). We will revisit the point regarding the intrinsic shapes of cores in the following subsection.\\\\
%The upshot of these observations is that a filament in relatively low-pressure environment tends to form broad cores irrespective of the choice of the initial linemass. Filaments with lower initial linemass ($f_{cyl}$ = 0.2), as in Cases 4 and 7 tend to form a fuzzy striated structure.\\\\
%This latter set of images shows clearly the propensity towards formation of prolate (i.e., \emph{broad}) cores via fragmentation of the initially sub-critical ($f_{cyl}$=0.5) fragment. W, the filament tends to . \\\\
Striations visible in, e.g., Case 4, are qualitatively similar to those reported in the Taurus molecular cloud (Goldsmith \emph{et al.} 2008). In our purely hydrodynamic realisations, striations are the result of amplification of density perturbations on the surface of the filament. These perturbations, however, do not condense into cores, owing to their relatively weak self-gravity. Unlike suggestions elsewhere in literature where the origin of striations has been attributed to magnetohydrodynamic waves or the Kelvin - Helmholtz instability (e.g., Heyer \emph{et al.} 2016), here we posit that even purely hydrodynamical processes can reproduce the appearance of striated structures in {\small MCs}. By contrast, the initially transcritical filament ($f_{cyl}$ = 0.9), in Case 9 forms \emph{pinched} cores as seen in the right-hand panel of Fig. 6(b). Interestingly, however, the similarly transcritical filament in Case 6 contracted to form a thin filamentary structure, but with two broad fragments, as seen in the right hand panel of Fig. 6(a). This structure, which looks similar to that observed in regions of a filamentary hub (see, e.g., Myers 2009), could help us reconcile the formation of hub structures. The contrasting end result of the initially transcritical filament in Cases 6 and 9 shows that fragmentation due to the \emph{Jeans-type} instability leading to the formation of pinched cores is unlikely in low-pressure environment, typically where $P_{{\small ICM}}/k_{B}\lesssim 10^{5}$ K cm$^{-3}$. In such low-pressure environments the \emph{collect-and-collapse} mode leading to broad cores dominates. \\\\
\begin{figure*}
\label{Fig 7}
  \vspace*{1pt}
  \includegraphics[angle=270,width=\textwidth]{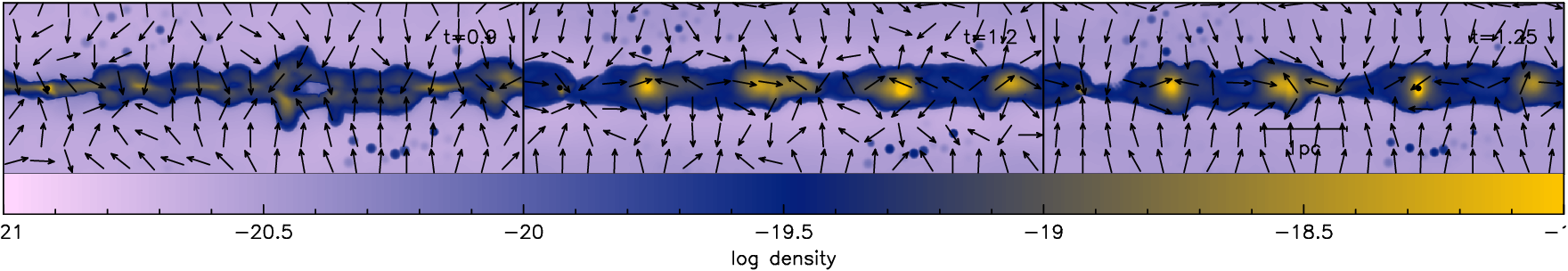}
  \caption{Same as the rendered density images shown earlier on various panels of Fig. 6, but now showing a latter epoch and the terminal epoch of the filament in Case 10. The overlaid arrows represent the local direction of gas-flow as usual.}
\end{figure*}
Finally, the later epochs and the terminal epoch of the initially transcritical fragment in Case 10 with the highest choice of external pressure, are shown in Fig. 7. From the parameters listed in Table 1 it is clear that the filament in this case is massive with a relatively large initial average density. As in other Cases, this filament also evolves through a series of contractions and expansions, albeit on a significantly shorter timescale due to the stronger self-gravity. Although we reported in Paper I that a filament in a high pressure environment eventually ruptures, we did not explicitly observe this behaviour in the present instance, at least until the realisation was allowed to continue (see the image on the right-hand panel of Fig. 7). The left-hand panel of Fig. 7, however, shows evidence for localised collapse in just one of the cores that formed as a result of fragmentation of the filament. As the filament contracts further before rebounding (as is visible in the image on the right-hand panel of Fig. 7), there is no further evidence of core formation in this case. \\\\
To sum up, realisations in this work show that an initially sub-critical fragment in a relatively low pressure environment ($P_{ext}/k_{B}\lesssim \mathrm{few\ times}\ 10^{4}$ K cm$^{-3}$) typically forms broad spheroidal cores, though occasionally a few could be spherical. An initially transcritical filamentary fragment in higher-pressure environment similar to that in the Solar neighbourhood, by contrast, evolves on a relatively shorter timescale and preferentially forms pinched cores via \emph{Jeans-like} fragmentation. At still higher pressures, typically upward of $\gtrsim 10^{6}$ K cm$^{-3}$, dense pockets form relatively quickly in a filament, but rupture (rather rebound) soon thereafter, an event which could explain the dearth of star formation in high pressure environs. From the forgoing it is clear that irrespective of the choice of external pressure as well as the initial linemass, the filament contracts radially before it eventually fragments due to the  growth of axial perturbations. In other words, even as the initially perturbed filament contracts radially, the fragmentation length scale is determined by the local physical conditions in the filament.
\subsubsection{External pressure and the intrinsic shape of cores}
It is tempting to describe broad cores as being prolate and pinched cores as being oblate. As described in \S 1 above, the task of determining the shape of a core is nontrivial. Observationally, it is subject to the inherent biases of the particular tracer being used. In this work, however, individual cores in the filament at its terminal epoch were identified by first scanning it for particles having density on the order of the maximum resolvable density defined in \S 2.2, the so-called seed particles. The immediate neighbourhood of these seed particles was then searched to identify the nearest neighbours with density within an order of magnitude of the peak density. The axial ratio of the cores so identified was then calculated. Listed in Table 2 are the mean axial ratios for cores identified in the filament in each realisation. \\\\
Evidently, the broad cores in cases where the external pressure $< 10^{5}$ K cm$^{-3}$ and the filament is initially sub-critical (i.e., in Cases 1, 2 \& 5) are prolate. Equivalently, the \emph{collect-and-collapse} mode of core formation preferably generates broad prolate cores. On the other hand, cores that form in the initially transcritical filaments in Cases 3, 6 and 9 due to the \emph{Jeans-like} compressional instability are pinched (and oblate) soon after their formation. As noted above, expansion of the natal filament causes these cores to become prolate, but they do not ever become broader than their natal filament. The axial ratio for these latter cases has been marked with an asterisk in Table 2. This observation can be readily corroborated by comparing the rendered density image shown on the lower right-hand panel of Fig. 2b and the image on the right hand panel of Fig. 3 with the rendered images shown in Fig. 4b. It must also be noted that the cores in the filament in Case 3 did not collapse further to form sink particles, unlike those in Case 9.
\begin{table}
\centering
\caption{Mean axial ratio for cores identified in each realisation. The mean axial ratio for cases in which initially pinched cores expanded to become prolate are marked with an $\ast$.}
\begin{tabular}{|l|c|c|}
\hline
Sr & $\frac{(P_{ext}/k_{B})}{[K \mathrm{cm}^{-3}]}$; $f_{cyl}$ & Axial ratio\\
No. & &    \\
\hline
1 & 6.0$\times 10^{3}$;0.2 & 1.65\\
2 & 6.0$\times 10^{3}$;0.5 & 1.2 \\
3 & 6.0$\times 10^{3}$;0.9 & 1.8* \\
4 & 7.3$\times 10^{4}$;0.2 & No cores\\
5 & 7.3$\times 10^{4}$;0.5 & 1.1\\
6 & 7.3$\times 10^{4}$;0.9 & 2.8*\\
7 & 6.5$\times 10^{5}$;0.2 & No cores\\
8 & 6.5$\times 10^{5}$;0.5 & 0.53\\
9 & 6.5$\times 10^{5}$;0.9 & 2.3* \\
10& 2.2$\times 10^{6}$;0.9 & 1.1*\\
\hline
\end{tabular}
\end{table}
%\footnotesize{*{\small \emph{See text below for discussion.}}}
%--------------------------------------------------
%====================================================
%\begin{figure*}
%  \begin{subfigure}{160mm}
%   \vspace*{1pt}   
%   \mbox{\includegraphics[angle=270,width=0.48\textwidth]{F1P1Seed1-Vpara.eps}
%        \includegraphics[angle=270,width=0.48\textwidth]{F2P2Seed1-Vpara.eps}}
%   \mbox{\includegraphics[angle=270,width=0.48\textwidth]{F3P3Seed1-Vpara.eps}
%         \includegraphics[angle=270,width=0.48\textwidth]{F3P4Seed1-Vpara.eps}}
%   \caption{.}
%  \end{subfigure}
%  \begin{subfigure}{160mm}
%   \vspace*{1pt}
%   \mbox{\includegraphics[angle=270,width=0.48\textwidth]{F1P1Seed1-Vperp.eps}
%         \includegraphics[angle=270,width=0.48\textwidth]{F2P2Seed1-Vperp.eps}}
%   \mbox{\includegraphics[angle=270,width=0.48\textwidth]{F3P3Seed1-Vperp.eps}
%         \includegraphics[angle=270,width=0.48\textwidth]{F3P4Seed1-Vperp.eps}}
%    \caption{of velocity gradient at different points along the length of the filament.}
%  \end{subfigure}
 % \caption{ Cases 1 and 5 on upper left and right-hand panels and Cases 9 and 10 on lower left and right-hand panels of (a) and (b).}
%\end{figure*}
\begin{figure}
\label{Fig 8}
  \begin{subfigure}{80mm}
   \vspace*{1pt}   
  \mbox{\includegraphics[angle=270,width=0.48\textwidth]{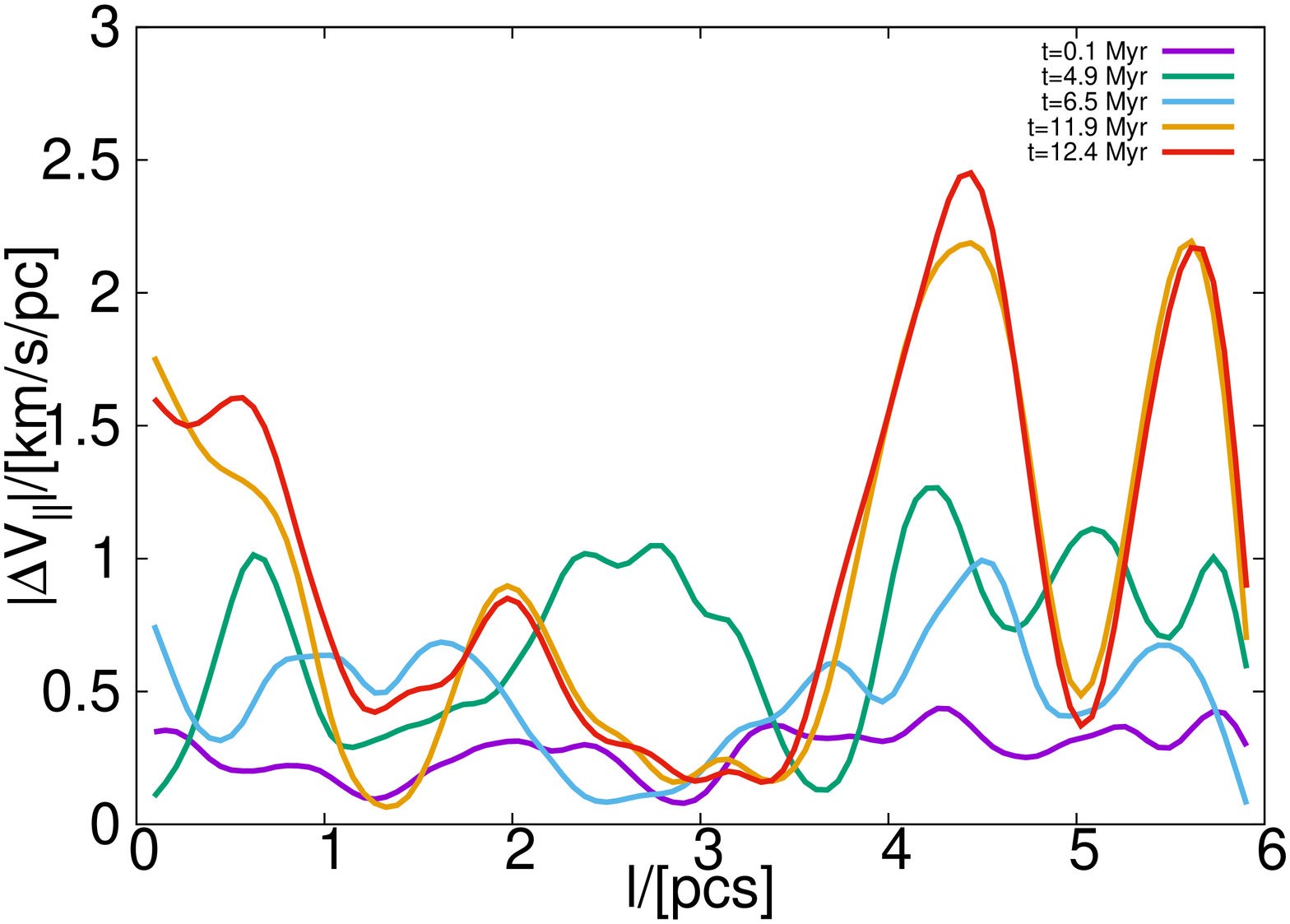}
        \includegraphics[angle=270,width=0.48\textwidth]{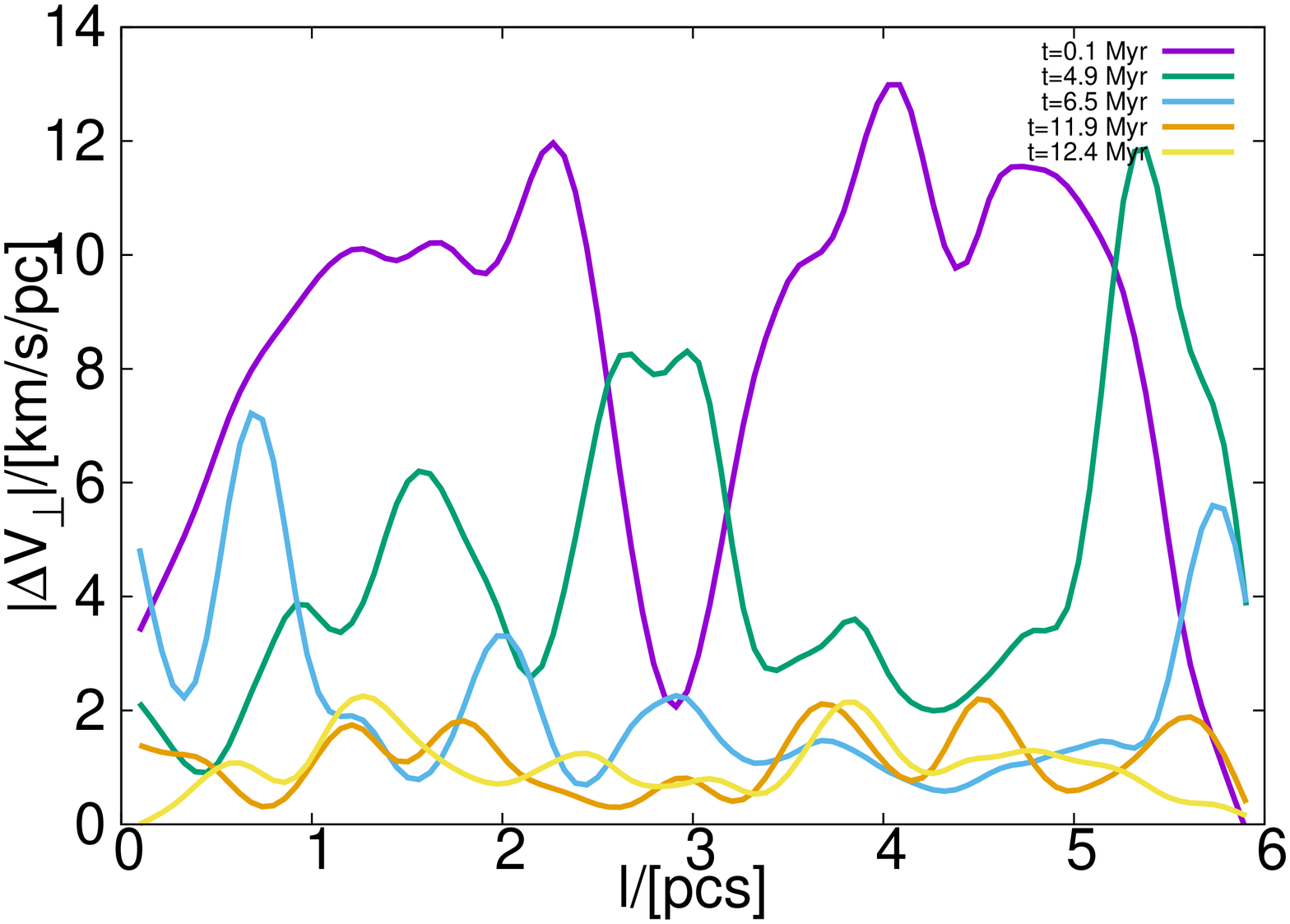}}
   \caption{Case 1 : $P_{ext}/k_{B}$ = 6.0x10$^{3}$ K cm$^{-3}$; $f_{cyl}$ = 0.2}
  \end{subfigure}
  \begin{subfigure}{80mm}
   \vspace*{1pt}
  \mbox{\includegraphics[angle=270,width=0.48\textwidth]{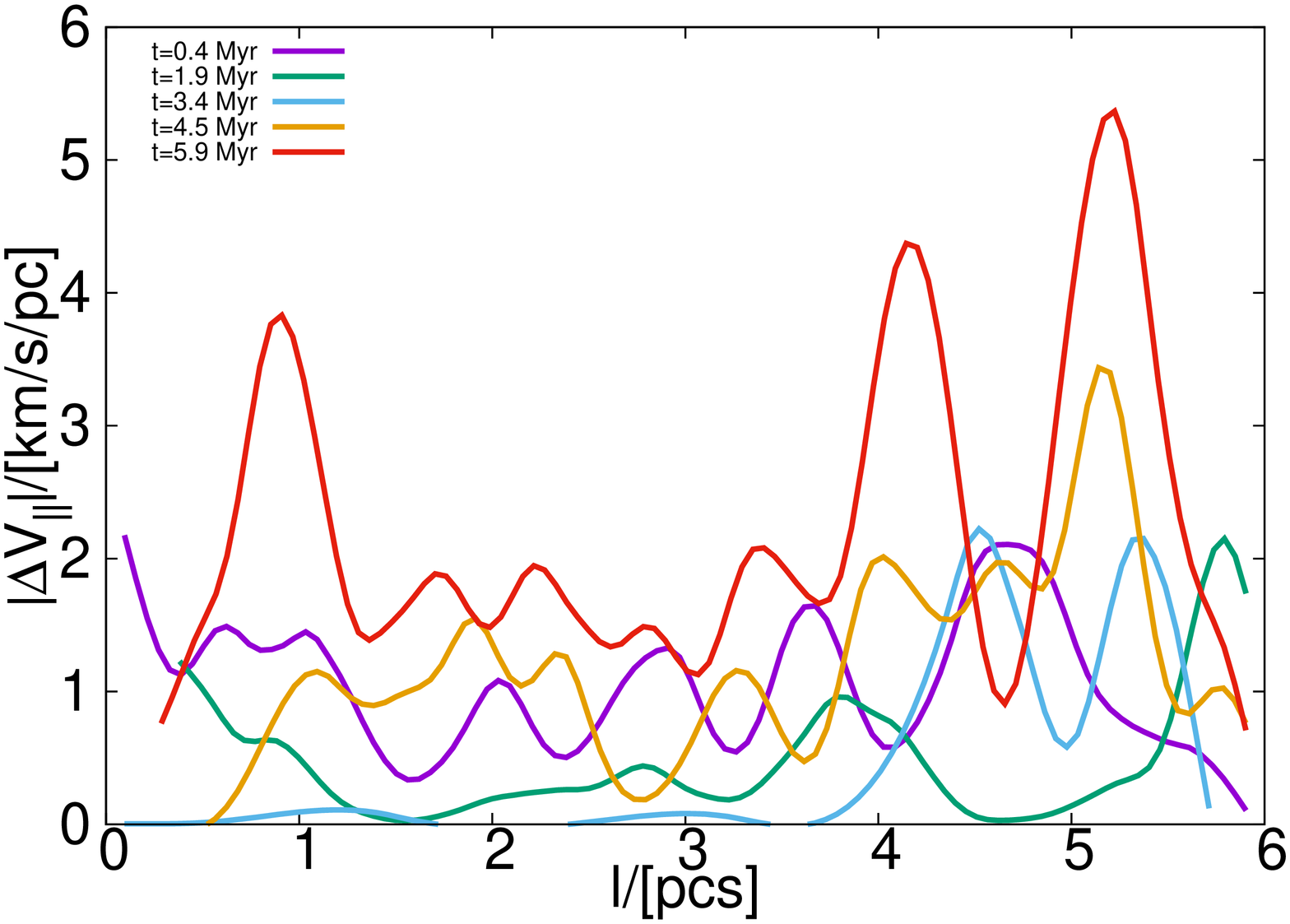}
        \includegraphics[angle=270,width=0.48\textwidth]{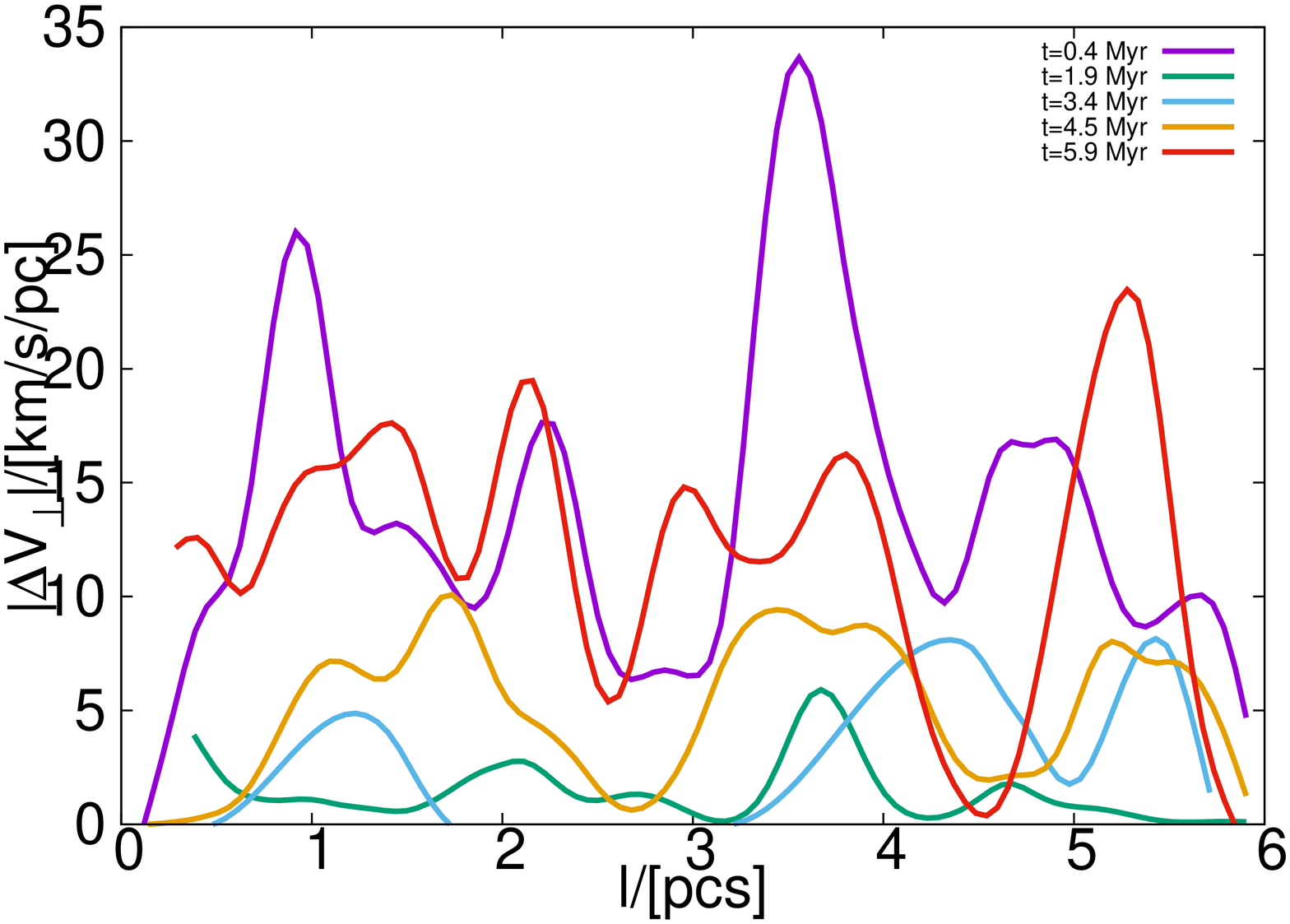}}
    \caption{Case 5 : $P_{ext}/k_{B}$ = 7.3x10$^{4}$ K cm$^{-3}$; $f_{cyl}$ = 0.5}
  \end{subfigure}
  \begin{subfigure}{80mm}
   \vspace*{1pt}
  \mbox{\includegraphics[angle=270,width=0.48\textwidth]{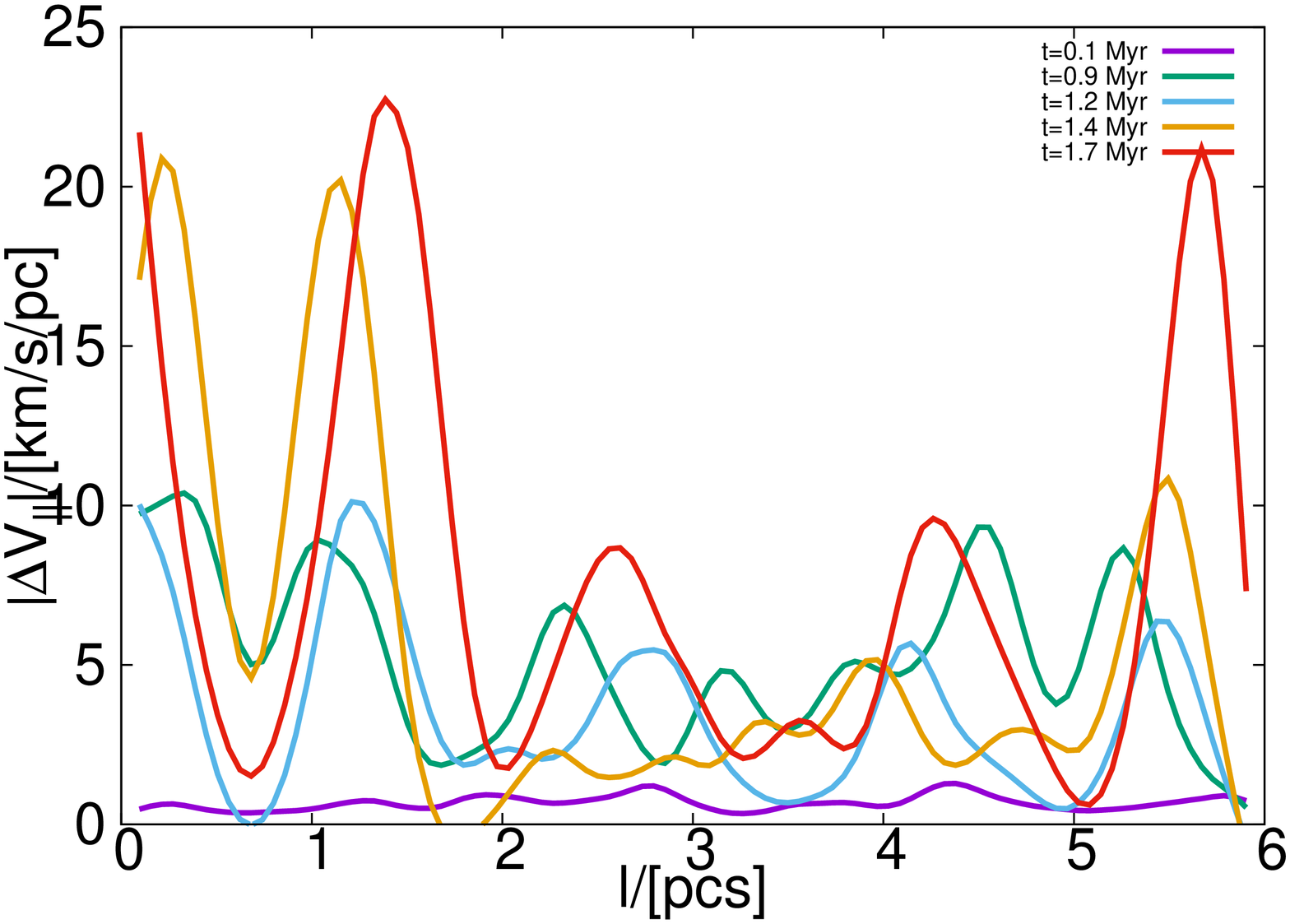}
        \includegraphics[angle=270,width=0.48\textwidth]{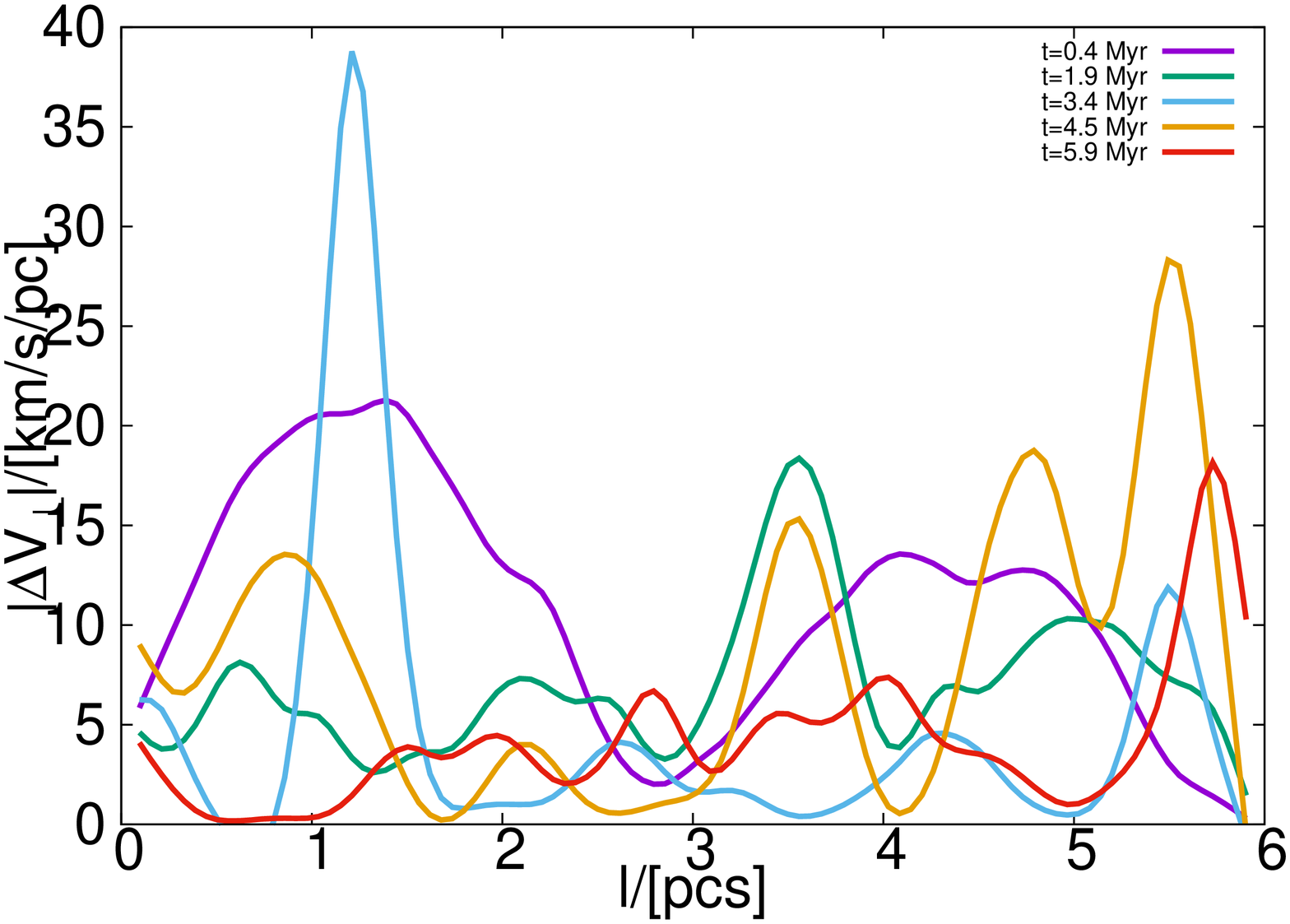}}
    \caption{Case 9 : $P_{ext}/k_{B}$ = 6.5x10$^{5}$ K cm$^{-3}$; $f_{cyl}$ = 0.9}
  \end{subfigure}
  \begin{subfigure}{80mm}
   \vspace*{1pt}
 \mbox{\includegraphics[angle=270,width=0.48\textwidth]{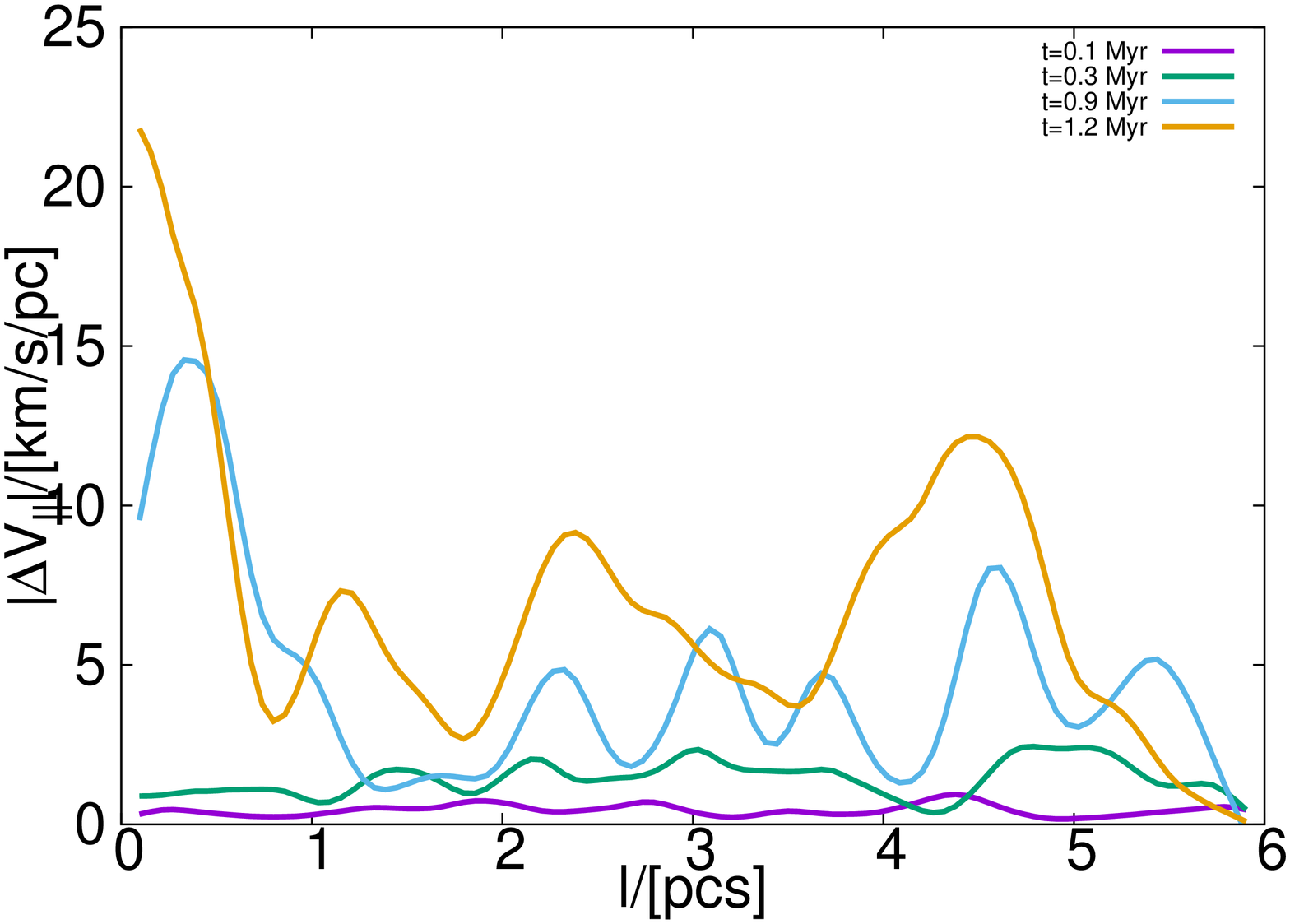}
       \includegraphics[angle=270,width=0.48\textwidth]{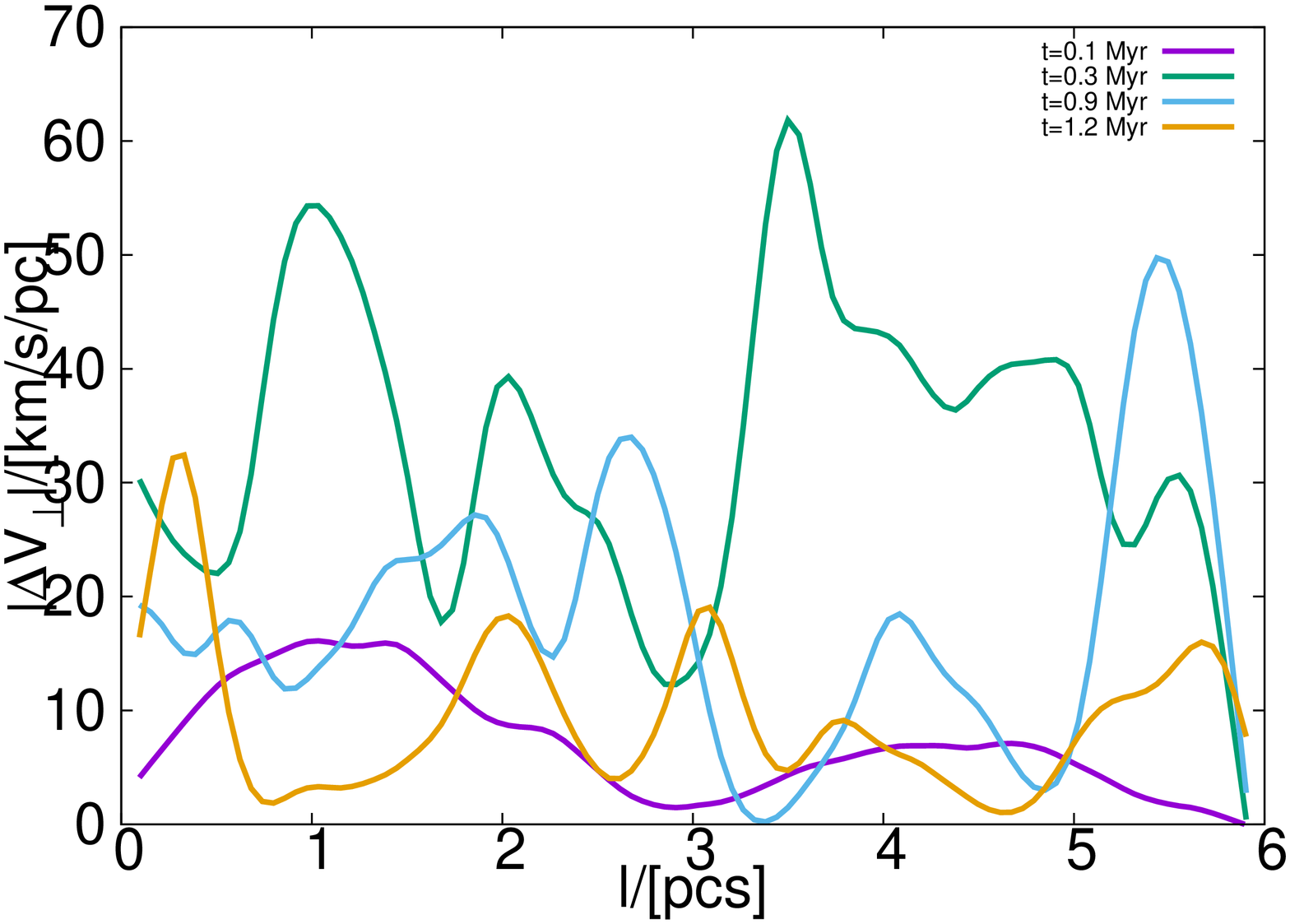}}
    \caption{Case 10: $P_{ext}/k_{B}$ = 2.2x10$^{6}$ K cm$^{-3}$; $f_{cyl}$ = 0.9}
  \end{subfigure}
  \caption{Respectively, the radial component and the axial component of velocity gradient at different points along the length of the filament ($l < L_{fil}$).}
\end{figure}
%=================================================
\subsection{Components of velocity gradient parallel and perpendicular to the axis of the fragment}
Quasi-periodic, oscillatory features in velocity fields are often reported towards filamentary clouds. These features are variously interpreted as signatures of localised collapse in density structures, or as evidence of gas-flow along the filament axis. In the interests of brevity, we show on various panels of Fig. 8(a) the component of the velocity gradient ($\Delta V_{\parallel}$) parallel to the filament axis for Cases 1, 5, 9 and 10 out of the entire set of realisations. This subset of Cases is chosen such that it spans the entire range of pressure and the initial filament linemass explored in this work. Quasi-oscillatory features in the axial component ($\Delta V_{\parallel}$) are easily identifiable in each case. Interestingly, while the  periodicity of $\Delta V_{\parallel}$ is not much affected by the magnitude of external pressure, its own magnitude, however, increases at higher pressure. Evidently, the magnitude of external pressure affects the gas dynamics within a filament.\\\\
Note that the peaks in the $\Delta V_{\parallel}$ profile do not always coincide with clearly distinguishable cores, except perhaps in Cases 1, 5 and 9 where they signify locally collapsing cores at the terminal epoch. Interestingly, however, such oscillatory features are visible over the entire course of evolution of the filament. Evidently, the associated peaks in the velocity gradient reflect the flow of gas into the crests and out of the troughs of the density perturbations that grow on the surface of the filament. These quasi-oscillatory features in the $\Delta V_{\parallel}$ profile are also interesting in view of the fact that the model filament in this work was not actively accreting gas. Instead, these oscillatory patterns are manifestations of features commonly associated with the shell instability. It is clear that the appearance of quasi-oscillatory features in $\Delta V_{\parallel}$ is an integral part of the evolutionary sequence of a filament, irrespective of whether a filament is actively accreting gas or not and irrespective of its initial linemass. \\\\
Various panels of Fig. 8(b) show plots of the radial component of the velocity gradient, i.e., that orthogonal to the filament axis, at different locations along its length. At the outset, it is clear that - \textbf{(i)} the magnitude of $\Delta V_{\perp}$ is significantly greater than that of $\Delta V_{\parallel}$, \textbf{(ii)} like $\Delta V_{\parallel}$, the magnitude of $\Delta V_{\perp}$ also increases with external pressure, and \textbf{(iii)} unlike $\Delta V_{\parallel}$, the oscillatory features of $\Delta V_{\perp}$ are aperiodic. In the absence of gas accretion, the radial component of velocity field is simply due to the  contraction/expansion cycle of the filament. It is therefore clear that the appearance of oscillatory $\Delta V_{\perp}$ features need not always be associated with accreting filaments. In fact, Chen \emph{et al.} (2020) suggested that such features in the radial profile of $\Delta V_{\perp}$ could originate due to inhomogeneous/anisotropic accretion flows as is indeed seen in numerical simulations (e.g., Chen \& Ostriker 2014). The observation here that $\Delta V_{\perp}$ is significantly greater than the parallel component of the velocity gradient is also consistent with inferences from observational surveys of filamentary clouds (e.g. Fern{\' a}ndez - L{\' o}pez \emph{et al.} 2014). The respective parallel and perpendicular components of the velocity gradient were calculated at different locations along the length of the fragment by averaging over contributions due to 500 nearest neighbours at each location. The number of bins (i.e., locations where respective components of the velocity gradient were calculated), along the length of the filament was chosen using the Sturges criterion for optimal bin-size (Sturges 1926).
\subsubsection{Radial variation of velocity gradients}
Fig. 9 shows the magnitude of the respective components of velocity gradients as a function of the radius for the same cases discussed above. As before, these plots have also been made at different epochs of evolution of the filament. There is no clear correlation between the respective components of velocity gradients as a function of radius. There is also no periodicity in the variation of the respective velocity components. Interestingly, in each of the four cases shown here, there are epochs where we see either a relatively steep rise in $\vert\Delta V_{\perp}\vert$ (associated with its contractional phase), or a decrease in its magnitude towards the filament axis (associated with its expansionary phase). Such intermittent behaviour of $\vert\Delta V_{\perp}\vert$ further emphasises the oscillatory nature of evolution of the filament. \\\\
\begin{figure}
\label{Fig 9}
  \begin{subfigure}{80mm}
   \vspace*{1pt}   
  \mbox{\includegraphics[angle=270,width=0.48\textwidth]{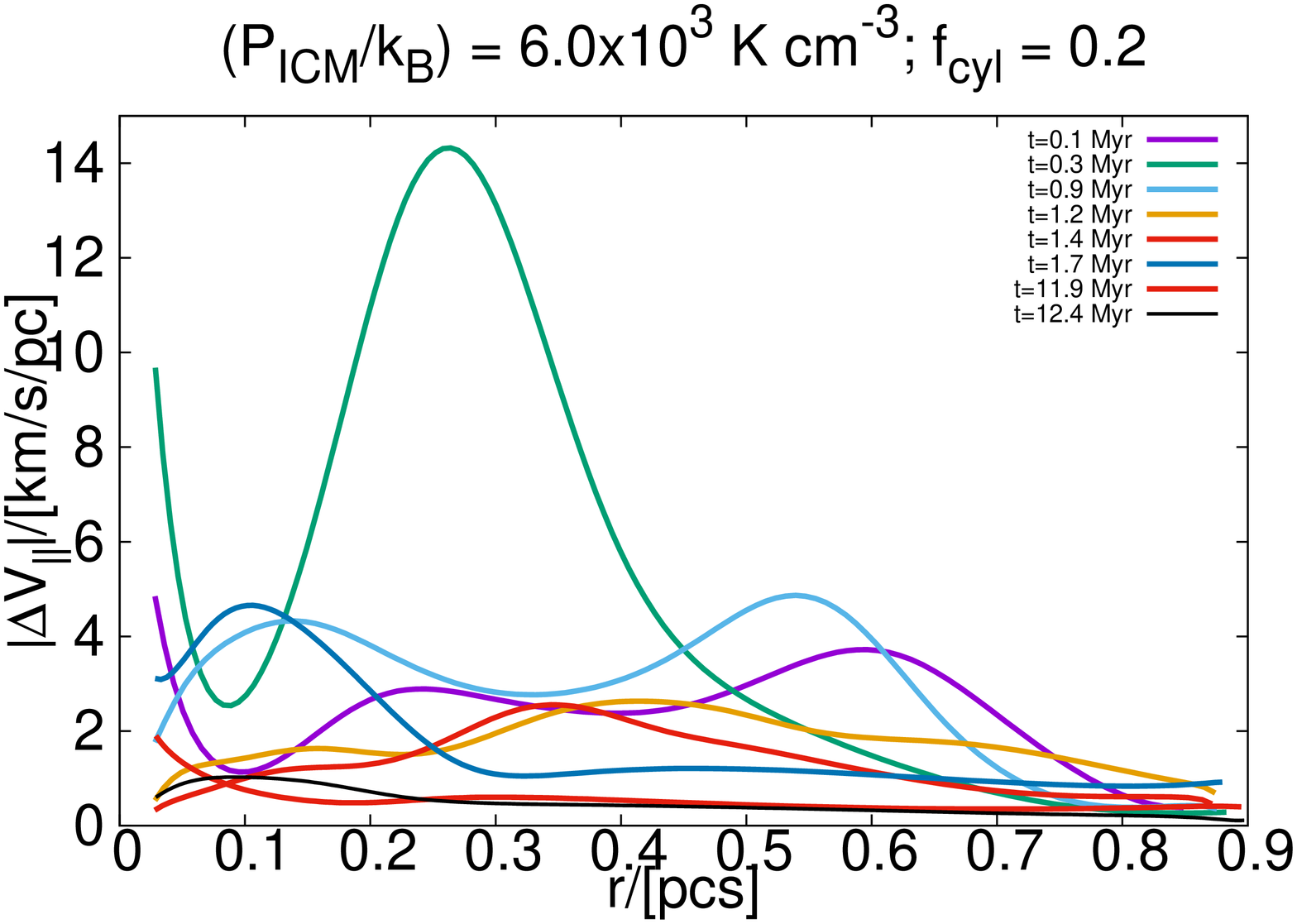}
        \includegraphics[angle=270,width=0.48\textwidth]{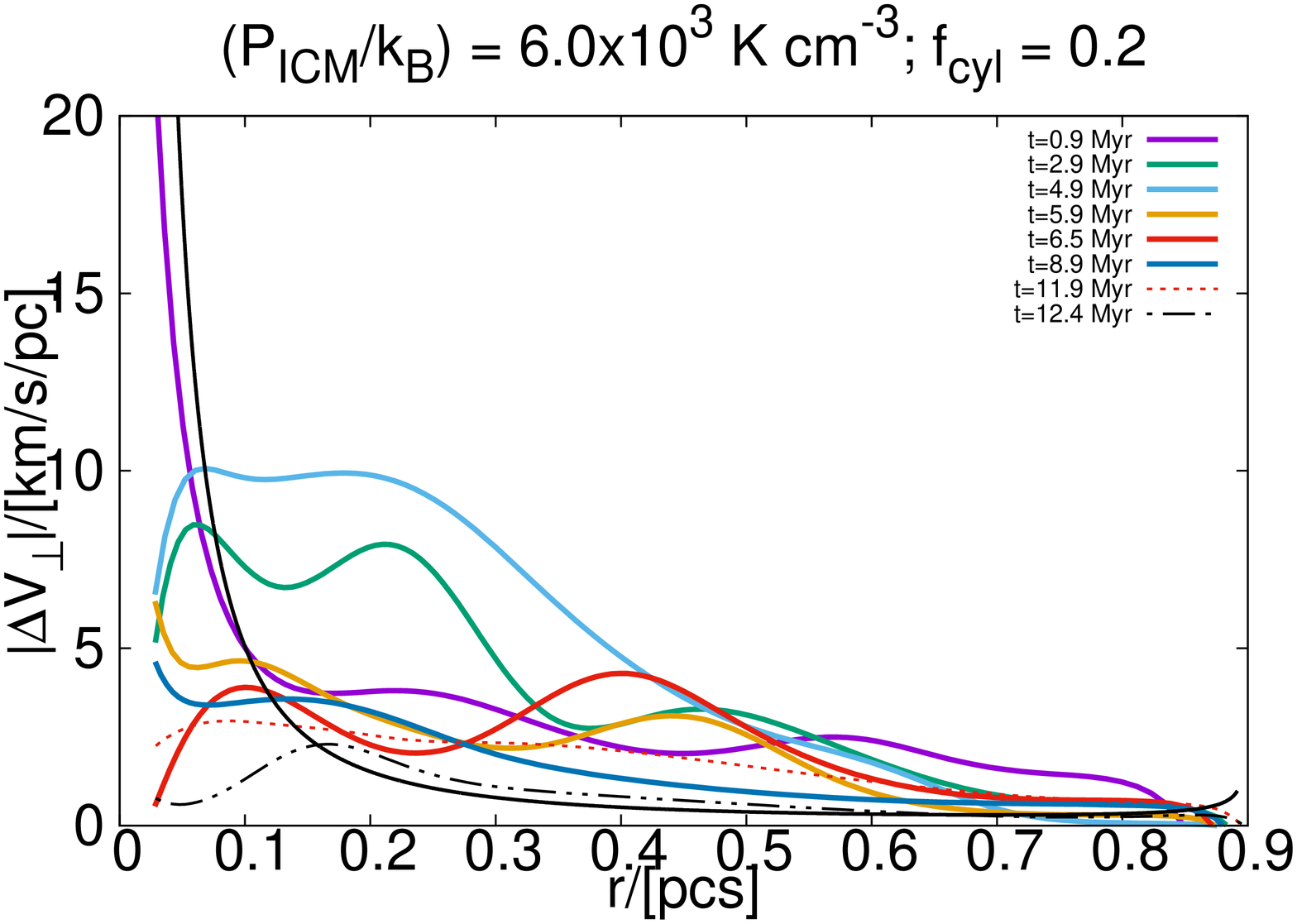}}
   \caption{Case 1}% :  $f_{cyl}$ = 0.2; $P_{ext}/k_{B}$ = 4x10$^{3}$ K cm$^{-3}$}
  \end{subfigure}
  \begin{subfigure}{80mm}
   \vspace*{1pt}
  \mbox{\includegraphics[angle=270,width=0.48\textwidth]{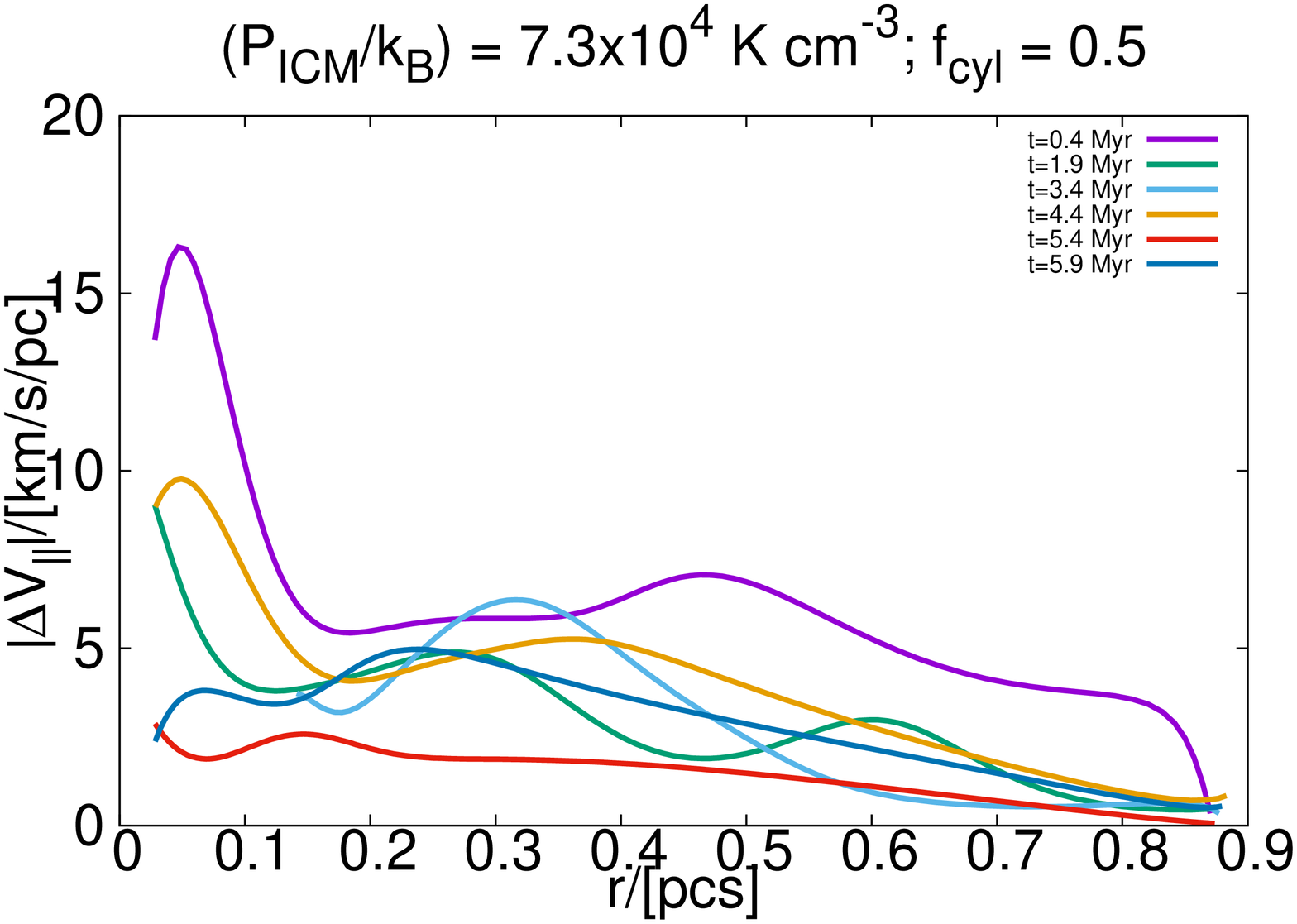}
        \includegraphics[angle=270,width=0.48\textwidth]{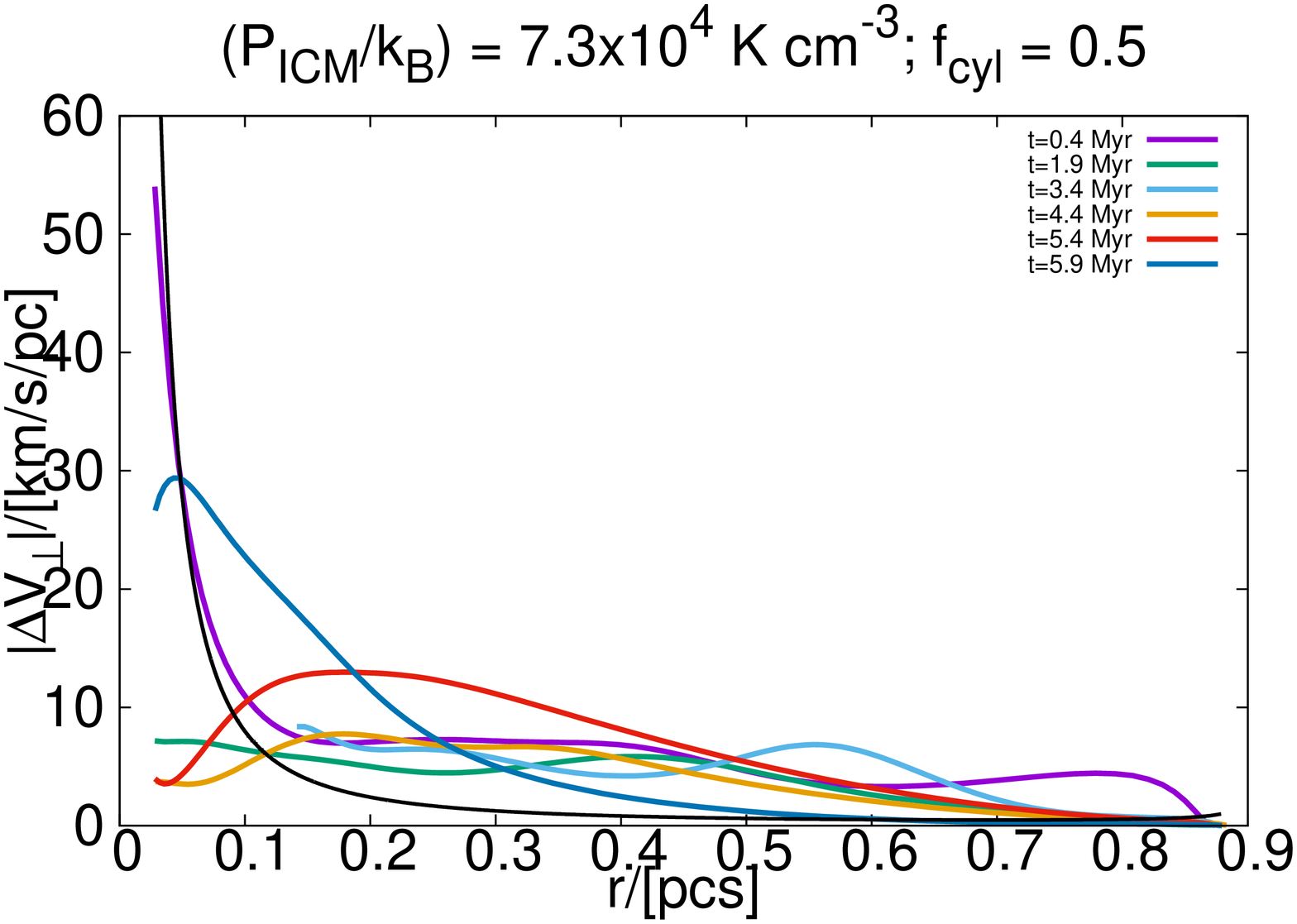}}
    \caption{Case 5}% :  $f_{cyl}$ = 0.5; $P_{ext}/k_{B}$ = 3x10$^{4}$ K cm$^{-3}$}
  \end{subfigure}
  \begin{subfigure}{80mm}
   \vspace*{1pt}
  \mbox{\includegraphics[angle=270,width=0.48\textwidth]{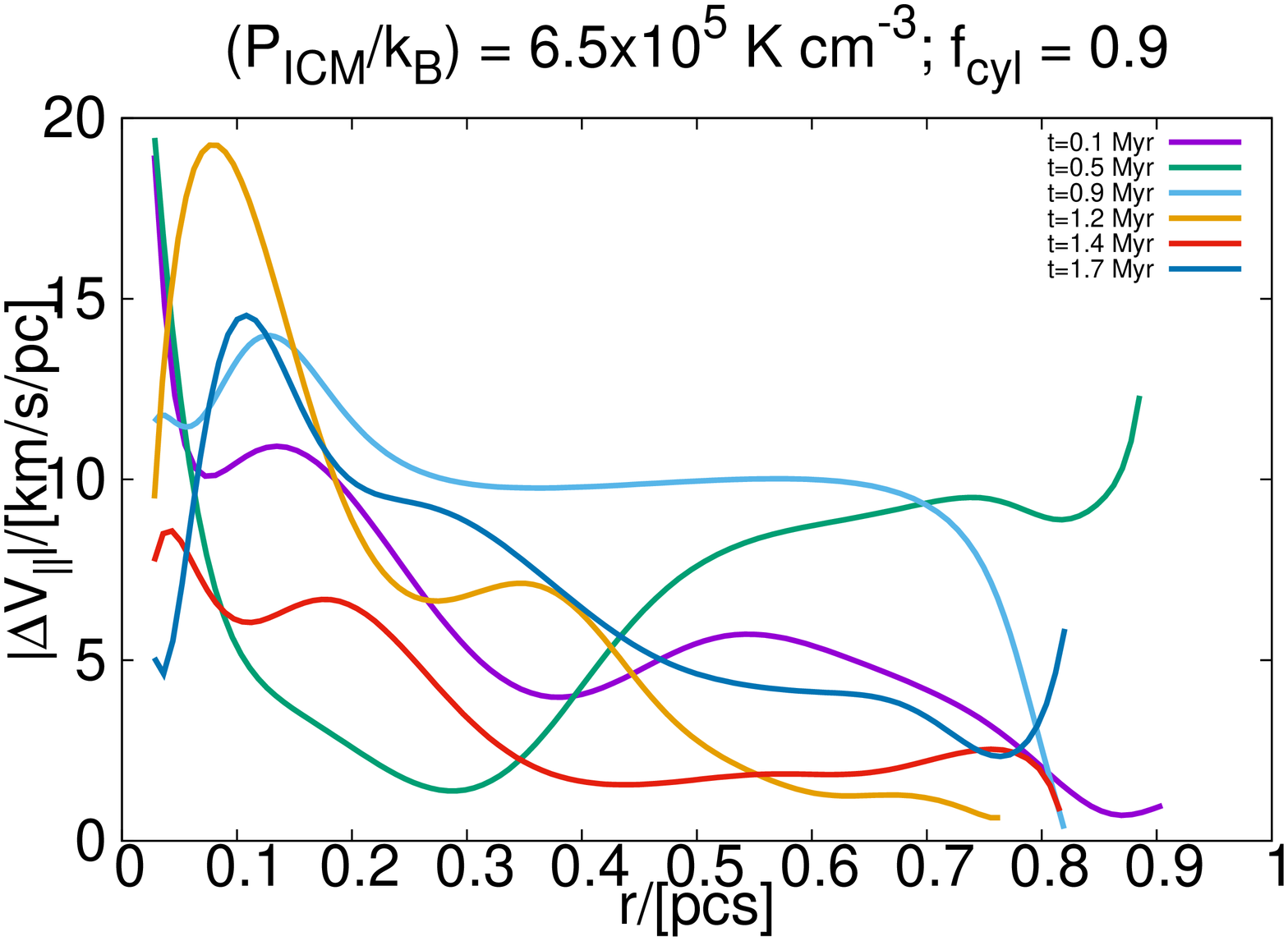}
        \includegraphics[angle=270,width=0.48\textwidth]{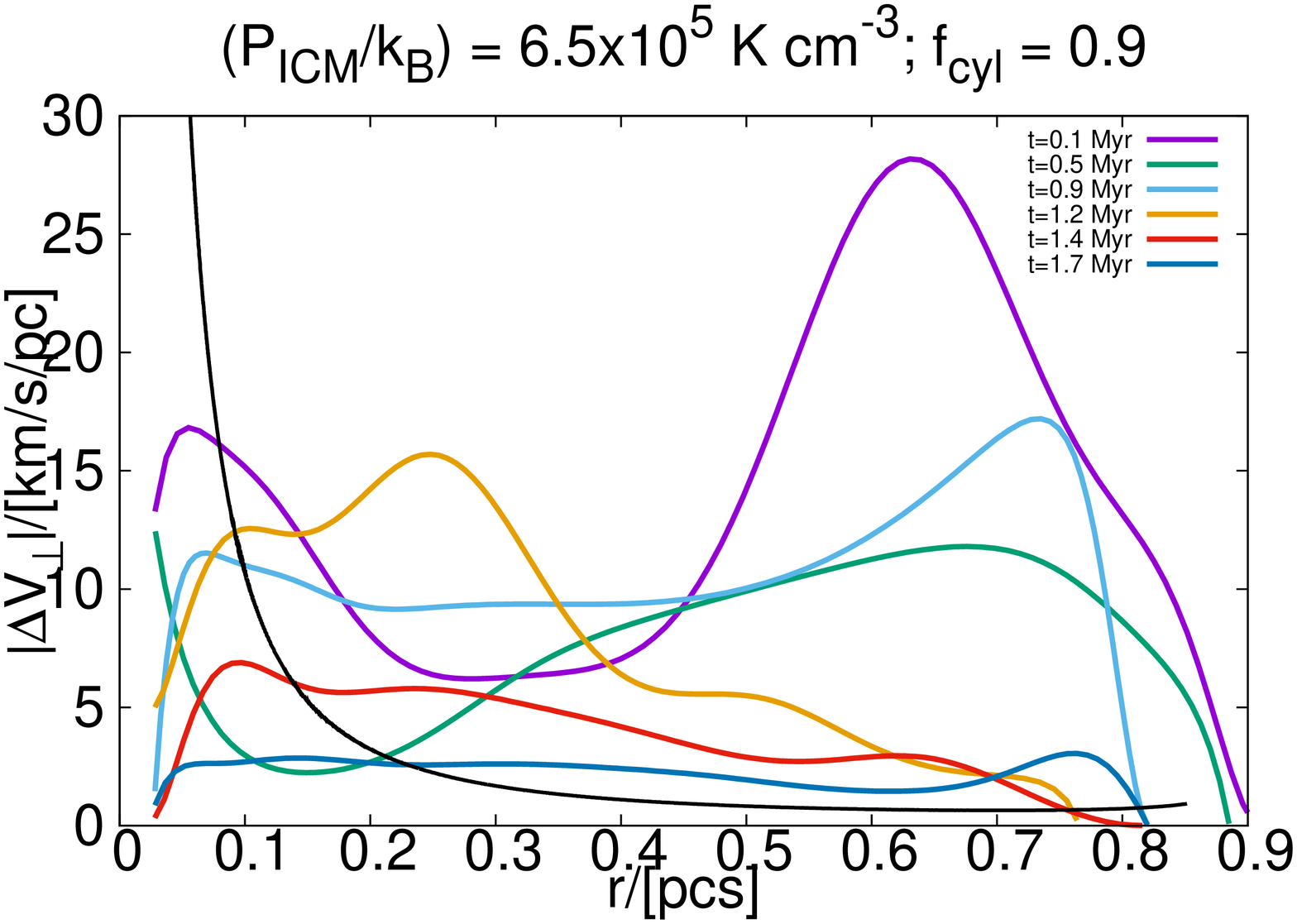}}
    \caption{Case 9}% :  $f_{cyl}$ = 0.9; $P_{ext}/k_{B}$ = 1.05x10$^{5}$ K cm$^{-3}$}
  \end{subfigure}
  \begin{subfigure}{80mm}
   \vspace*{1pt}
 \mbox{\includegraphics[angle=270,width=0.48\textwidth]{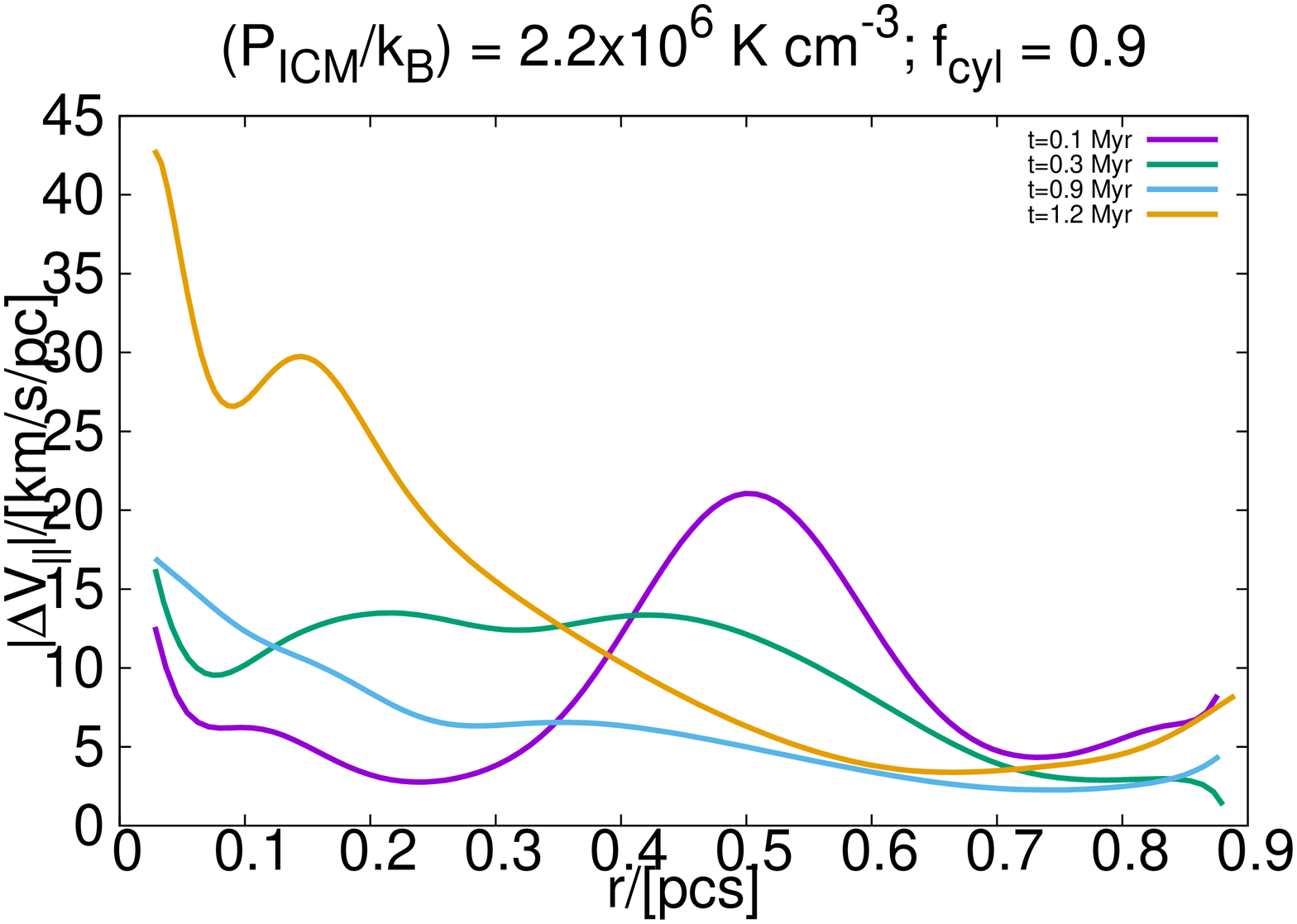}
       \includegraphics[angle=270,width=0.48\textwidth]{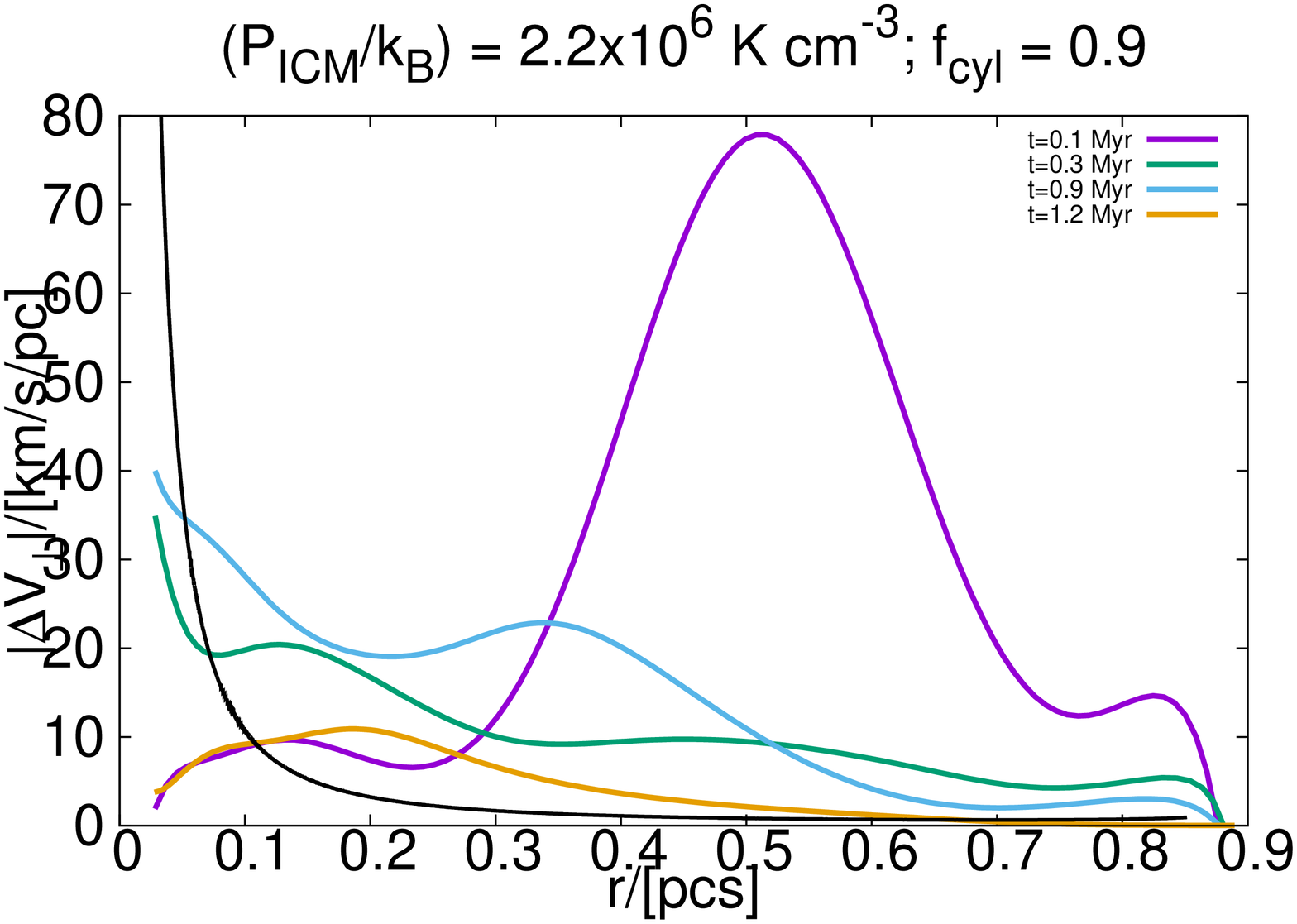}}
    \caption{Case 10} %: $f_{cyl}$ = 0.9; $P_{ext}/k_{B}$ = 3x10$^{6}$ K cm$^{-3}$}
  \end{subfigure}
  \caption{Same as plots shown on various panels of Fig. 8, but now showing the  respective velocity gradients along the radius of the filament. The black line represents Eqn. (7) for respective choices of linemass, $f_{cyl}$.}
\end{figure}
%The observed steep rise in the magnitude of $\Delta V_{\perp}$ towards the axis at some epochs is associated with the contractional phase, while the decrease in its magnitude with the expansionary phase.
\begin{figure*}
\label{Fig 10}
  \begin{subfigure}{160mm}
  \centering  
  \vspace*{1pt}   
   \includegraphics[angle=270,width=\textwidth]{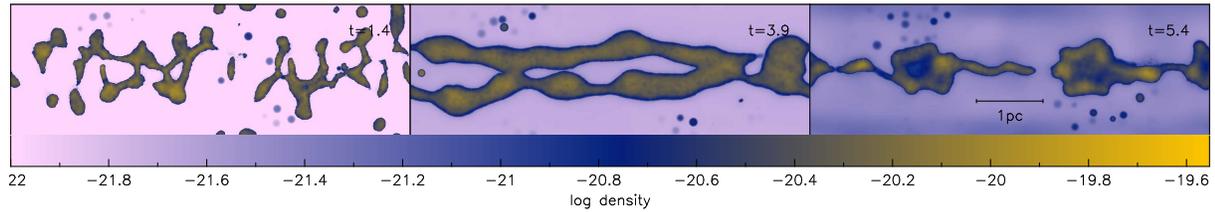}
   \caption{Rendered density plots similar to those shown earlier in Fig. 2. Images from the left to the right-hand panel correspond to Cases 4, 5 and 6 where $f_{cyl}$ = 0.2, 0.5 and 0.9 respectively.}
  \end{subfigure}
  \begin{subfigure}{160mm}
  \centering     
  \vspace*{1pt}  
  \includegraphics[angle=270,width=\textwidth]{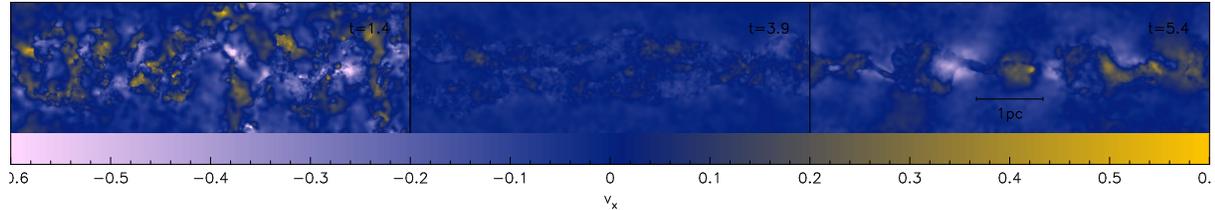}
  \caption{Velocity rendered position - position images for the same filamentary fragments as those in the panel above. The $V_{x}$ component of velocity in units of km s$^{-1}$ has been colour coded in this set of images.}
  \end{subfigure}
  \caption{Rendered density images and velocity rendered position - position images which show that a typical filament is not a singular cylindrical entity, but a composition of multiple velocity coherent entities.}
  \end{figure*}
%--------------------------------------------------- 
Analytic estimates of $\Delta V_{\perp}$ are fairly straightforward. If $M_{l}$ is the initial linemass of a filamentary fragment, then the acceleration due to self-gravity at any point $r < r_{fil}$ within it is, $\ddot{r} = \frac{-2GM_{l}}{r}$. Integrating yields the radial velocity component -
\begin{equation}
  v_{r}\equiv \dot{r} = \int_{r_{fil}}^{r_{0}} \ddot{r}\ dt\equiv 2\Big(GM_{l}\ln\Big(\frac{r_{fil}}{r_{0}}\Big)\Big)^{1/2},
\end{equation}
where $r_{0}$ is the initial position of a parcel of gas within the filamentary fragment. The radial gradient of $v_{r}$ is then simply,
\begin{equation}
\Big\vert\frac{dv_{r}}{dr}\Big\vert = \frac{r_{fil}}{r_{0}^{2}}\Big(\frac{2}{GM_{l}}\ln\Big(\frac{r_{fil}}{r_{0}}\Big)\Big)^{-1/2}.
\end{equation}
Plots on the right hand of each panel in Fig. 9 point to an interesting contrast in the behaviour of $\vert\Delta V_{\perp}\vert$ for external pressure less than $10^{5}$ K cm$^{-3}$ (as in Cases 1 and 5), and for such pressures greater than a few $\times 10^{5}$ K cm$^{-3}$ (as in Cases 9 and 10). \\\\
As is evident from these respective plots, in Cases 1 and 5 the observed $\vert\Delta V_{\perp}\vert$ is consistent with the prediction of Eqn. (7) at small radii later in the evolutionary sequence of the filament. By contrast, in the other two Cases (viz., Cases 9 and 10), however, consistency between the observed $\vert\Delta V_{\perp}\vert$ and that predicted by Eqn. (7) is only visible at the earlier epochs of filament evolution. Irrespective of the mechanism that causes the filament to contract and its central density to be accordingly enhanced, it is clear that a contracting filament is never in freefall, as was also noted in an earlier work (Anathpindika \& Freundlich 2015). The velocity gradient was calculated by using the simple trapezoidal rule to solve the integral in Eqn. (7). In their recent analysis of data for filaments in {\small NGC1333}, Chen \emph{et al.} (2020) also observed trends of decreasing magnitude of radial velocity gradients towards filament axis which suggests that none of those filaments are in freefall either. Plots on the right-hand panels of Fig. 9 suggest that the behaviour of $\vert\Delta V_{\perp}\vert$ for a filament in the field could vary depending on its evolutionary stage. Interestingly, the radial behaviour of $\Delta V_{\parallel}$, shown on the left-hand panel of Fig. 9 for the respective cases are mutually similar, apart from the fact that the magnitude of $\Delta V_{\parallel}$ increases with increasing external pressure.
%at different stages of its evolution.}\\\\
%Observations here are therefore consistent with the findings of Chen \emph{et al.} (2020) in the sense that the behaviour of the respective components of velocity is mutually similar. Crucially though, our realisations show that the direction and magnitude of these velocity components alters significantly over the course of evolution of a filamentary fragment, and as a function of external pressure.
%=================
%======================================================\\
%\begin{figure*}
%  \begin{subfigure}{80mm}
%    \centering
%   \vspace*{1pt}   
%  \mbox{\includegraphics[angle=270,width=0.48\textwidth]{F1P1-veldisp-linemass.eps}
%        \includegraphics[angle=270,width=0.48\textwidth]{F2P2-veldisp-linemass.eps}}
%   \caption{Correlation between the velocity dispersion and the linemass for respectively Cases 1 and 5.}
%  \end{subfigure}
%  \begin{subfigure}{80mm}
%    \centering
%   \vspace*{1pt}
%  \mbox{\includegraphics[angle=270,width=0.48\textwidth]{F3P3-veldisp-linemass.eps}
%        \includegraphics[angle=270,width=0.48\textwidth]{F3P4-veldisp-linemass.eps}}
% \caption{Same as the plots in panel (a) above, but now for respectively Cases 9 and 10.}    \end{subfigure}
%\end{figure*}
\subsection{Velocity coherent structures}
In the foregoing discussion we saw that the filament evolved via numerous cycles of radial contraction and that it was also simultaneously susceptible to the growth of several oscillatory features on its surface due to the rapid amplification of density perturbations. This is true irrespective of the initial linemass and the magnitude of external pressure. These dynamical processes cause rapid mixing between layers of gas within the fragment which generates sub-structure. Shown in Fig. 10(a), for instance, are rendered density images of the mid-plane of the filament in Cases 4, 5, and 6 at intermediate epochs when sub filamentary structures, also known as \emph{fibres}, to use the parlance introduced by Hacar \& Tafalla (2011) and Hacar \emph{et al.} (2013), are visible. In other words, the filament no longer appears like a singular feature and instead, more than one elongated features are now visible along the original axis of the filament. This substructure is more readily visible from the coeval rendered images in Fig. 10(b) that show the local velocity-field in these regions of the filament.\\\\
%In fact, the velocity coherent structures in Cases 4 and 5 shown on the left and the central panels of Figs. 10(a) and 10(b), respectively, also show evidence of core-formation via the \emph{collect-and-collapse} mode discussed earlier in \S 3.1. On the other hand, \emph{fibres} in Case 6 ($f_{cyl}$ = 0.9), where the filamentary fragment was initially transcritical and shown in the right panel of Figs. 10 (a) and 10 (b), appear tightly bound together with some evidence of formation of oblate cores via gravitational fragmentation. 
%======================================================
\begin{figure}
\label{Fig 11}
  \begin{subfigure}{80mm}
   \vspace*{1pt}   
  \mbox{\includegraphics[angle=270,width=0.48\textwidth]{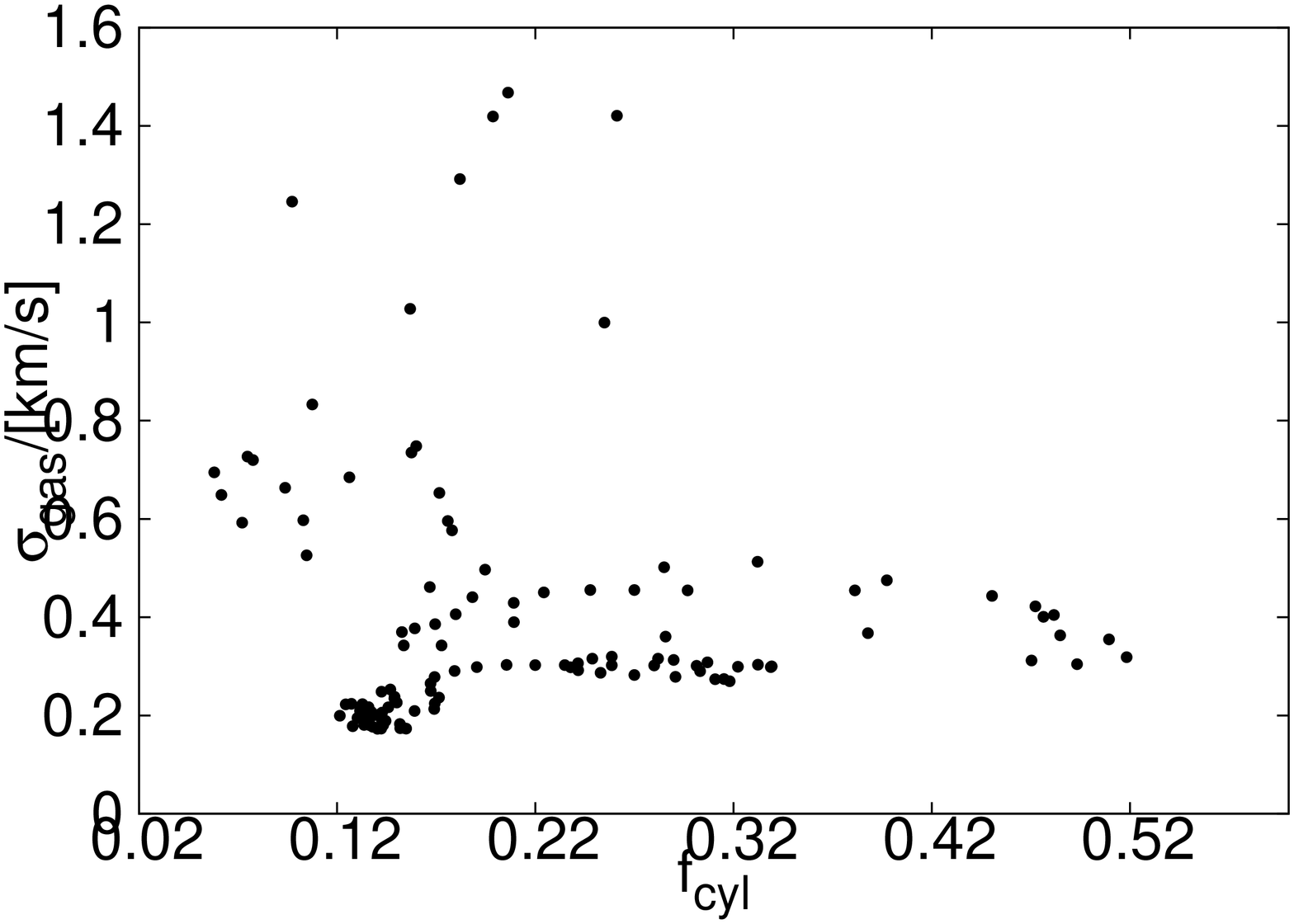}
        \includegraphics[angle=270,width=0.48\textwidth]{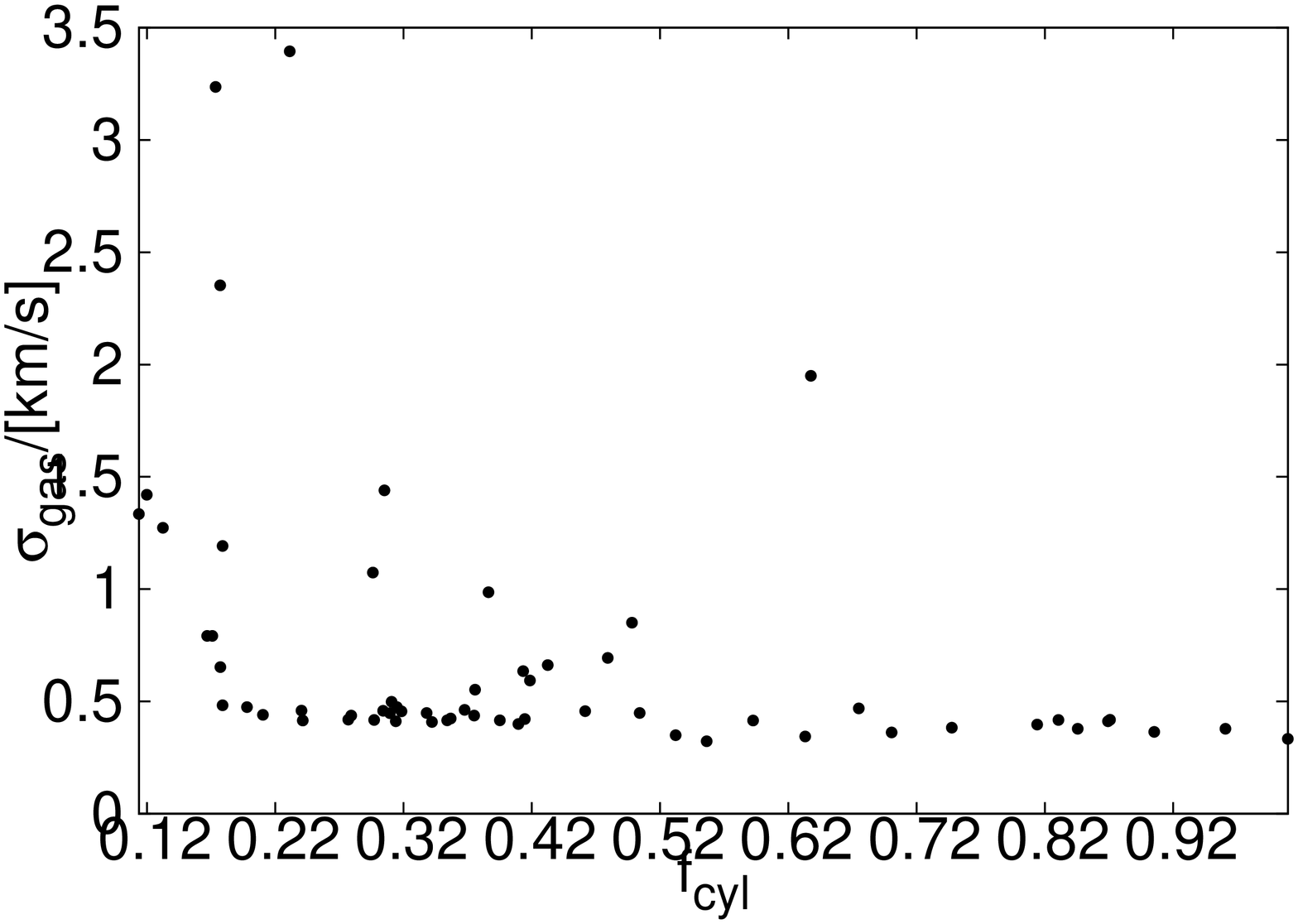}}
   \caption{Correlation between the velocity dispersion and the linemass for respectively Cases 1 and 5 shown on the left and right-hand panel.}
  \end{subfigure}
  \begin{subfigure}{80mm}
   \vspace*{1pt}
  \mbox{\includegraphics[angle=270,width=0.48\textwidth]{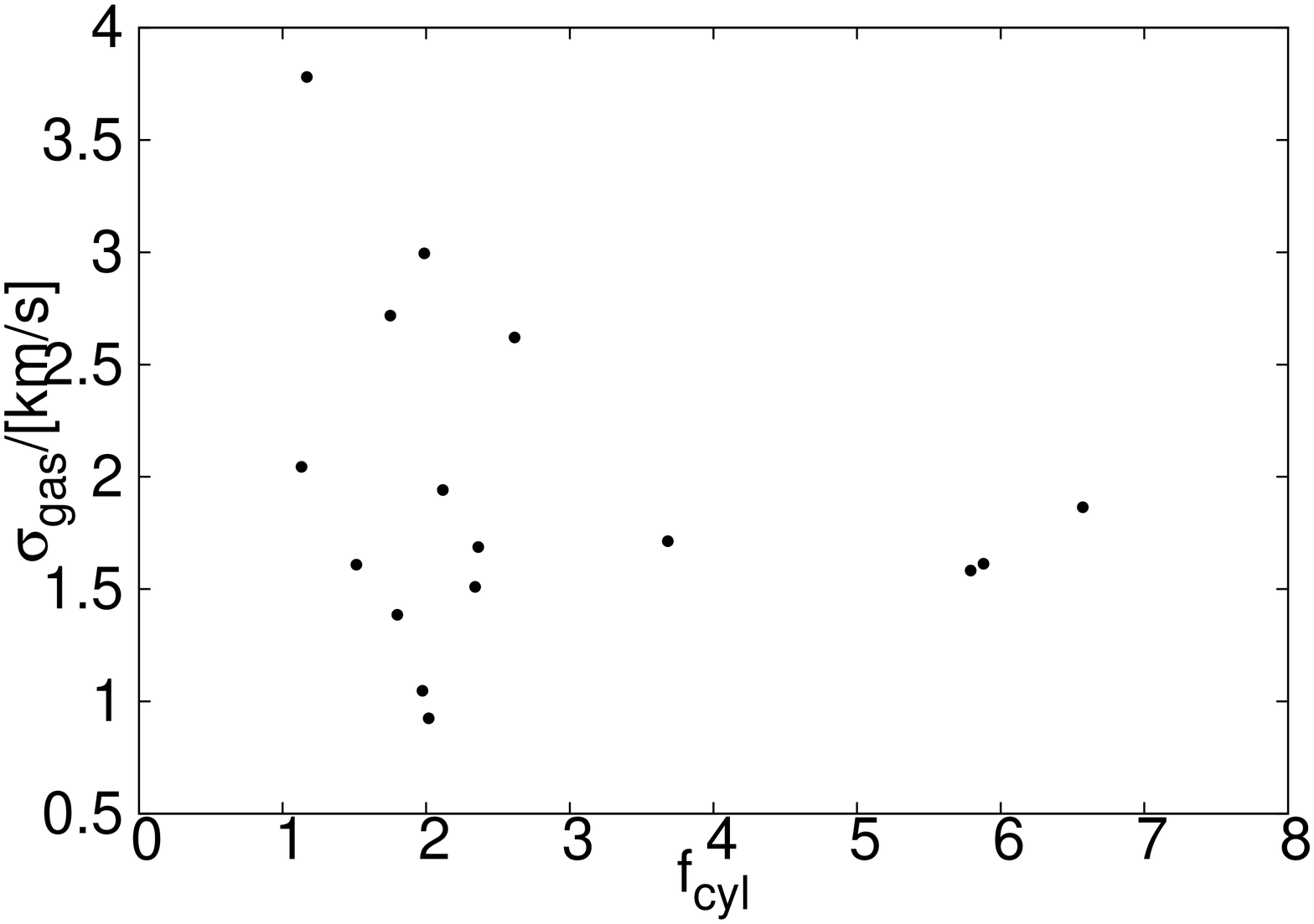}
        \includegraphics[angle=270,width=0.48\textwidth]{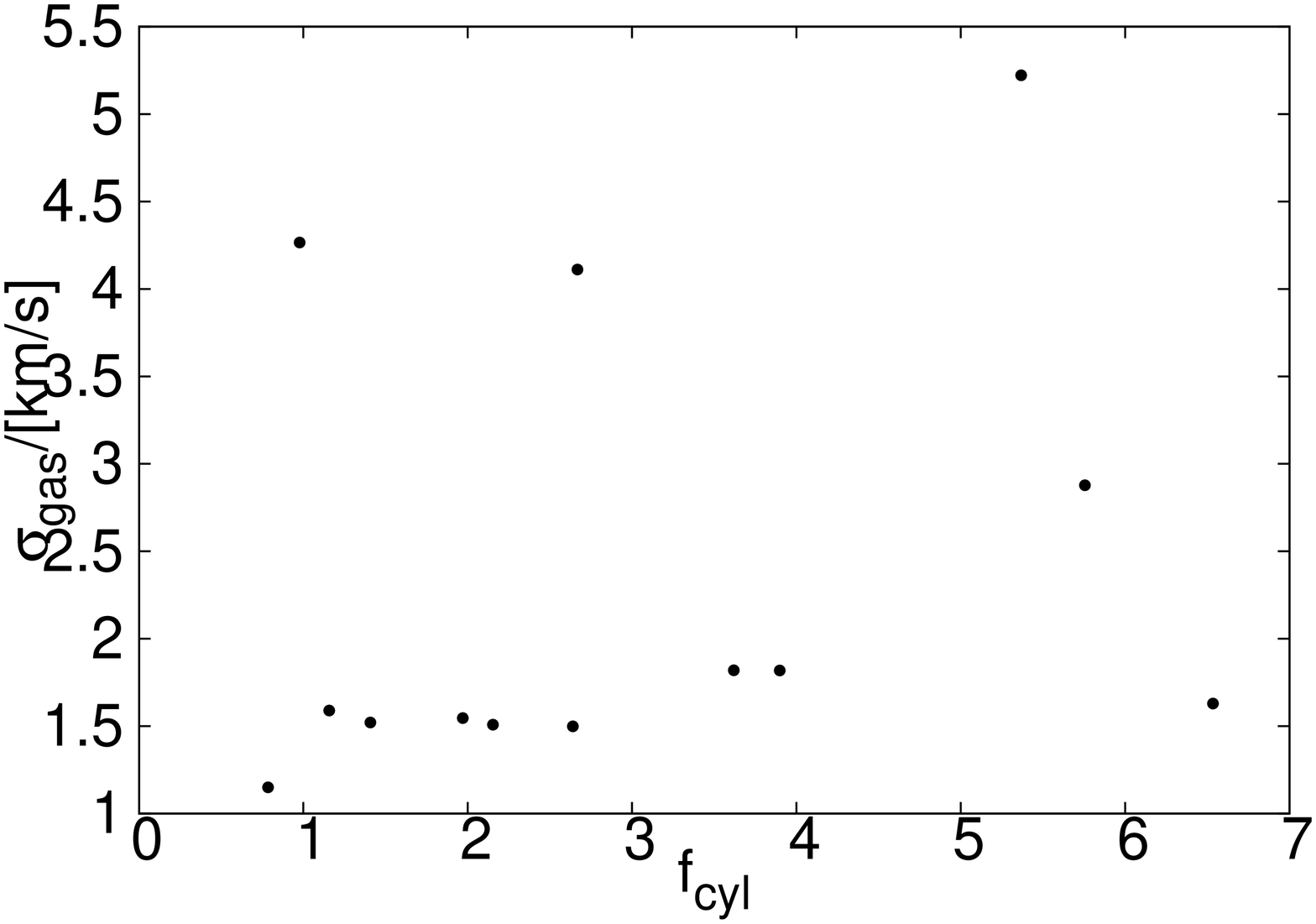}}
    \caption{Same as the plots in panel (a) above, but now for respectively Cases 9 and 10 shown on the left and right-hand panel.}
  \end{subfigure}
  \begin{subfigure}{80mm}
   \vspace*{1pt}
  \mbox{\includegraphics[angle=270,width=0.48\textwidth]{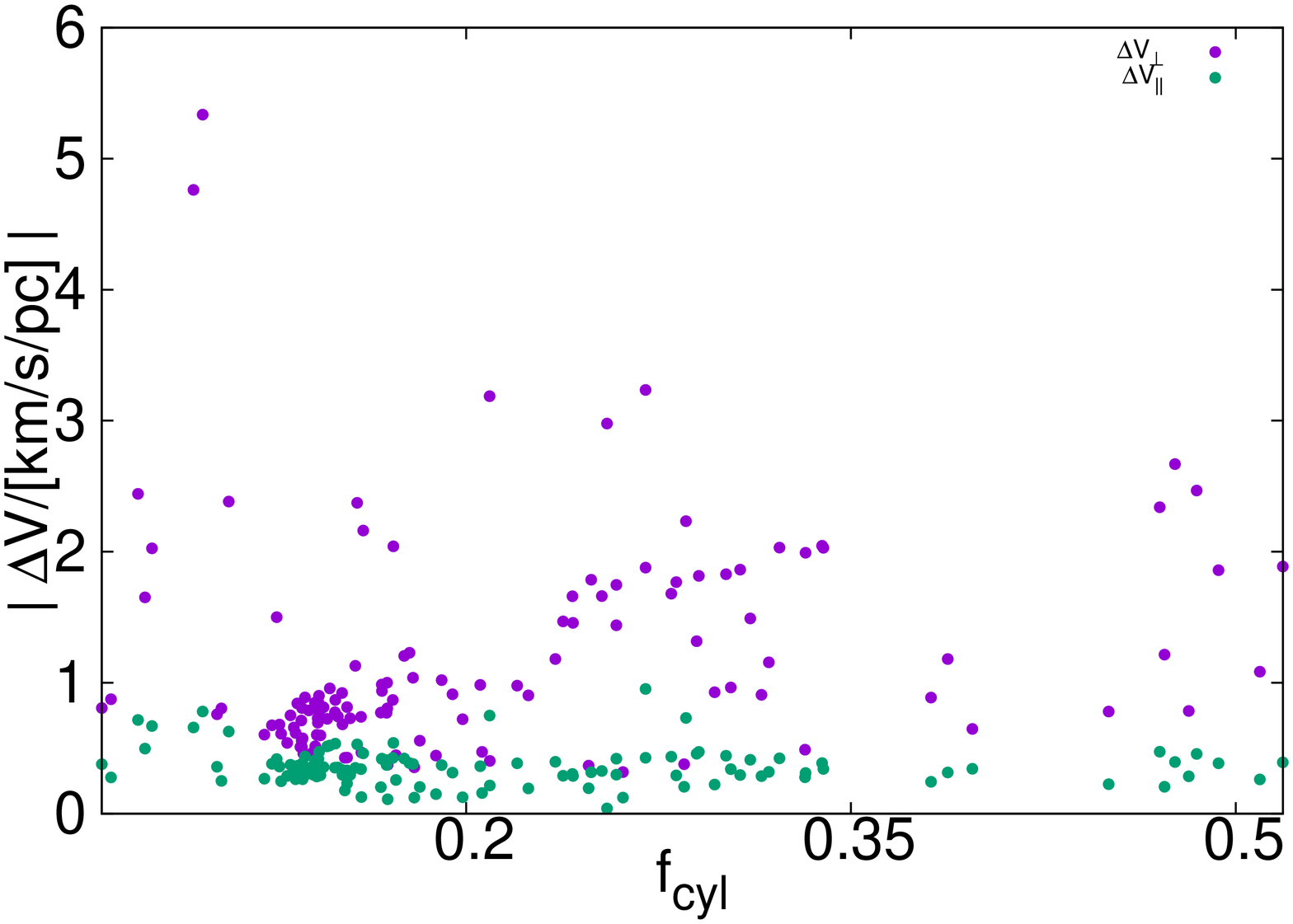}
        \includegraphics[angle=270,width=0.48\textwidth]{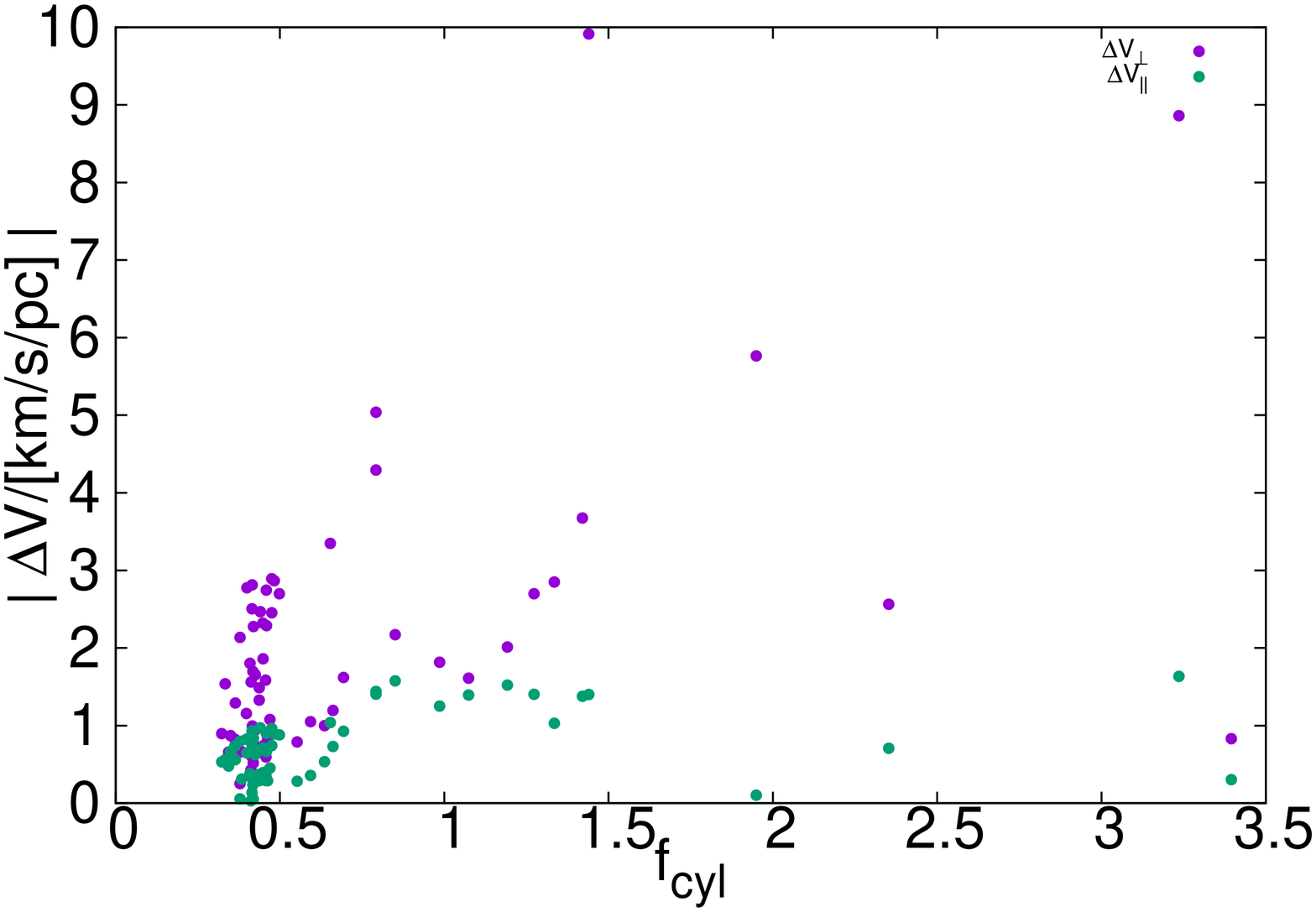}}
    \caption{Correlation between the parallel and radial component of velocity gradient and the line mass for respectively, Cases 1 and 5 shown on the left and right-hand panel.}
  \end{subfigure}
  \begin{subfigure}{80mm}
   \vspace*{1pt}
 \mbox{\includegraphics[angle=270,width=0.48\textwidth]{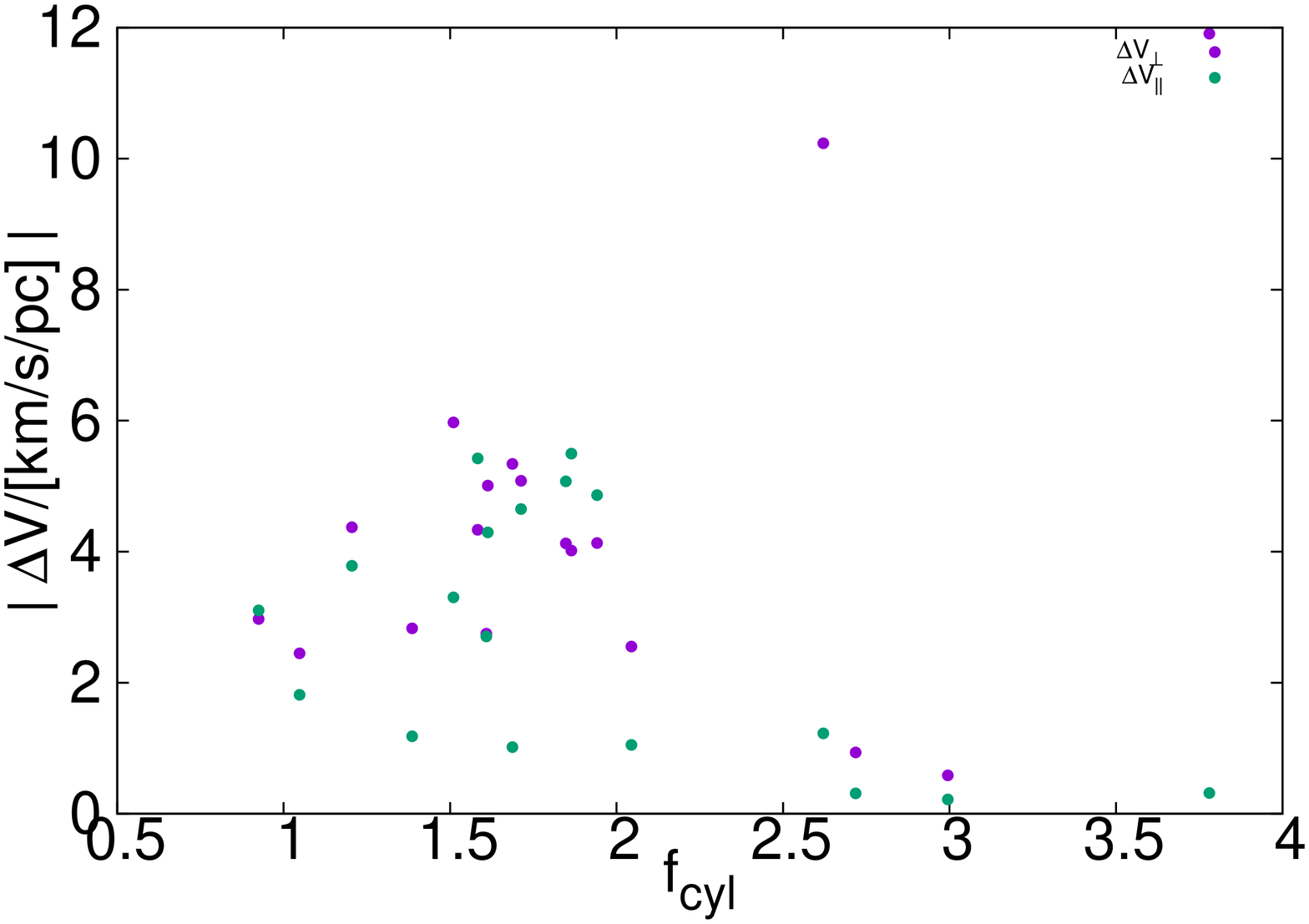}
       \includegraphics[angle=270,width=0.48\textwidth]{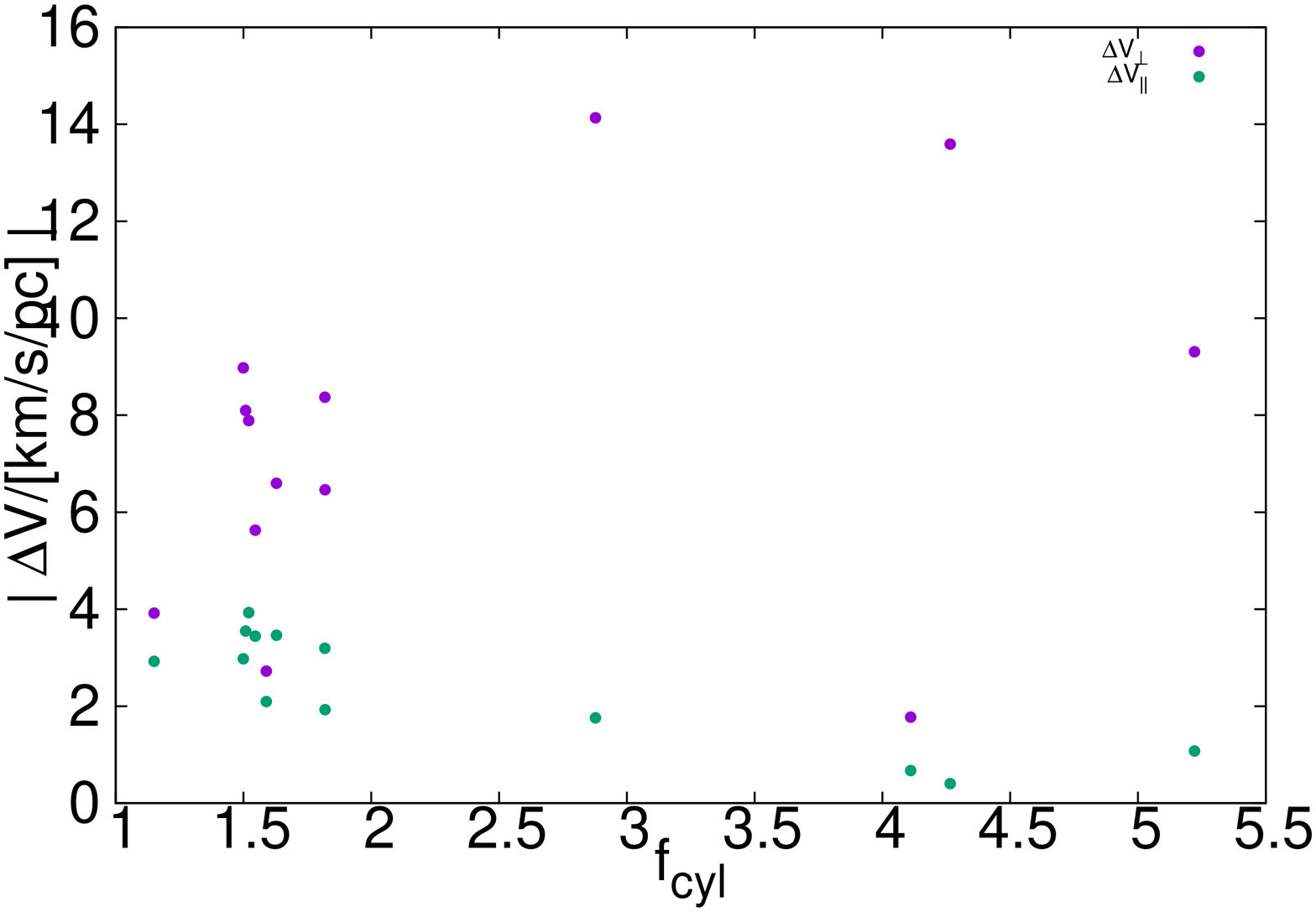}}
    \caption{Same as the plots in panel (c) above, but now for respectively, Cases 9 and 10 shown on the left and right-hand panel.}
  \end{subfigure}
  \caption{Correlations between the velocity dispersion, the parallel and the radial component of velocity gradient and the linemass of the filamentary fragment.}
\end{figure}
%==============================================================
%==================================
\begin{figure}
\label{Fig 12}
  \vspace{1pt}
  \centering
  \includegraphics[angle=270,width=0.48\textwidth]{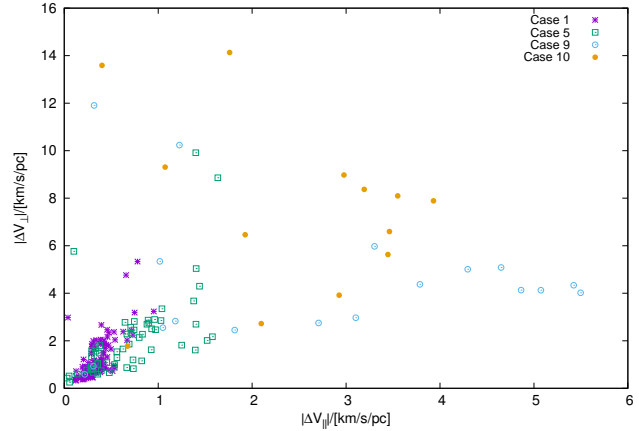}
  \caption{Same as plots in Fig. 11, but now showing the correlation between parallel and radial components of the velocity gradient.}
\end{figure}
\subsection{Correlations between linemass, velocity dispersion and the components of velocity gradient}
In view of the findings by Klessen \& Hennebelle (2010) from their numerical work, Arzoumanian \emph{et al.} (2013) argued that the observed correlation between the velocity dispersion $\sigma_{gas}$, and the linemass of thermally super-critical filaments in the Serpens South region, was evidence for filament growth due to accretion. In other words, the accreting gas not only injects turbulence in filaments but also makes them super-critical over the course of their evolution. Interestingly, however, Arzoumanian \emph{et al.} (2013) did not observe a similar correlation for sub-critical filaments which within the Klessen \& Hennebelle (2010) paradigm could imply these filaments are not accreting, and therefore are relatively stable. \\\\
Before proceeding to examine correlations between various physical quantities, let us first describe the calculation of the filament linemass, $M_{l}$, at each epoch of its evolution. We recall from Paper I that the filament linemass was calculated at different locations along its length as, $M_{l}(l) = \rho_{l}r_{l}^{2}$. Here $\rho_{l}$ and $r_{l}$ are respectively the mean density, and the effective filament radius at a location $l$ along its length. The number of intervals ($N_{int}$) along the length of the filament were calculated using the usual Sturges criterion (Sturges 1926). The mean filament linemass was calculated as $M_{l}=\frac{1}{N_{int}} \sum_l M_{l}(l)$. Note that the observed temporal variation in linemass of a filament - despite the filament not accreting gas from its surroundings - is essentially a reflection of the corresponding variation in its radial density. While plots shown on various panels of Fig. 9 show clearly that the radial extent of the filament (i.e., its outer radius, $r_{fil}$), does not vary much over the course of its evolution, the radial distribution of gas density within it, however, varies significantly over the course of its evolution. Thus during the contractional phase, for instance, the density at small radii close to the central axis is significantly larger than that further out. Hence, the linemass, $M_{l}(l)$, as calculated above is dominated by contributions from close to the central axis. The converse is true during the relaxation phase of the filament. \\\\
To calculate the density, $\rho_{l}$, we divide the transverse section of the filament at each axial location, $l$, into $N_{shell}$ number of concentric rings. Gas particles are then distributed over these concentric shells. Thus, a shell with higher density is one with a relatively large mass or equivalently, one in which a higher number of gas particles, are concentrated. Naturally then, a shell with little or no mass, i.e., one with few or no particles in it contributes little, or not at all to the calculation of mean density at a location $l$ along the filament length. The effective filament radius $r_{l}$ at this location is also then smaller than $r_{fil}$. Thus the filament linemass increases when the filament is centrally condensed and has a smaller effective radius $r_{l}$
. \\\\
The density at an axial location $l$ is then, $\rho_{l}\equiv\frac{1}{N_{shell}}\Sigma_{j}\Big(\frac{m_{j}(r_{j} < r_{fil})}{vol_{j}}\Big)$. The subscript $j$ on the right hand side of this expression runs over the number of shells at each axial location $l$; $m_{j}$ and $vol_{j}$ are respectively the mass of each concentric shell at a radial location $r_{j} < r_{fil}$, and the volume of this shell. By choice, each shell has the same volume. The number of concentric rings is also chosen according to the Sturges criterion. Note, the mass of a shell is calculated merely by summing over the masses of individual particles in it. Thus it is important to bear in mind that the mean linemass, $M_{l}$, is not calculated by using the average density and the mean radius of the filament at a given epoch. \\\\
Now Fig. 11(a) shows the $\sigma_{gas} - f_{cyl}$ correlation for Cases 1 and 5 where the filament was initially sub-critical. Recall that $f_{cyl}$ is the ratio of the linemass $M_{l}$, to the critical linemass $M_{l_{crit}}$ (calculated for the average gas temperature of the filament at that epoch). Meanwhile, Table 3 lists the Spearman rank coefficient, $r_{s}$ for various correlations derived here. Column 3 of Table 3 shows, $\sigma_{gas}$ and $f_{cyl}$ are weakly correlated in Cases 1 and 10 while they are anticorrelated in Cases 5 and 9. This observed anticorrelation in the latter two Cases is, however, inconsistent with the deductions made by Arzoumanian \emph{et al.} (2013). Importantly, the nature of the $\sigma_{gas} - f_{cyl}$ correlation does not exhibit any clear dependence on the magnitude of external pressure.\\\\
\begin{table}
\centering
\caption{Spearman rank coefficient ($r_{s}$) for respective data plotted in Figs. 11 and 12}
\begin{tabular}{|l|c|c|c|c|c|}
\hline
Sr & $\frac{(P_{ext}/k_{B})}{[K \mathrm{cm}^{-3}]}$ & $(r_{s})^{a}$ & $(r_{s})^{b}$ & $(r_{s})^{c}$ & $(r_{s})^{d}$ \\
No. & &  & & &  \\
\hline
1 & 6.0$\times 10^{3}$ (Case 1)& 0.29 & -0.62 &-0.92 & 0.67 \\
2 & 7.3$\times 10^{4}$ (Case 5)& -0.61 & -0.24 & -0.19 & 0.24\\
3 & 6.5$\times 10^{5}$ (Case 9)& -0.23 & 0.38 & 0.72 & 0.30 \\
4 & 2.2$\times 10^{6}$ (Case 10)& 0.46 & 0.33 & -0.03 & -0.14 \\
\hline
\end{tabular}
$ ^{a} (\sigma_{gas} - f_{cyl}), ^{b} (\Delta V_{\perp} - f_{cyl}), ^{c} (\Delta V_{\parallel} - f_{cyl}), ^{d} (\Delta V_{\parallel}-\Delta V_{\perp}) $
\end{table}
While there is \emph{ambiguity in the nature of} the $\sigma_{gas} - f_{cyl}$ correlation, 
Figs. 11(a) and (b) show clearly that the magnitude of $\sigma_{gas}$ increases with an increase in external pressure. This behaviour highlights the impact of varying environmental conditions, i.e., merely the magnitude of external pressure without having to invoke additionally the magnetic field. Within the paradigm of filament assembly due to accretional flows, one may also expect the radial component of the velocity gradient to be correlated with its linemass as is indeed reported by Chen \emph{et al.} (2020) in the case of filaments detected in {\small NGC 1333}. Figs. 11(c) and (d) and the corresponding Spearman rank coefficient listed in Column 4 of Table 3, however, show for the above four cases that $\vert \Delta V_{\perp}\vert$  actually moderately anticorrelates with the linemass in Cases 1 and 5, while they moderately correlate in the other two cases, viz., Cases 9 and 10. Similarly, Column 5 of Table 3 shows, $\vert\Delta V_{\parallel}\vert$ is anticorrelated, also moderately, with the linemass in Cases 1, 5 and 10. This correlation, however, is relatively strong in Case 9. Nevertheless, Figs 11(c) and (d) again highlight the fact that $\vert\Delta V_{\perp}\vert$ increases with increasing external pressure, which once again underscores the impact of ambient environment. \\\\
Finally, the correlation between $|\Delta V_{\parallel}|$ and $|\Delta V_{\perp}|$ shown in Fig. 12 is merely an extension of the results discussed above. The plot shows that the axial component of the velocity gradient is at least weakly correlated with the radial component of the velocity gradient for most of the Cases. The Spearman rank coefficient for the respective realisations is listed in Column 6 of Table 3. Given the representative nature of the choice of various parameters involved in this work, however, we should not read too much into the actual magnitudes of velocity dispersion and the respective velocity gradients, but instead look at the strength of respective correlations as reflected by the Spearman rank coefficient. The upshot of the various plots shown in Figs. 11 and 12 is that 
the respective correlations do not exhibit any clear dependence on the magnitude of external pressure. The observed moderate correlation between these quantities here demands a more realistic initial model configuration to probe the filament assembly model via inflowing gas.
%$|\Delta V_{\parallel}|$ and $|\Delta V_{\perp}|$ and $\sigma_{gas}$ increase with increasing $P_{ext}$. 
\subsubsection{Radial density profile of filaments}
As has been noted in \S 3.1 above, the filament contracts radially irrespective of its initial linemass and the external pressure. Shown on various panels of Fig. 14 are radial density profiles of the filaments at different epochs of their evolution across the realisations developed in this work. We observe that the density profiles (as a function of cylindrical radius $r$) are consistent with the Plummer distribution given by 
\begin{equation}
\rho(r) = \frac{\rho_{c}}{[1 + (r/r_{flat})^{2}]^{\frac{p}{2}}},
\end{equation}
where $\rho_{c}$ is the central density and $r_{flat}$ is the radius of the flat inner portion of the density profile, as is often the case with filaments observed in nearby star-forming {\small MCs} (e.g., Arzoumanian \emph{et al.} 2011). We note that sometimes during a contractional phase, density profiles slightly steeper than the Plummer profile are also observed, as seen in the central panel of Fig. 14. Interestingly, however, the observed radial density profiles are generally shallower than the Ostriker (1964) model ($p$=4) for an isothermal filament, which is consistent with our findings reported in Paper I. \\\\
Although there is a slight increase in the peak density with increasing external pressure, the density profiles do not exhibit steepening. We suspect this behaviour stems from the interplay in the filament between self-gravity and the buoyancy provided by thermal pressure and the pressure due to injected turbulence. In the foregoing, we noted that a higher external pressure injects a stronger turbulence so that the density profile remains shallow even in Cases 9 and 10, for example, where the self-gravity was the strongest in the current ensemble of realisations. 
\subsection{Variation of the random number generator seed}
We repeated the entire set of realisations  listed in Table 1 with a new seed for the random number generator to generate the turbulent velocity fields. In the interest of brevity, however, shown in Fig. 13 is the rendered image of the terminal epoch of the filamentary fragment for a repetition of Case 10 only. This image is qualitatively similar to the one shown in Fig. 7 for Case 10 where we saw that the filament eventually ruptured and ended up as a set of elongated fragments, of which some show evidence of localised collapse. Thus varying the random number generator does not have any bearing on the evolution of the filament.
\begin{figure}
\label{Fig 13}
  \vspace*{1pt}
  \includegraphics[angle=270,width=0.48\textwidth]{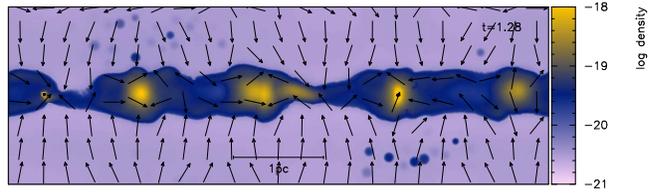}
  \caption{Same as the rendered density image shown earlier on the right - hand panel of Fig. 7, but now showing the terminal epoch of the fragment in a repetition of Case 10 with a different random number generator. As usual, the thick black blobs represent position of the sink particles.}
\end{figure}
 %and that are at least mutually weakly correlated which lends credence to the 
\begin{figure*}
\label{Fig 14}
\begin{subfigure}{160mm}
  \vspace{1pt}
  \mbox{\includegraphics[angle=270,width=0.33\textwidth]{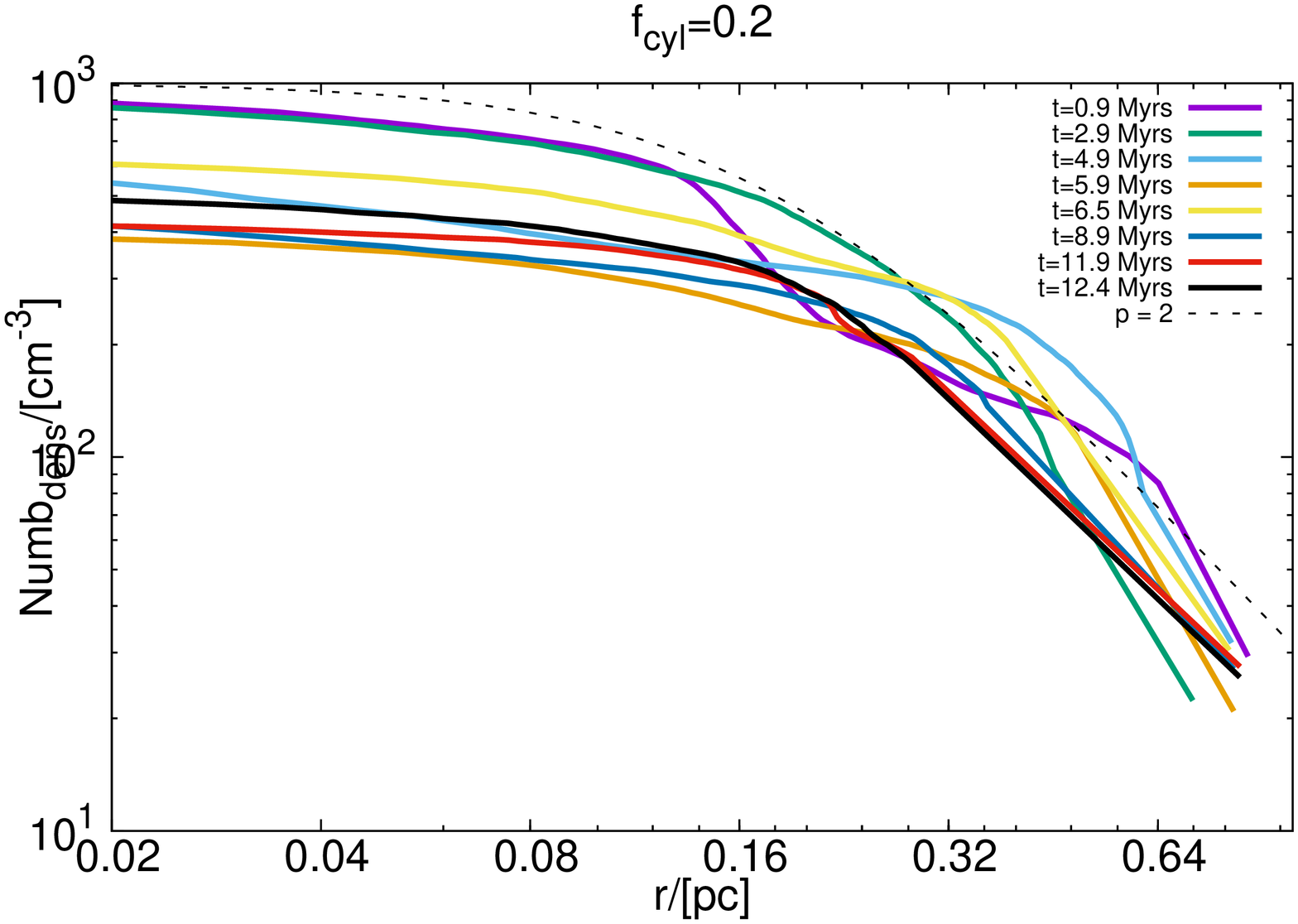}
        \includegraphics[angle=270,width=0.33\textwidth]{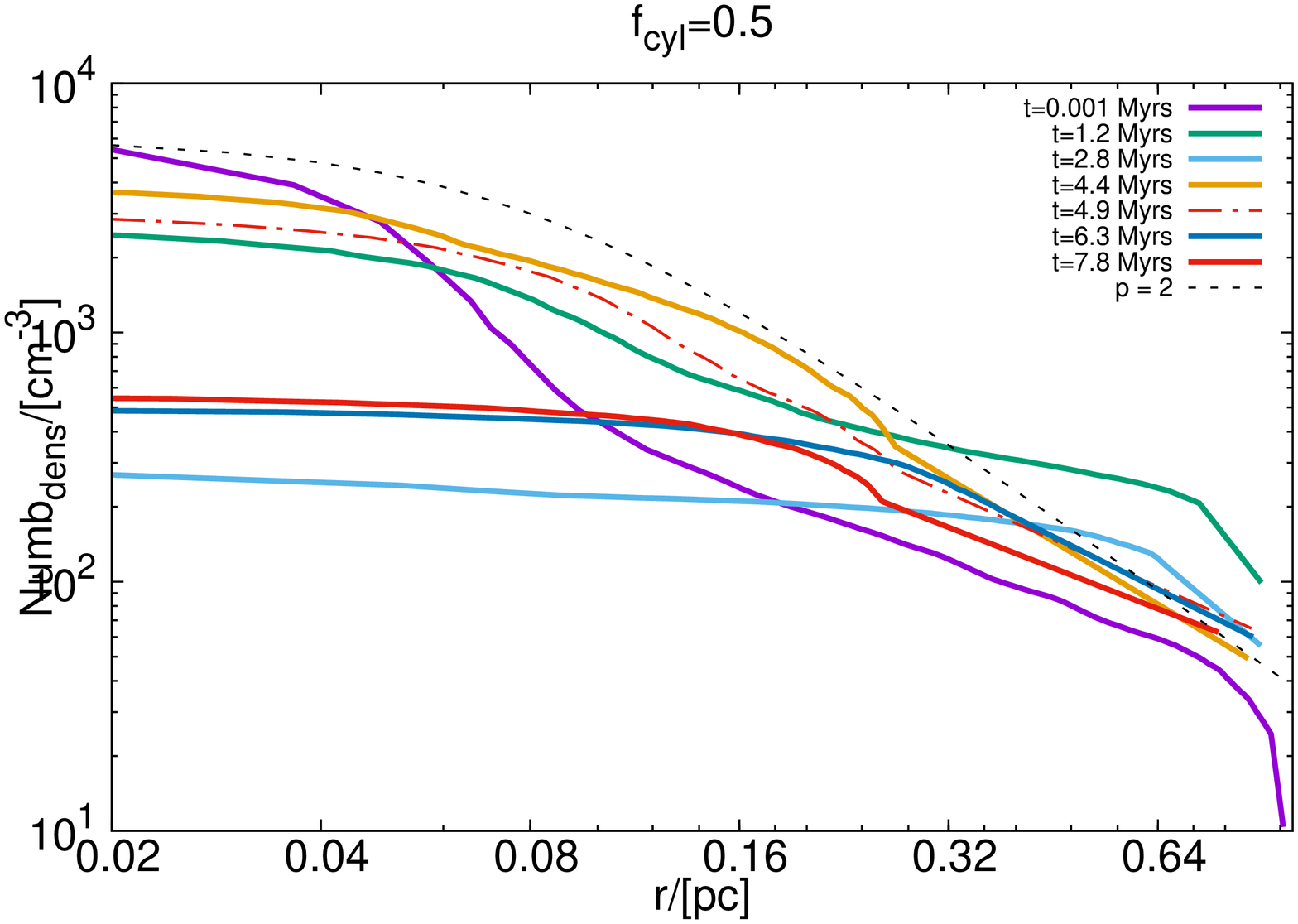}
        \includegraphics[angle=270,width=0.33\textwidth]{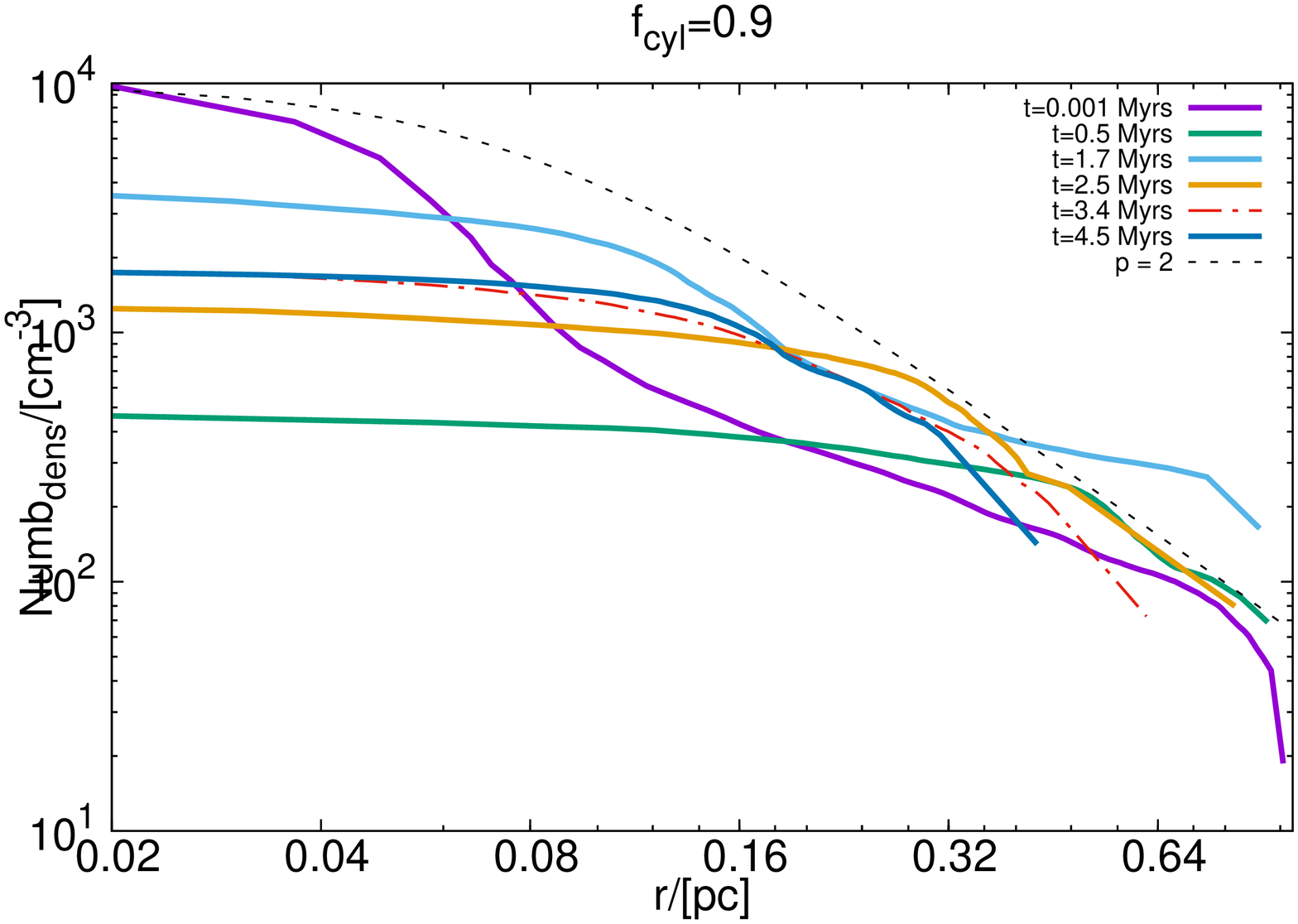}}
  \caption{$\frac{P_{ext}}{k_{B}} = 6\times 10^{3}$ K cm$^{-3}$}
\end{subfigure}
\begin{subfigure}{160mm}
  \vspace{1pt}
  \mbox{\includegraphics[angle=270,width=0.33\textwidth]{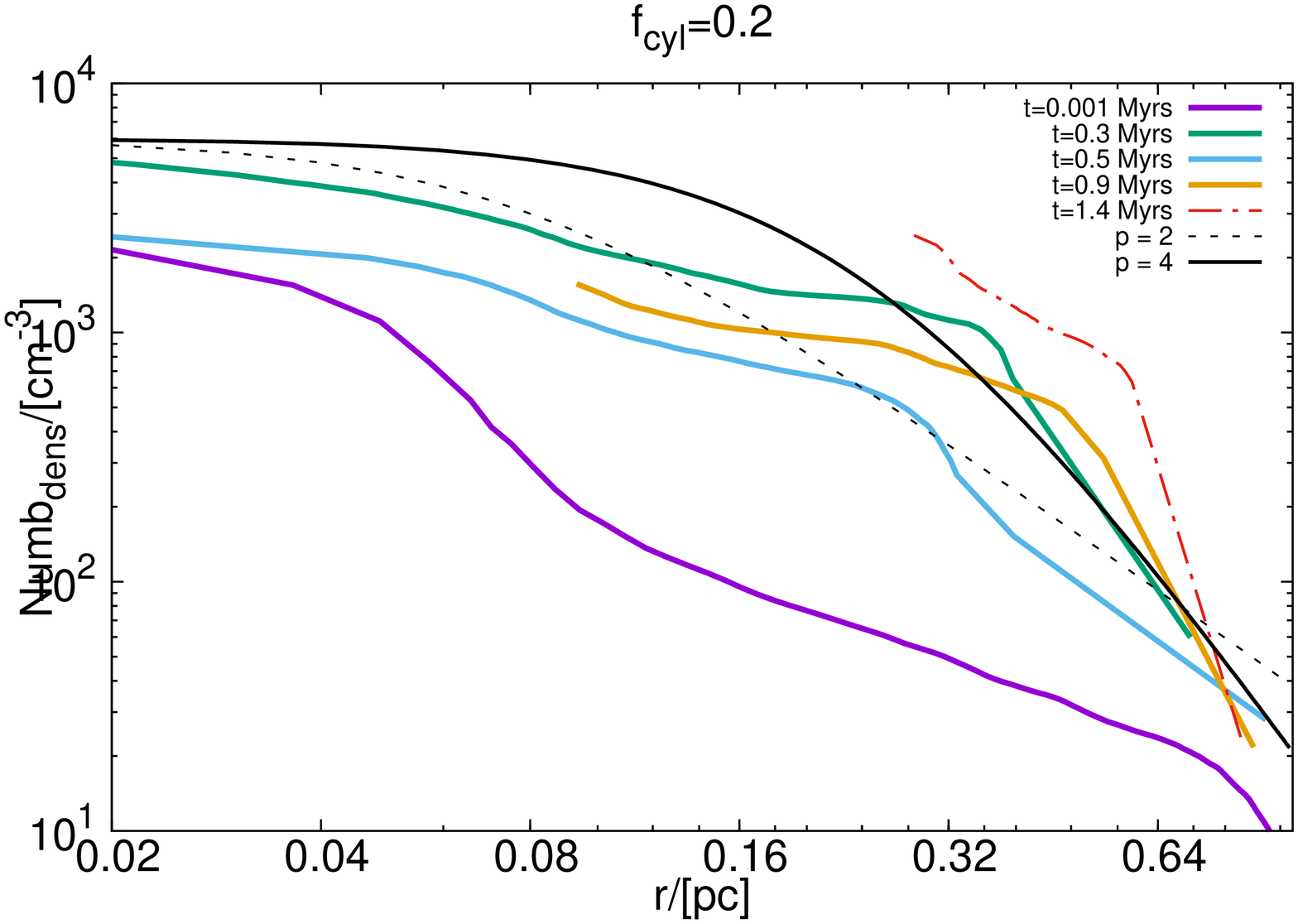}
        \includegraphics[angle=270,width=0.33\textwidth]{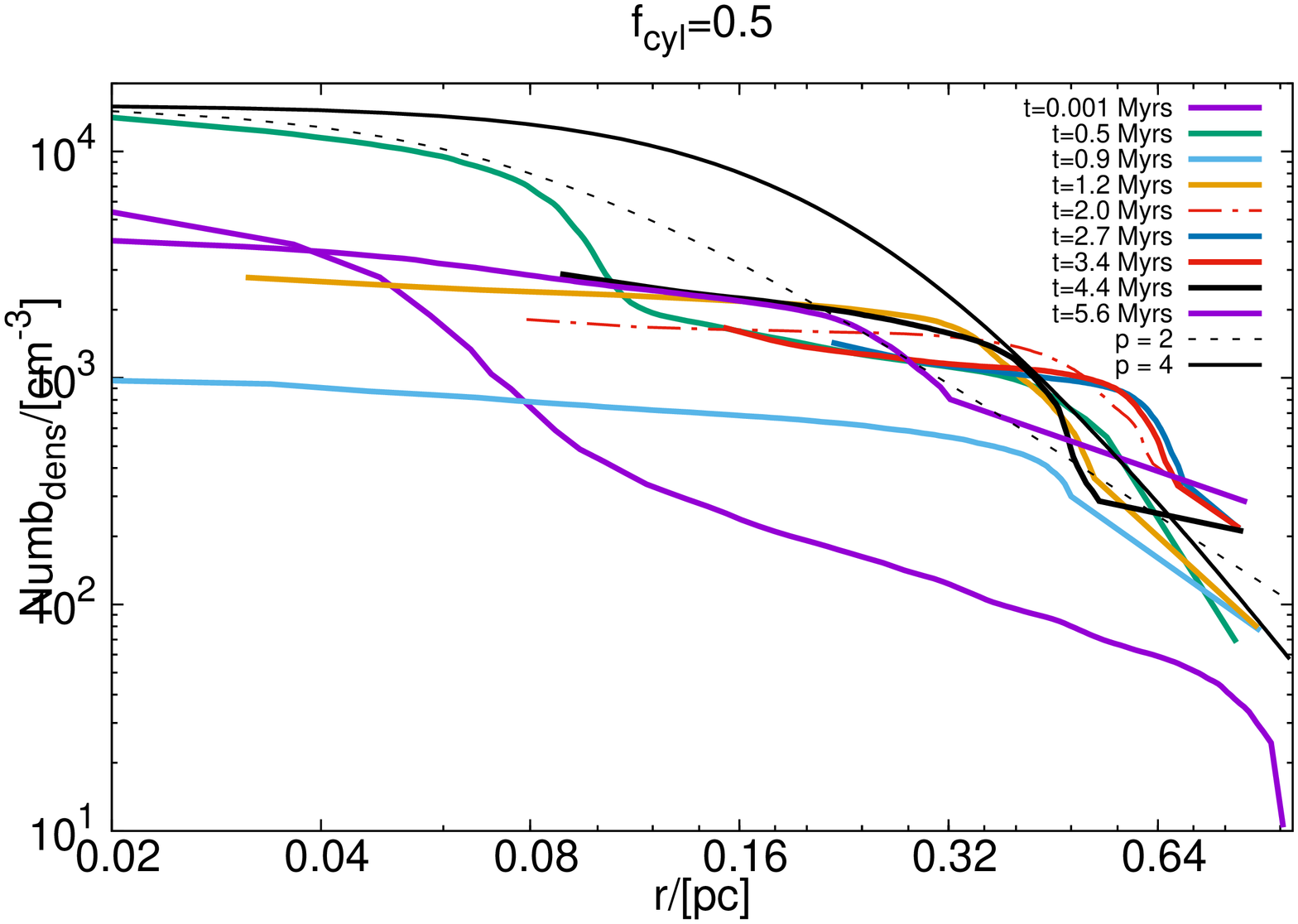}
        \includegraphics[angle=270,width=0.33\textwidth]{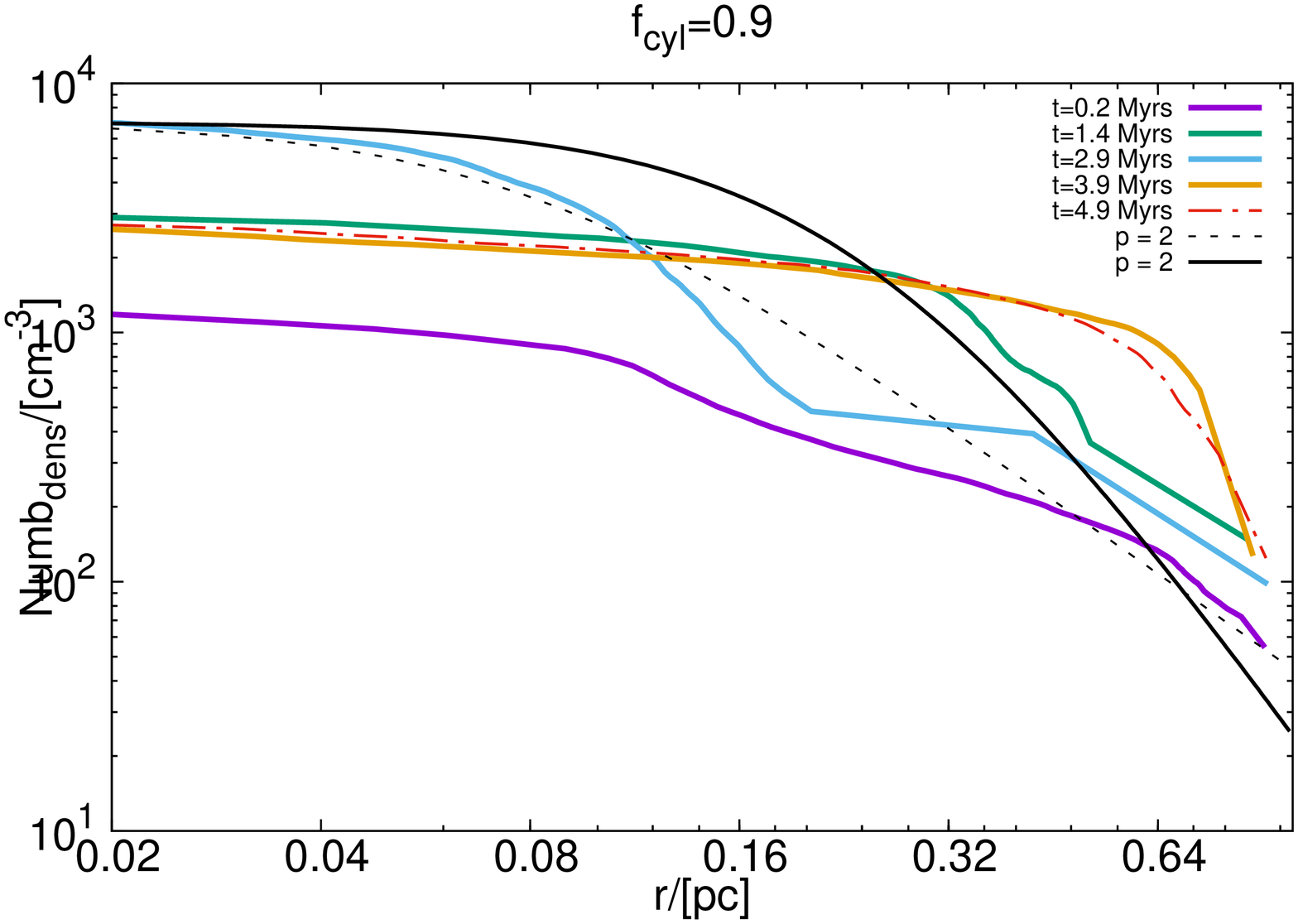}}
  \caption{$\frac{P_{ext}}{k_{B}} = 7.3\times 10^{4}$ K cm$^{-3}$}
\end{subfigure}
\begin{subfigure}{160mm}
  \vspace{1pt}
  \mbox{\includegraphics[angle=270,width=0.33\textwidth]{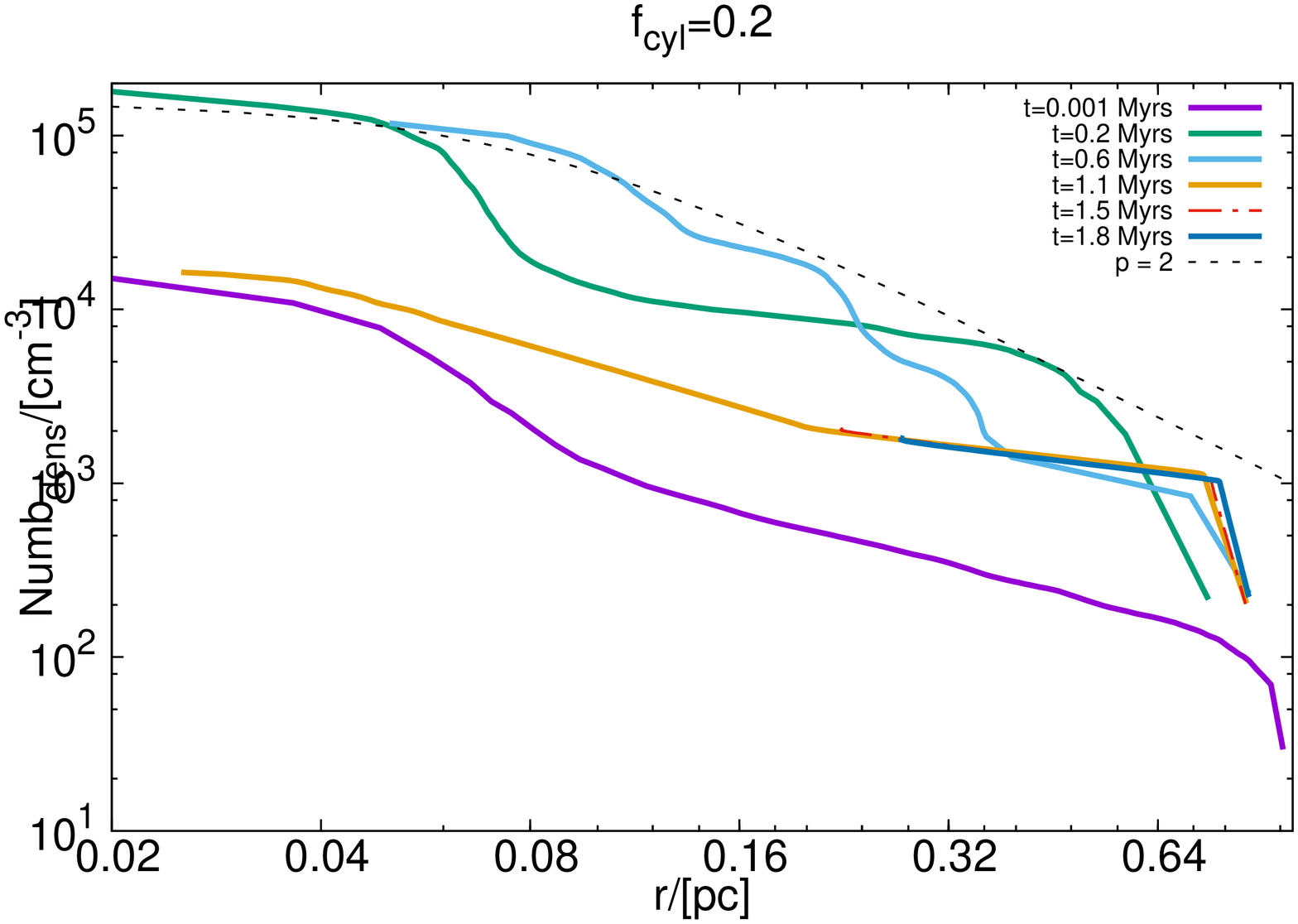}
        \includegraphics[angle=270,width=0.33\textwidth]{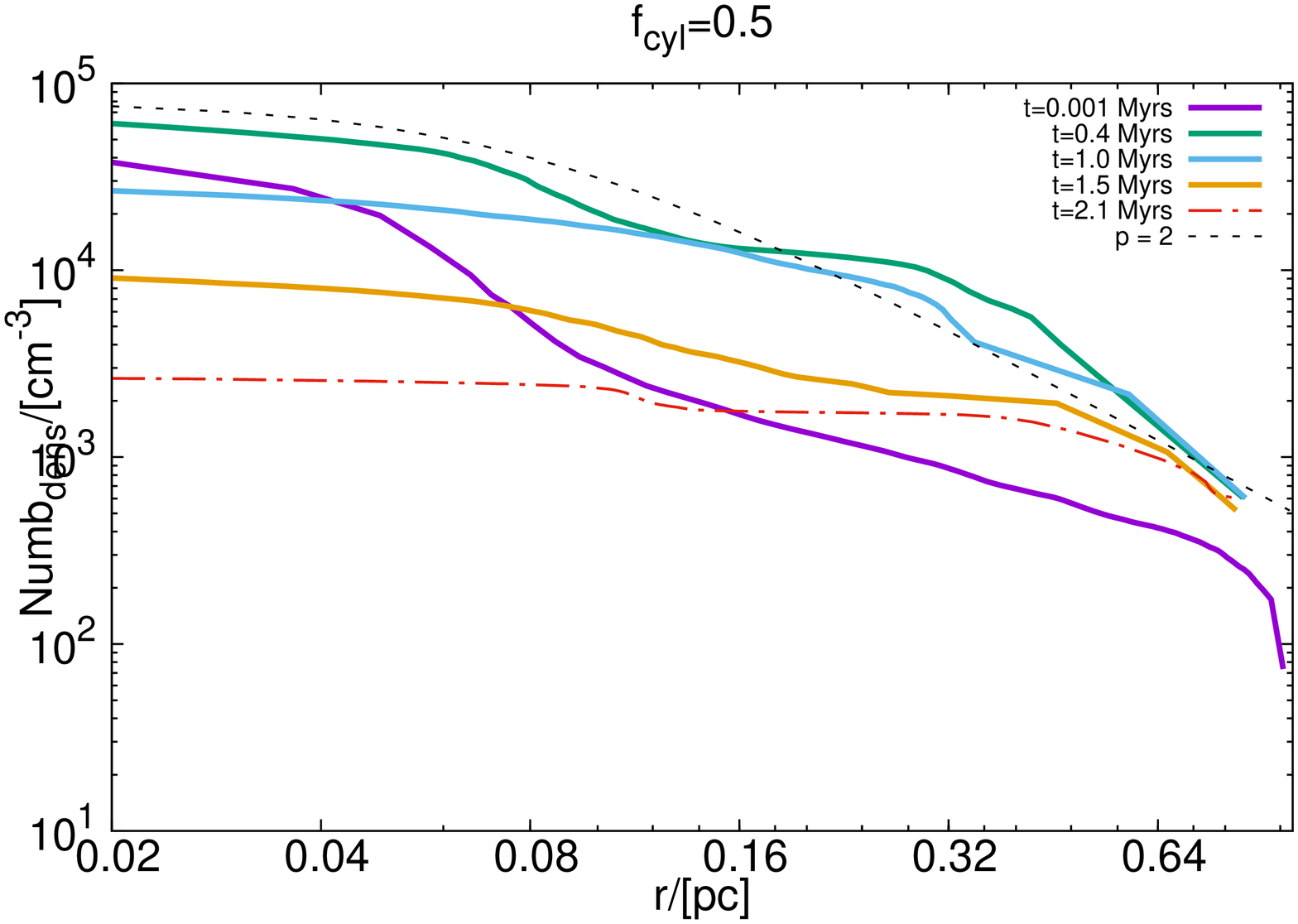}
        \includegraphics[angle=270,width=0.33\textwidth]{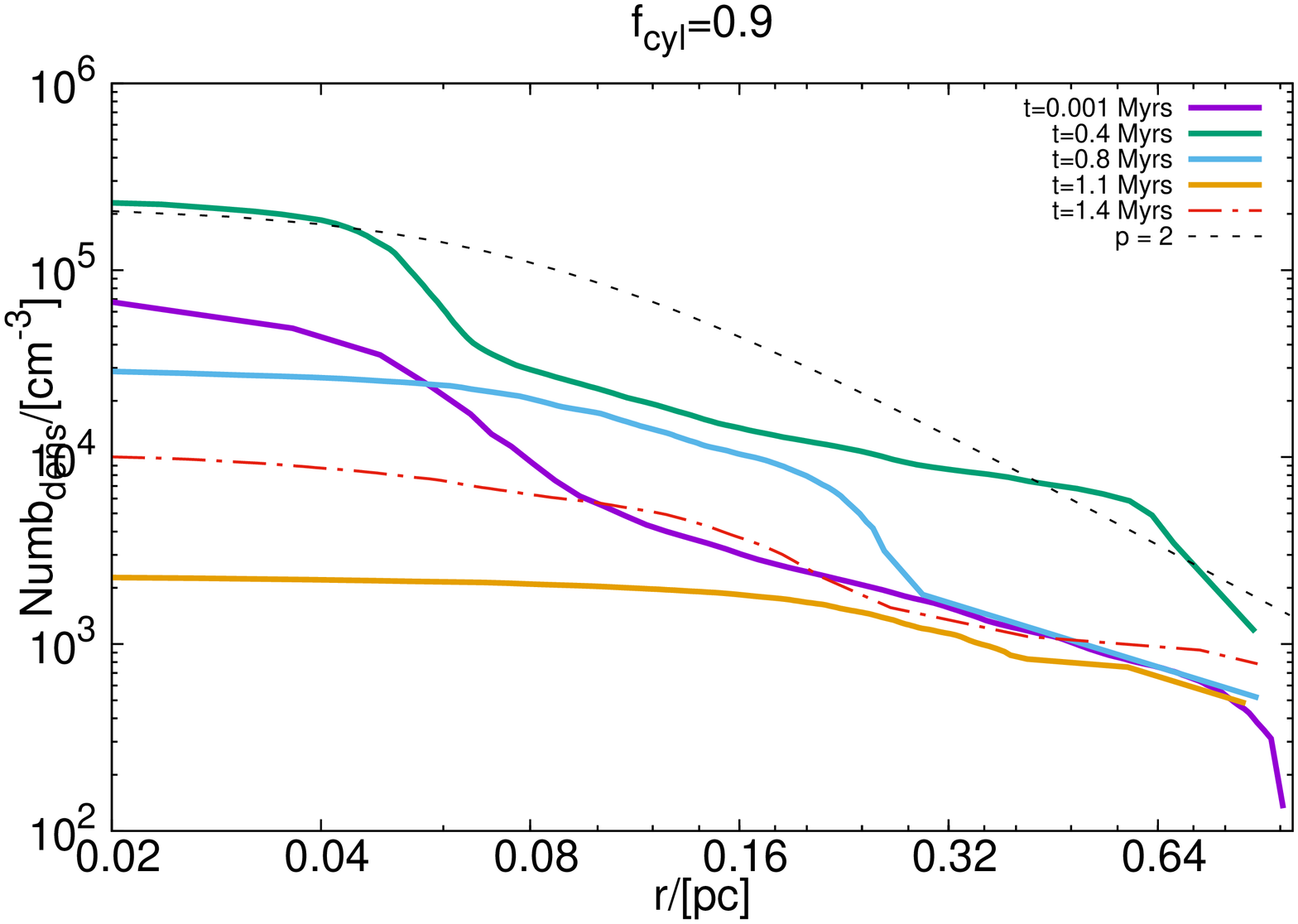}}
  \caption{$\frac{P_{ext}}{k_{B}} = 6.5\times 10^{5}$ K cm$^{-3}$}
\end{subfigure}
\caption{Radial density profile of the filaments at different epochs of their evolution for different choices of external pressure.}
\end{figure*}
%=====================================================
%=============================================
\section{Discussion}
\subsection{Core-formation via fragmentation of the filament / morphology of cores}          In Paper I, we saw that the filament evolution process is much more complicated than suggested by analytic works (e.g., Nagasawa 1987, Inutsuka \& Miyama 1992). \\\\
First, the work by Nagasawa (1987) identifies two possible modes of filament fragmentation - either the \emph{sausage-type} instability (also called deformation instability, when the filament radius is smaller than its scale height), or the Jeans-like \emph{compressional} instability (when the filament radius is greater than its scale height). This work further suggests that the so-called deformation instability occurs when the contribution due to external pressure dominates the net energy budget of the filament. When the contribution due to $P_{ext}$ is relatively small such that the filament evolution is governed by the interplay between its self-gravity and the thermal energy, however, the compressional instability dominates. \\\\
Second, in the Inutsuka \& Miyama (1992) paradigm, an initially super-critical filament confined by a relatively small external pressure rapidly collapses towards its axis without fragmenting. The thin post-collapse filament then fragments due to the compressional instability. Transcritical filaments confined by a relatively large external pressure, fragment to form approximately spherical cores in this paradigm. Thus the fragmentation timescale of extended objects such as filaments relative to the timescale of their global collapse is absolutely crucial while studying the formation of cores. In a more recent contribution to the subject, Pon \emph{et al.} (2011) showed analytically that the growth of small localised perturbations is especially favoured over a global collapse in filamentary geometry. Similarly while the work by Fischera \& Martin (2012), where filaments were idealised as infinitely long cylinders clearly demonstrated the pressure dependence of various physical parameters, the impact of external pressure on filament fragmentation and on the core morphology was far from clear.\\\\
In reality, the picture of filament evolution is much more complex because it is naturally affected by the filament-linemass, the magnitude of external pressure and the gas thermodynamics in a filament. Contrary to the suggestions by Pon \emph{et al.} (2011), in the present work we observe that the initially perturbed filament contracts radially irrespective of its linemass or the magnitude of external pressure. This behaviour occurs because dynamic cooling of gas within the filament and dissipation of the initial turbulence disturbs the equilibrium of the initial set-up. The filamentary fragment then continues to contract until it reaches a temporal maximum in density and rebounds thereafter. This cycle continues till a stable equilibrium state is found during which time instabilities grow along its axis (see also for e.g., Anathpindika \& Freundlich 2015). \emph{The choice of the initial linemass, $f_{cyl}$, and the magnitude of $P_{ext}$ do not have much bearing on the course of evolution of the filament, though they alter the timescale of core formation, and the morphology of the cores that eventually form.} \\\\
As can be seen through the series of rendered density images in Figs. 3 and 6(a) and 6(b) it is evident that the initially sub-critical fragment confined by a relatively small external pressure evolves on a fairly long timescale before cores begin to form along its length. This timescale reduces significantly for filaments experiencing external pressures roughly on the order of that in the Solar neighbourhood. Also, at a given external pressure cores form faster in the initially transcritical filament as is evident from the images corresponding to Cases 3, 6 and 9 in Figs. 3, 6(a) and (b). For external pressure $\lesssim 10^{4}$ K cm$^{-3}$ (as in Cases 1 and 4), broad cores form via the so-called \emph{Collect-and-Collapse} mode on a timescale comparable to or greater than the \emph{e-folding} time defined by Eqn. (6). At higher pressure, comparable to that in the Solar neighbourhood, the filament evolves on a timescale relatively shorter than the \emph{e-folding} time and the resulting pinched cores form via Jeans-like fragmentation as is evident from the images in Fig. 6(b). \\\\
While the cores that formed form via the \emph{collect-and-collapse} mode are generally prolate, barely a couple of them appear approximately spherical. This observation is consistent with the finding of Heigl \emph{et al.} (2018) who demonstrated that a sub-critical filament favours the formation of broad prolate cores, while pinched oblate cores are likely to be spawned by a filament that is supercritical. Indeed, we observe the formation of pinched cores via the fragmentation of an initially transcritical filament confined by a pressure similar to that of the Solar neighbourhood, i.e., $\sim 10^{5}$ K cm$^{-3}$ (as in Cases 8 and 9). In fact, images corresponding to Case 9, for example, in Figs. 4(b) and 6(b), respectively, show that although the \emph{compressional instability} initially leads to the formation of pinched oblate cores, they soon become prolate as their natal filaments expand. Owing to this expansion, while individual cores do become prolate, as also reflected by the axial ratio (see the axial ratio for Cases 3, 9 and 10) listed in Table 2, they do not become broader than their natal filaments. \emph{In view of this observation, we conclude that cores towards supercritical filaments in the field must be smaller than their natal filaments (i.e., pinched) but they could appear oblate, or sometimes even prolate. If indeed true, a majority of the cores in the field should be prolate, but broad in low pressure environments and pinched in environments similar to the Solar neighbourhood.}\\\\
\begin{figure}
\label{Fig 15}
  \vspace{1pt}
  \centering
  \includegraphics[angle=270,width=0.48\textwidth]{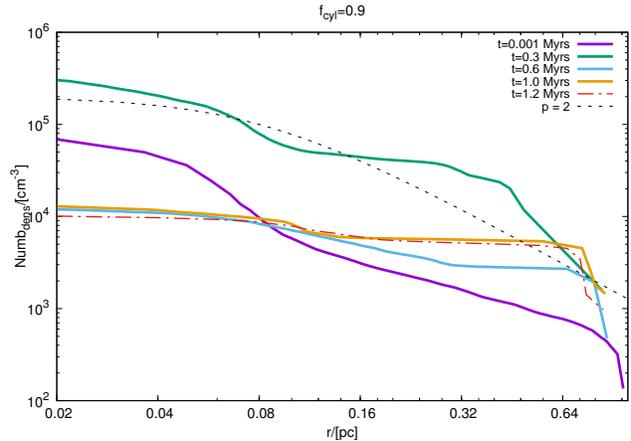}
  \caption{Same as the plots in Fig. 14, but now for $\frac{P_{ext}}{k_{B}} = 2.2\times 10^{6}$ K cm$^{-3}$.}
\end{figure}
Irrespective of the initial choice of linemass and the external pressure, corrugations appear on the surface of the filament. In the case of filaments confined by a relatively small pressure ($P_{ext}/k_{B}\lesssim 10^{4}$ K cm$^{-3}$), these perturbations on the filament surface grow steadily via the accumulation of mass due to the lateral flow of gas within the filament. For an intermediate range of pressure ($10^{4} \le P_{ext}/k_{B}\lesssim 10^{6}$ K cm$^{-3}$), radial contraction of the fragment dominates the amplification of perturbations on the filament surface. Consequently, the filament fragments due to growth of the Jeans-like compressional instability leading to oblate (i.e., \emph{pinched}) cores. On the contrary, such perturbations on an initially sub-critical filament experiencing an intermediate range of pressure ($P_{ext}/k_{B}\sim \mathrm{few}\ x10^{4}$ K cm$^{-3}$) grow rapidly to form wispy, striated structures. For even higher external pressure ($P_{ext}/k_{B}\gtrsim 10^{6}$ K cm$^{-3}$), the surface perturbations amplify rapidly. Although a few cores do form shortly thereafter, only some collapse while others are unstable and rebound (See images on the panels of Fig. 7). \emph{This observation could perhaps help reconcile the observed poverty of star formation in higher pressure environments. We also argue that the so-called geometric fragmentation suggested by Gritschender \emph{et al.} (2017) is a manifestation of the collect-and-collapse mode of core-formation observed here in the case of an initially sub-critical filament in low-pressure environment.} \\\\
Interestingly, from Fig. 5 we note that one of the prolate cores extracted from the fragmented filament in Case 1 has a radial density distribution that resembles the density profile of a thermally supported {\small BE} sphere. In the \emph{collect-and-collapse} mode of core formation/evolution, we envisage such a core steadily acquires mass from its natal filament and then slowly proceeds to collapse. A core that forms in this manner will remain starless for as long as its mass is less than the Jeans mass, or as long as its interior does not cool sufficiently via some mechanism such as dust coagulation (e.g., Anathpindika 2016). \emph{In this manner, we can reconcile the dynamical stability of cores over several free-fall times, or equivalently, the starlessness of these cores without invoking an additional buoyant force such as a strong magnetic field.}
%\begin{figure}
%  \begin{subfigure}{80mm}
%    \centering
%   \vspace*{1pt}   
%   \includegraphics[angle=270,width=\textwidth]{F1P123Seed1-finrendimg.eps}
%   \caption{Filament at the epoch when respective realisations for Cases 1, 2 and 3 were terminated ($f_{cyl}$ = 0.2).}
%  \end{subfigure}
%  \begin{subfigure}{80mm}
%    \centering
%   \vspace*{1pt}
%   \includegraphics[angle=270,width=\textwidth]{F2P123Seed1-finrendimg.eps}
% \caption{Same as the plots in panel (a) above, but now for respectively Cases 4, 5 and 6 ($f_{cyl}$ = 0.5).}    \end{subfigure}
%    \begin{subfigure}{80mm}
 %   \centering
%   \vspace*{1pt}
%   \includegraphics[angle=270,width=\textwidth]{F3P123Seed1-finrendimg.eps}
%    \caption{Same as the plots in panels (a) and (b) above, but now for respectively Cases 7, 8 and 9 ($f_{cyl}$ = 0.9).}
%  \end{subfigure}
 % \caption{Rendered density images showing the mid-plane of the filament at the epoch when the respective realisations were terminated.}
%\end{figure}
\subsubsection{Core morphology} Some observational evidence suggests that oblate pinched cores are likely to be preferentially found in super-critical filaments. Marsh \emph{et al.} (2014) and Palmeirim \emph{et al.} (2013), for example, found this to be the case with cores observed towards some filaments in the Taurus {\small MC}. On the contrary, N$_{2}$ H$^{+}$ and C${}^{18}\textrm{O}$ observations of cores towards the {\small L1495} and {\small B213} filaments also in the Taurus region revealed cores broader than their natal filaments (Hacar \emph{et al.} 2013; Tafalla \& Hacar 2015). In view of the fact that different chemical species have different critical densities (e.g., Tielens 2005), observational inferences about core morphology are likely to be tracer dependent. As was also noted in \S 1, observational studies have been largely unable to establish a clear connection between core-morphology and the ambient physical conditions of filaments.\\\\
In this work, we not only observe that cores are generally spheroidal which is consistent with various observational studies noted in \S 1, but also that prolate broad cores tend to form in filaments that are initially thermally sub-critical and confined by a relatively small external pressure (in the range between a few times $10^{3}$ K cm$^{-3}$ and $\lesssim 10^{4}$ K cm$^{-3}$ as in Cases 1, 2, 4, 5 and 8). Pinched cores on the other hand typically form via the Jeans-type fragmentation due to the compressional instability in initially transcritical filaments confined by an intermediate range of pressure similar to that typically found in nearby {\small MCs}. Interestingly, these pinched cores are not always oblate, but do sometimes become prolate as the natal filament expands. This behaviour is readily evident from the mean axial ratio for cores listed in Table 2 above. \emph{While our conclusions here are consistent with those from other related works, the results discussed here demonstrate the additional constraint of the magnitude of external pressure.}
  \subsubsection{Constraining core-formation}
In an earlier work, Anathpindika \& Whitworth (2008) reported that 72\% of the outflows in some nearby filamentary clouds were within 45$^{\circ}$ of being orthogonal to their natal filaments, while only 28\% were within 45$^{\circ}$ of being parallel to their natal filaments. Furthermore, in view of the fact that the outflows spawned by a core are approximately aligned with the direction of its angular momentum, i.e., orthogonal to the plane of an accretion disk within, they postulated the cores that spawned the outflows had their respective angular momenta roughly orthogonal to the axis of the respective filaments. Indeed, this observation supported the picture in which a network of filaments forms out of the fragmentation of a pressure-confined layer of gas that itself formed out of convergent gas-flows. In this scenario, cores that form along the filament acquire mass as gas is channelled along the filament axis. Cores also acquire a net angular momentum roughly orthogonal to this axis as oppositely directed gas streams accumulate in the local potential well of a putative core. A core thus derives its angular spin from the torque so generated by the oppositely flowing axial gas-flows while spiralling into the potential well; see also Whitworth \emph{et al.} (1995). Rendered images on panels of our Fig. 4(b) in fact show the oppositely moving streams accumulating in a putative core. The collapse of such a core would naturally launch an outflow approximately orthogonal to the axis of its natal filament. \\\\
As has already been noted in the foregoing sections the model filament evolves through a series of radial contractions and expansions, and some expansion can be evident even after fragmentation. It is plausible therefore that such expansion may somewhat alter the orientation of putative outflows relative to the natal filament which could reconcile the inference drawn by Anathpindika \& Whitworth (2008). Interestingly, in this work we have seen that some gas-flow is always aligned along the axis of the filament irrespective of the choice of the initial linemass and the external pressure. Evidently, the nature of gas-flow within a filament must be determined by the manner in which a filament is assembled.\\\\
Works such as those by Banerjee \emph{et al.} (2006), Banerjee \& Pudritz (2007), and Offner \emph{et al.} (2008) suggest an alternative scenario in which the large-scale angular momentum is aligned parallel to the filament axis so that cores spawned by such filaments acquire an angular momentum in the same direction. One can envisage such a possibility in case of filaments assembled via large-scale shear that will impart a net angular momentum to the resulting filaments. In this scenario, the direction of the angular momentum vector would be aligned with the filament-axis and gas feeding putative cores will spiral-in instead of flowing laterally along the filament-axis. Consequently, the angular spin (and by extension, the direction of any putative outflows) of such cores will also be aligned with the axis of the natal filament. In such cases, one might expect a roughly cyclic variation (as may be inferred from Doppler shifts), in the component of velocity gradient orthogonal to the filament-axis. \\\\
Such investigation, however, is beyond the remit of the current work. \emph{The upshot of our work here is that appearance of the deformation instability on the surface of filaments does not alter the axial nature of gas-flow in them. Consequently, cores spawned by these filaments acquire angular momentum approximately orthogonal to the filament axis. This observation therefore favours a scenario in which filaments themselves form via fragmentation of gas-layers assembled due to converging gas-flows.}
\subsection{Velocity gradients : Their periodicity and correlations}
As already seen from the various rendered density images above, the filamentary fragment is attended by the so-called \emph{deformation} instability (Nagasawa 1987), that causes its surface to become corrugated. The crests along these corrugations are the sites where gas accumulates. The process of gas accumulation is accompanied with momentum transfer as the density crests create pressure imbalances which amplifies the perturbations in turn. The greater the external pressure, the faster the process unfolds. Naturally then the amplitude of oscillations is larger for a higher external pressure. Recall that the process is similar to that in the Thin Shell instability (Vishniac 1983). It is interesting to note here that although works, for example, by Nagasawa (1987) and Motiei \emph{et al.} (2021) discuss  stabilisation of filaments by external pressure, the pressure in those calculations is due to the interstellar magnetic field which tends to dampen the growth of gravitational and hydrodynamic instabilities. Conversely, our present work is primarily concerned with the fragmentation of filaments due to growth of the latter type of instability and the impact of interstellar pressure on it.\\\\
Interestingly, not all these density crests condense out as cores. Nonetheless, signatures of these density inhomogeneities along the surface are reflected in the local velocity field, especially in the axial component of the velocity gradient ($\Delta V_{\parallel}$), as is visible from the plots in Fig. 8(a). Such oscillatory features have also often been reported in typical filamentary clouds. Kirk \emph{et al.} (2013), for instance, reported such features for filaments in the Serpens South region. Similar features were also reported by Hacar \emph{et al.} (2013) and Tafalla \& Hacar (2015) in the Taurus {\small L1495/B213} filament, and by Fernand{\' e}z-L{\' o}pez \emph{et al.} (2014) for filaments in the Serpens South region. Multicomponent fits for {\small L1517} in Taurus - Auriga (Hacar \& Tafalla 2011) also showed similar quasiperiodic oscillatory features. Such oscillations may represent lateral gas-flows, parallel to the axis of the filament (e.g., Dhabal \emph{et al.} 2018). \\\\
In a more recent contribution, Chen \emph{et al.} (2020) reported similar quasi-oscillatory velocity gradient features in filaments identified in the {NGC 1333} of the Perseus {\small MC}. In view of the fact that these oscillatory features did not always coincide with regions where condensations occurred along the length of those filaments, they ruled out any association between the occurrence of these quasi-oscillatory features and the gravitational instability leading to the formation of cores. Instead, they conjectured that such features were driven by magnetic waves such as those explored by Tritsis \& Tassis (2016) and Offner \& Liu (2018), for instance. \emph{On the contrary, even the relatively simple hydrodynamic model discussed in this work can reconcile the quasi-oscillatory features. Our realisations demonstrate clearly that such oscillatory features are triggered due to the growth of density perturbations on the surface of the filament as gas flows laterally along the filament axis.}\\\\
In fact, observations from our realisations suggest that the appearance of these oscillatory features must be integrally associated with the filament evolution process. \emph{In view of the discussion in the preceding subsection about possible constraints on core-formation, we suggest that quasi-oscillatory features of the axial and the radial components of the velocity gradient along the filament axis are a useful diagnostic that can be used to understand the physical processes that could have assembled it.} Furthermore, as can be seen from the plots in Figs. 8(a) and 8(b), not only is the magnitude of $\Delta V_{\perp}$ significantly larger than that of $\Delta V_{\parallel}$, but the profile for $\Delta V_{\perp}$ shows comparatively few oscillatory features, irrespective of the magnitude of $P_{\small ICM}$ and the initial linemass. This behaviour is again consistent with similar observational inferences (e.g. Palmeirim \emph{et al.} 2013, Dhabal \emph{et al.} 2018, Chen \emph{et al.} 2020). \\\\
%It is believed that $\Delta V_{\perp}$ is an excellent proxy of ongoing accretion. Interestingly, however,  found no convincing evidence to suggest that filaments in {\small NGC 1333} were indeed accreting gas from their surrounding environment. \\\\
\emph{Our realisations here also show that the few oscillatory features visible in $\Delta V_{\perp}$ can appear even in a purely hydrodynamic realisation of a non-accreting filament merely by virtue of its radial contraction/expansion over the course of its evolution. It is therefore possible to even reconcile similar signatures for filaments in {\small NGC1333}, which may not be necessarily  accreting gas (Chen \emph{et al.} 2020).} Besides the apparent ubiquity of quasi-oscillatory features in either component of the velocity gradient along the length of the filament, we also observe an environmental dependence of the magnitude of these respective components of the velocity gradients. The increasing magnitude of the respective components of velocity gradients is readily evident from the plots in Figs. 8(a) and (b). \\\\
Various panels in Fig. 9 show the radial dependencies of these respective components of velocity gradients. While it is observed that $\Delta V_{\perp}$ increases steeply towards the central axis of the filament irrespective of the magnitude of the external pressure, $P_{\small ICM}$, and its initial linemass, $f_{cyl}$, the contracting filament in the realisations with $\Big(\frac{P_{\small ICM}}{k_{B}}\Big) > 10^{5}$ K cm$^{-3}$ (i.e., Cases 9 and 10 in Fig. 9), is also associated with a readily discernible signature of an inwardly directed compressional wave during the early epochs of its evolution. As is also well known from existing literature (e.g., Inutsuka \& Miyama 1992), the contracting filament in these realisations does not collapse to a singularity despite the steep increase in $\Delta V_{\perp}$ towards its central axis. \\\\
Furthermore, it is also interesting to note that the nature of the respective components of the velocity gradients changes over the course of evolution of the filamentary fragment. The contractional phase is associated with an increasing $\Delta V_{\perp}$ (as well as the $\Delta V_{\parallel}$), towards the central axis of the filament. The situation reverses following the expansion phase. Chen \emph{et al.} (2020) observed similar features in the radial dependence of the respective velocity gradients across some filaments identified in {\small NGC 1333}. \emph{Although they were not able to fully reconcile these observations because of the absence of any convincing evidence to show these filaments were accreting gas, our realisations here associate such oscillatory features in the radial direction with the radial oscillations of a filament. We predict that a wider survey will inevitably show a variety in the radial dependency of $\Delta V_{\perp}$ across filaments because individual filaments in such a survey are likely at different stages of their evolution.} \\\\
A variety of correlations between the respective components of velocity gradients, the velocity dispersion ($\sigma_{gas}$), and the linemass ($f_{cyl}$), are seen in Figs. 11 and 12. Contrary to the observational inferences by Arzoumanian \emph{et al.} (2013), we do not observe here strong correlations between these physical quantities. In fact, we do not even see any clear dependence of these respective correlations on the magnitude of external pressure. Table 3 lists the Spearman rank coefficients for the four representative cases analysed. \\\\
%\emph{Limitations} 
This absence of a linemass-dispersion correlation is probably due to the absence of injected turbulence that usually happens through accretion of gas. We observe a similarly weak correlation between the respective components of velocity gradient. While there is no doubt that the ambient environment affects the magnitude of these respective physical quantities, their association with the linemass is not well established in this work due to the idealised nature of our initial model configuration. Although we observe considerable variation in linemass over the course of evolution of the filament, especially in realisations where it was initially transcritical, there being no mechanism to inject turbulence externally, i.e., the fragment was not allowed to accrete gas, we could not probe if $\Delta V_{\perp}$ is in fact associated with the physical properties of accreted gas.\\\\
\subsubsection{Density profile of filaments}
Filaments with a shallow density profile are ubiquitously found in nearby star forming {\small MCs}. The radial density profiles plotted in Figs. 14 and 15 show that such relatively shallow density profiles can be reconciled with relatively simple hydrodynamic models such as those discussed in this work. In particular, we observed a Plummer-like density profile even in the more massive filaments such as those in Cases 7 - 10 with external pressure in the range between a few times $10^{5}$ K cm$^{-3}$ to a few times $10^{6}$ K cm$^{-3}$.
\section{Conclusions}
In this work, we demonstrated that the magnitude of ambient pressure clearly has a bearing on the morphology of cores as well as on the timescale of their formation. Below are our principal conclusions - 
\begin{enumerate}
  \item Broad prolate cores are more likely to form due to fragmentation of initially sub-critical filaments in low pressure environs.  Pinched cores, on the other hand, form more readily in initially transcritical filaments experiencing relatively high pressure, similar to the magnitude found in the Solar neighbourhood.
  \item Broad cores are observed to form in the initially sub-critical filament via the so-called \emph{collect-and-collapse} mode on a fairly long timescale, roughly an order of magnitude larger than the free-fall time of the filament, but comparable to, or even greater than,the \emph{e-folding} timescale. This observation therefore naturally reconciles the starless nature of some cores and their apparent longevity without having to invoke the magnetic field or any other  support mechanism. 
  \item While broad cores are generally observed to be prolate, pinched cores are oblate soon after their formation but can sometimes become prolate as their natal filament expands. Such cores, however, never become broader than their natal filaments. We therefore suggest that a majority of cores in environments similar to the Solar neighbourhood could be pinched, yet prolate. 
  \item The filament evolves in a quasistatic manner through a series of radial oscillations. In spite of this motion, some gas-flow within the filament remains axial so that the momentum fed to the cores is also similarly in the axial direction. Since a core derives its angular spin from the torque generated by the inflowing gas-flow, naturally then any putative outflows from a core assembled in this manner should be approximately orthogonal to the natal filament.
  \item An initially sub-critical filament experiencing pressure similar to that in the Solar neighbourhood does not form cores and eventually ends up as wispy striations. Density perturbations on the surface of the filament in this case are simply not strong enough to condense out as cores. 
  \item The formation of velocity coherent structures is an integral part of the filament-evolution process. Indeed, some of these sub-filaments also show evidence of core formation.
  \item Although the velocity dispersion in the filament is directly correlated with the magnitude of external pressure, we do not observe a strong correlation between the linemass, the respective components of velocity gradient, and the velocity dispersion. This behaviour, however, is probably due to the non-accreting nature of our test filament.
\end{enumerate}

\textbf{ORCID iDs} \\
Sumedh V. Anathpindika\orcidA{https://orcid.org/0000-0002-8884-806X}\\
James Di Francesco\orcidB{https://orcid.org/0000-0002-9289-2450}\\

%\begin{figure*}
%\vbox to 220mm{\vfil Landscape figure to go here. This figure was
%not part of the original paper and is inserted here for
%illustrative purposes.\\ See the author guide for details (section
%2.2 of \verb|mn2eguide.tex|) on how to handle landscape figures or
%tables. \caption{} \vfil} \label{landfig}
%\end{figure*}
%\begin{enumerate}
%  \item the envelope is spherically symmetric,
%  \item the IR-emitting grains are predominantly of the same kind, and
%  \item in the infrared the absorption efficiency $Q_{\rmn{abs}}
%        (\nu)\propto\nu$,
%\end{enumerate}
\section*{Acknowledgements}
This project was initiated with funds made available under the From {\small GMCs} to stars project ({\small GMCS}/000304/2014), funded by the Department of Science \& Technology, India. Simulations discussed in this work were developed using supercomputing facilities made available by WestGrid (www.westgrid.ca) and Compute Canada Calcul Canada (www.computecanada.ca). We gratefully acknowledge comments of an anonymous referee that helped clarify better some of the results discussed in this work. 

\section*{Data Availability Statement}
No valid data repositories exist as the data generated by numerical simulations discussed in this work are too big to be shared. Instead, we discuss in detail the numerical methods and the initial conditions used to generate these data sets. The initial conditions file and the script of the numerical code can be made available to researchers on reasonable request.

\appendix

%\section[]{Large gaps in L\lowercase{y}${\balpha}$ forests\\* due to fluctuations in line distribution}

%\newpage
%
%\begin{figure}
%\vspace{11pc}
%\caption{$P(>x_{\rmn{gap}})$ as a function of $x_{\rmn{gap}}$ for,
% from left to right, $N=160$, 150, 140, 110, 100, 90, 50, 45 and~40.
% Compare this with \protect\citet{b15}.}
%\label{appenfig}
%\end{figure}

\bsp

\label{lastpage}


\begin{thebibliography}{99}
\bibitem[\protect\citeauthoryear{Alves}{2001}]{b1} Alves, J., Lada, C., \& Lada, E. 2001, Nature, 409, 159
\bibitem[\protect\citeauthoryear{Anathpindika}{2008}]{b2} Anathpindika, S. \& Whitworth, A., 2008, A\&A, 487, 605
\bibitem[\protect\citeauthoryear{Anathpindika}{2011}]{b3} Anathpindika, S. V., 2011a, MNRAS, 405, 1431
\bibitem[\protect\citeauthoryear{Anathpindika}{2011}]{b4} Anathpindika, S. V., 2011b, New Astronomy, 16, 323
\bibitem[\protect\citeauthoryear{Anathpindika}{2015}]{b5} Anathpindika, S \& Freundlich, J., 2015, PASA, 32, 7
\bibitem[\protect\citeauthoryear{Anathpindika}{2016}]{b6} Anathpindika, S., 2016, PASA, 33, 34
\bibitem[\protect\citeauthoryear{Anathpindika}{2021}]{b7} Anathpindika, S. V \& Di Francesco, J., 2021, MNRAS, 502, 564A (Paper I)
\bibitem[\protect\citeauthoryear{Andre}{2010}]{b8} Andr{\' e}, Ph., Men'schikov, A., Bontemps, S., K$\ddot{\mathrm{o}}$nyves, V., Motte, F., Schneider, N \emph{et al.}, 2010, A\&A, 518, L102
%\bibitem[\protect\citeauthoryear{Arzoumanian}{2011}]{b9} Arzoumanian, D., Andr ́{\' e}, Ph., Peretto, N \& K $\ddot{\mathrm{o}}$nyves, V., 2013, A\&A, 553, A119
\bibitem[\protect\citeauthoryear{Attwood}{2007}]{b10} Attwood, R. E., Goodwin, S. P \& Whitworth, A. P., 2007, A\&A, 464, 447
\bibitem[\protect\citeauthoryear{Banerjee}{2006}]{b11} Banerjee, R., Pudritz, R \& Anderson, D. W., 2006, MNRAS, 373, 1091
\bibitem[\protect\citeauthoryear{Banerjee}{2007}]{b12} Banerjee, R \& Pudritz, R., 2007, ApJ, 660, 479
\bibitem[\protect\citeauthoryear{Basu}{2004}]{b13} Basu S \& Ciolek G. E., 2004, ApJ, 607, L39
\bibitem[\protect\citeauthoryear{Bate}{1997}]{b70} Bate, M \& Burkert, A., 1997, MNRAS, 288, 1060
\bibitem[\protect\citeauthoryear{Bate}{2015}]{b14} Bate, M \& Keto, E., 2015, MNRAS, 449, 2643
\bibitem[\protect\citeauthoryear{Bodenheimer}{1992}]{b33} Bodenheimer, P., 1992, in Tenorio-Tagle, G., Prieto M., S{\' a}nchez F., Eds., Stare Formation in Stellar Systems. Cambdridge Univ. Press, Cambridge, p. 1 
\bibitem[\protect\citeauthoryear{Boss}{1993}]{b15} Boss, A., 1993, in Sahade J., McCluskey, G. E., Konodo T., Eds, The Realm of Interacting Binary Stars. Kluwer, Dordrecht, p. 355
\bibitem[\protect\citeauthoryear{Ciolek}{2006}]{b16} Ciolek G. E \& Basu S., 2006, ApJ, 652, 442
\bibitem[\protect\citeauthoryear{Chen}{2014}]{b17} Chen, Che-Yu \& Ostriker, E. C., 2014, ApJ, 785, 69
\bibitem[\protect\citeauthoryear{Chen}{2020}]{b18} Chen Chun-Yuan, M., Di Francesco, Rosolowsky, E., Keown, J, Pineda, J \emph{et al.}, 2020, ApJ, 891, 84

\bibitem[\protect\citeauthoryear{Clarke}{1992}]{b68} Clarke, C. J., 1992, in Duquennoy, A., Major M., Eds., Binaries as Tracers of Stellar Formation. Cambridge University Press, Cambridge, p. 38 
\bibitem[\protect\citeauthoryear{Csengeri}{2013}]{b19} Csengeri, T., Bontemps, S., Schneider, N., Motte, F \emph{et. al.,} 2013, ApJL, 740, L5
\bibitem[\protect\citeauthoryear{Curry}{2002}]{b20} Curry, C. L., 2002, ApJ, 576, 849
\bibitem[\protect\citeauthoryear{Dhabal}{2018}]{b21}Dhabal, A., Mundy, L. G., Rizzo, M. J., Storm, S., \& Teuben, P. 2018, ApJ, 853, 169
\bibitem[\protect\citeauthoryear{Fernandez}{2014}]{b22} Fern{\' a}ndez - L{\' o}pez, M., Arce, H. G., Looney, L., Mundy, L. G \emph{et al.}, 2014, ApJL,790, L19 
\bibitem[\protect\citeauthoryear{Fischera}{2012}]{b72} Fischera, J \& Martin, P. G., 2012, A\&A, 542, A77
\bibitem[\protect\citeauthoryear{Friesen}{2002}]{b23} Friesen, R. K., Medeiros, L., Schnee, S. \emph{et al.}, 2013, MNRAS, 436, 1513
\bibitem[\protect\citeauthoryear{Gammie}{2003}]{b24} Gammie C. F., Lin Y.-T., Stone J. M \& Ostriker E. C., 2003, ApJ, 592, 203
\bibitem[\protect\citeauthoryear{Goldsmith}{2008}]{b25} Goldsmith, P. F., Heyer, M., Narayanan, G.,  Snell, R., Li, Di \& Brunt, C., 2008, ApJ, 680, 428
\bibitem[\protect\citeauthoryear{Goodwin}{2002}]{b26} Goodwin, S., Ward-Thompson, D \& Whitworth, A., 2002, MNRAS, 330, 769
\bibitem[\protect\citeauthoryear{Gritschneder}{2017}]{b27} Gritschneder, M., Heigl, S \& Burkert, A., 2017, ApJ, 834, 202
\bibitem[\protect\citeauthoryear{Hacar}{2011}]{b28} Hacar, A \& Tafalla, M., 2011, A\&A, 533, A34
\bibitem[\protect\citeauthoryear{Hacar}{2013}]{b29} Hacar, A., Tafalla, M., Kauffmann, J \& Kov{\ 'a}cs, A., 2013, A\& A, 554, A55
\bibitem[\protect\citeauthoryear{Heyer}{2016}]{b30} Heyer, M., Goldsmith, P., Yildiz, U., Snell, R., Falgarone, E \& Pineda, J., 2016, MNRAS, 461, 3918 
\bibitem[\protect\citeauthoryear{Heigl}{2016}]{b71}Heigl, S., Gritschneder, M \& Burkert, A., 2016, MNRAS, 463, 4301
\bibitem[\protect\citeauthoryear{Heigl}{2018}]{b31} Heigl, S., Gritschneder, M \& Burkert, A., 2018, MNRAS, 481, L1
\bibitem[\protect\citeauthoryear{Hubber}{2006}]{b32} Hubber, D., Goodwin, S \& Whitworth, A., 2006, A\&A, 450, 881
%\bibitem[\protect\citeauthoryear{Hubber}{2011}]{b33} Hubber, D., Batty, C., McLeod, A \& Whitworth, A., 2011, A\&A, 529, 28
\bibitem[\protect\citeauthoryear{Inutsuka}{1992}]{b34}Inutsuka, S \& Miyama, S., 1992, ApJ, 388, 392
\bibitem[\protect\citeauthoryear{Jackson}{2010}]{b35} Jackson, J. M., Finn, S. C., Chambers, E.T., \emph{et al.} 2010, ApJ, 719, L185
\bibitem[\protect\citeauthoryear{Jones}{2001}]{b36} Jones, C. E., Basu, S \& Dubinski, J., 2001, ApJ, 551, 387
\bibitem[\protect\citeauthoryear{Jones}{2002}]{b37} Jones, C. E \& Basu, S., 2002, ApJ, 569, 280
\bibitem[\protect\citeauthoryear{Kirk}{2005}]{b38} Kirk, J. M., Ward-Thompson, D., \& Andr{\' e}, Ph. 2005, MNRAS,360, 1506
\bibitem[\protect\citeauthoryear{Kirk}{2013}]{b39}Kirk, H., Myers, P. C., Bourke, T. L., \emph{et al.} 2013, ApJ, 766, 115
\bibitem[\protect\citeauthoryear{Klessen}{2010}]{b40} Klessen, R \& Hennebelle, P., 2010, A\&A, 520, A17
\bibitem[\protect\citeauthoryear{Lee}{1999}]{b41} Lee, C. W \& Myers, P. C. 1999, ApJS, 123, 233
\bibitem[\protect\citeauthoryear{Li}{2004}]{b42} Li P. S., Norman M. L., Mac Low M.-M. \& Heitsch F., 2004, ApJ, 605, 800
\bibitem[\protect\citeauthoryear{Maret}{2007}]{b43} Maret, S., Bergin, E. \& Lada, C. 2007, ApJ, 670, L25
\bibitem[\protect\citeauthoryear{Marsh}{2014}]{b44} Marsh, K A., Griffin, M. J., Palmeirim, P., Andr{\' e}, Ph, Kirk, J \emph{et al.}, 2014, MNRAS, 439, 3683 
\bibitem[\protect\citeauthoryear{Menschikov}{2010}]{b45} Men'schikov, A., Andr{\' e}, Ph., Didelon, P \emph{et al.}, 2010, A\&A, 518, L103
\bibitem[\protect\citeauthoryear{Molinari}{2010}]{b46} Molinari, S., Swinyard, B., Bally, J \emph{et al.}, 2010, A\& A, 518, L100
\bibitem[\protect\citeauthoryear{Motiei}{2021}]{b73} Motiei, M., Hosseinirad, M \& Abbassi, S., 2021, MNRAS, 502, 6188 
\bibitem[\protect\citeauthoryear{Myers}{1991}]{b47} Myers, P. C., Fuller, G. A., Goodman, A. A \& Benson, P. J., 1991, ApJ, 376, 561
\bibitem[\protect\citeauthoryear{Myers}{2009}]{b47} Myers, P. C., 2009, ApJ, 700, 1609
\bibitem[\protect\citeauthoryear{Nagasawa}{1987}]{b48} Nagasawa, M., 1987, Prog. in Th. Phys., 77, 635
\bibitem[\protect\citeauthoryear{Offner}{2008}]{b49} Offner, S. S. R., Klein, R \& McKee, C., 2008, ApJ, 686, 1174
\bibitem[\protect\citeauthoryear{Offner}{2018}]{b50} Offner, S. S. R \& Liu, Y., 2018, Nature Astronomy, 2, 896
\bibitem[\protect\citeauthoryear{Ostriker}{1964}]{b72} Ostriker, J., 1964, ApJ, 140, 1056
\bibitem[\protect\citeauthoryear{Palmeirim}{2013}]{b51} Palmeirim, P., Andr{\' e}, Ph., Kirk, J., Ward-Thompson, D \emph{et al.}, 2013, A\& A, 550, A38
\bibitem[\protect\citeauthoryear{Peretto}{2014}]{b52}Peretto, N., Fuller, G. A., Andr{\' e}, Ph, Arzoumanian, D., Rivilla, V. M \emph{et al.}, 2014, A\&A, 561, 83
\bibitem[\protect\citeauthoryear{Poidevin}{2014}]{b53} Poidevin, F. \emph{et al.}, 2014, ApJ, 791, 43
\bibitem[\protect\citeauthoryear{Pon}{2011}]{b54} Pon, A., Johnstone, D \& Heitsch, F., 2011, ApJ, 740, 88
%\bibitem[\protect\citeauthoryear{Price}{2008}]{b55} Price, D., 2008, Jour. of Comp. Phys., 227, 10040
\bibitem[\protect\citeauthoryear{Ryden}{1996}]{b56} Ryden, B. S., 1996, BAAS, 28, 886 
\bibitem[\protect\citeauthoryear{Schneider}{1981}]{b57} Schneider, S \& Elmegreen, B. G., 1979, ApJSS, 41, 87
\bibitem[\protect\citeauthoryear{Sturges}{1926}]{b58}Sturges, H. A., 1926, Jour. of American Stat. Assc., 65 - 66
\bibitem[\protect\citeauthoryear{Tafalla}{1998}]{b59} Tafalla, M., Mardones, D., Myers, P. C., Caselli, P., Bachiller, R \& Benson, P. J., 1998, ApJ, 504, 90
\bibitem[\protect\citeauthoryear{Tafalla}{2015}]{b60} Tafalla, M \& Hacar, A., 2015, A\& A, 574, A104
\bibitem[\protect\citeauthoryear{Tassis}{2007}]{b61} Tassis, K., 2007, MNRAS, 379, L50
\bibitem[\protect\citeauthoryear{Tassis}{2009}]{b62} Tassis, K., Dowell, C. D., Hildebrand, R. H., Kirby, L \& Vaillancourt, J. E., 2009, MNRAS, 399, 1681
\bibitem[\protect\citeauthoryear{Tielens}{2005}]{b63} Tielens, A. G. G. M., 2005, \emph{The Physics and Chemistry of the Interstellar Medium}, Cambridge University Press, Cambridge
\bibitem[\protect\citeauthoryear{Tassis}{2015}]{b64} Tritsis A., Panopoulou G. V., Mouschovias T. C., Tassis K \& Pavlidou V., 2015, MNRAS, 451, 4384
\bibitem[\protect\citeauthoryear{Tritsis}{2016}]{b65} Tritsis A. \& Tassis, K., 2016, MNRAS, 462, 3602
\bibitem[\protect\citeauthoryear{Vishniac}{1983}]{b66} Vishniac, E., 1983, ApJ, 274, 152
\bibitem[\protect\citeauthoryear{Whitworth}{1995}]{b67} Whitworth, A., Chapman, S., Bhattal, A \emph{et al.}, 1995, MNRAS, 277, 727
\bibitem[\protect\citeauthoryear{Wuensch}{2001}]{b68} Wuench, R \& Palou$\breve{\mathrm{s}}$, J., 2001, A\&A, 374, 746
\end{thebibliography}
\end{document}